\documentclass[12pt,twoside,english]{extarticle}
\usepackage{lmodern}
\usepackage{helvet}
\usepackage[T1]{fontenc}
\usepackage[latin9]{inputenc}
\usepackage{geometry}
\usepackage{color}
\usepackage{babel}
\usepackage{float}
\usepackage{mathrsfs}
\usepackage{amsmath}
\usepackage{amsthm}
\usepackage{amssymb}
\usepackage{graphicx}
\usepackage{setspace}
\usepackage[unicode=true,pdfusetitle,
 bookmarks=true,bookmarksnumbered=false,bookmarksopen=false,
 breaklinks=true,pdfborder={0 0 0},pdfborderstyle={},backref=false,colorlinks=true]
 {hyperref}

\makeatletter

\providecommand{\tabularnewline}{\\}

\numberwithin{equation}{section}
\numberwithin{table}{section}
\numberwithin{figure}{section}
\usepackage[natbibapa]{apacite}
\theoremstyle{plain}
\newtheorem{assumption}{\protect\assumptionname}
\theoremstyle{plain}
\newtheorem{thm}{\protect\theoremname}[section]
\theoremstyle{plain}
\newtheorem{lem}{\protect\lemmaname}[section]

\usepackage{graphicx}
\usepackage{amsmath}
\usepackage{dcolumn}
\usepackage{listings}
\usepackage{color}
\usepackage{setspace}
\usepackage[normalem]{ulem}  
\definecolor{hellgelb}{rgb}{1,1,0.8}
\definecolor{colKeys}{rgb}{0,0,1}
\definecolor{colIdentifier}{rgb}{0,0,0}
\definecolor{colComments}{rgb}{1,0,0}
\definecolor{colString}{rgb}{0,0.5,0}

\usepackage{lmodern}
\usepackage[T1]{fontenc}
\usepackage{geometry}
\usepackage{fancyhdr}
\usepackage{amsthm}
\usepackage{thmtools}
\usepackage{amsmath}
\usepackage{amssymb}
\usepackage{esint}
\usepackage{multirow}
\usepackage{mathrsfs} 
\usepackage{textgreek}
\usepackage{amsfonts}
\usepackage{amssymb}
\usepackage{mathrsfs} 

\usepackage[USenglish]{isodate}
\usepackage{chngcntr}

\numberwithin{equation}{section}
\numberwithin{table}{section}
\numberwithin{assumption}{section}
\counterwithout{figure}{section}
\counterwithout{table}{section}

\makeatother

  \providecommand{\assumptionname}{Assumption}

  \providecommand{\lemmaname}{Lemma}

  \providecommand{\theoremname}{Theorem}
 
 \providecommand{\theoremname}{Theorem}
 
\usepackage{thmtools}

\newtheoremstyle{MyTheoremstyle}
  {\topsep} 
  {\topsep} 
  {} 
  {} 
  {\bfseries} 
  {.} 
  {.90em} 
  {} 
\theoremstyle{MyTheoremstyle} 
\theoremstyle{MyTheoremstyle} 
\theoremstyle{MyTheoremstyle} 
\theoremstyle{MyTheoremstyle} 
\theoremstyle{MyTheoremstyle}

\declaretheoremstyle[
    headfont=\bfseries,
    notefont=\normalfont,
    bodyfont=\itshape,
    headpunct=\newline,
    headformat={%
        \makebox{\NAME\ \NUMBER\ }{\NOTE}%
    },
]{theorem}

\newlength{\spacelength}
\declaretheoremstyle[
    headfont=\bfseries,
    notefont=\normalfont,
    bodyfont=\itshape,
    headpunct=\newline,
    headformat={%
        \makebox[0pt][l]{\NAME\ \NUMBER\ }\hskip-\spacelength{\NOTE}%
    },
]{theore}

 \usepackage{fancyhdr}
\usepackage{booktabs}
\usepackage{float}
\usepackage{graphicx}
\usepackage{epstopdf}
\usepackage{morefloats}
\usepackage[referable]{threeparttablex}
\usepackage{footnote}

\usepackage{setspace} 
\usepackage{graphicx,color}

\usepackage{listings}
\RequirePackage[T1]{fontenc}
\RequirePackage{ae,fancyvrb}
\DefineVerbatimEnvironment{Sinput}{Verbatim}{fontshape=sl}
\DefineVerbatimEnvironment{Soutput}{Verbatim}{}
\DefineVerbatimEnvironment{Scode}{Verbatim}{fontshape=sl}

\title{\bf Theory of Evolutionary Spectra for Heteroschesdasticity and Autocorrelation Robust Inference in possibly Misspecified and Nonstationary Models}
\author{
\textsc{\textcolor{MyBlue}{Alessandro Casini}}\thanks{Department of Economics and Finance, University of Rome Tor Vergata, Via Columbia 2, Rome 00133, IT. 
Email: 
\texttt{\textcolor{MyBlue}{{alessandro.casini@uniroma2.it}}}.} 
\\
\small{University of Rome Tor Vergata}
}

\date{\small{\today} 
}

 \makeatletter

\numberwithin{equation}{section}
\makeatother

\usepackage{babel}
\addto\captionsenglish{%
}

\usepackage{color}
\usepackage{xcolor}

\renewcommand*{\thesection}{\arabic{section}}

\definecolor{MyRed}{rgb}{0.8,0,0}
\definecolor{MyBlue}{rgb}{0,0,0.7}
\definecolor{Green}{rgb}{0,0.5,0}
\definecolor{hellgelb}{rgb}{1,1,0.8}
\definecolor{colKeys}{rgb}{0,0,1}
\definecolor{colIdentifier}{rgb}{0,0,0}
\definecolor{colComments}{rgb}{1,0,0}
\definecolor{colString}{rgb}{0,0.5,0}

\definecolor{MyLightRed}{rgb}{2.2,0.2,0.4} 
\definecolor{MyLightRed2}{rgb}{0.6,0.2,0.3} 
\definecolor{MyLightRed2temp}{rgb}{0.6,0.2,0.3}
\definecolor{MyLightRed3}{rgb}{0.8,0.1,0.1} 
\definecolor{MyRed}{rgb}{0.7,0.0,0}

\definecolor{MyLigthBlue13}{rgb}{0,0.2,0.7}
 \definecolor{MyLigthBlack}{rgb}{0.2,0.25,0.3} 

\hypersetup{%
  bookmarks=true
  pdftitle = {Title Here},
  pdfsubject = {},
  pdfkeywords = {Keywords},
  pdfauthor = {Alessandro Casini},}
\hypersetup{ 
  colorlinks = {true}, 
  citecolor = {MyBlue},
  linkcolor = {MyRed},
  filecolor= {Green},	
  urlcolor = {MyBlue},
  hyperindex = {true},
  linktocpage = {true},
  linkbordercolor ={1 0 0},
 citebordercolor ={0 1 1},
 urlbordercolor ={0 1 1},
hypertexnames=false
}

\usepackage{bookmark}

\usepackage[bottom]{footmisc}
\usepackage{multibib}
\newcites{ReferencesSupp}{References}

\makeatother

\providecommand{\assumptionname}{Assumption}
\providecommand{\lemmaname}{Lemma}
\providecommand{\theoremname}{Theorem}

\begin{document}
\pagebreak{}

\setcounter{page}{0}

\raggedbottom
\title{\textbf{The Fixed-$\boldsymbol{b}$ Limiting Distribution and the
ERP of HAR Tests Under Nonstationarity}\thanks{I am grateful to Pierre Perron for his support and advice. I thank
the Editor and an anonymous referee for useful comments. I thank Andrew
Chesher, Jungbin Hwang, Oliver Linton, Ulrich M\"{u}ller, Julius
Vainora, Daniel Whilem and seminar participants at University College
London, University of Cambridge and University of Connecticut for
comments.}}
\maketitle
\begin{abstract}
{\footnotesize{}We show that the  limiting distribution of HAR test
statistics under fixed-$b$ asymptotics is not pivotal when the data
are nonstationary (i.e., time-varying autocovariance structure). It
takes the form of a complicated function of Gaussian processes and
depends on the second moments of the relevant series (e.g., of the
regressors and errors for the case of the linear regression model).
Hence, fixed-$b$ inference methods based on stationarity are not
theoretically valid in general. The nuisance parameters entering the
fixed-$b$ limiting distribution can be consistently estimated under
small-$b$ asymptotics but only with nonparametric rate of convergence.
We show that the error in rejection probability (ERP) is an order
of magnitude larger than that under stationarity and is also larger
than that of HAR tests based on HAC estimators under conventional
asymptotics. These theoretical results reconcile with recent finite-sample
evidence  showing that existing fixed-$b$ HAR tests can perform
poorly when the data are nonstationary. They can be conservative under
the null hypothesis and have non-monotonic power under the alternative
hypothesis irrespective of how large the sample size is. Based on
the new nonstationary fixed-$b$ distribution, we propose a feasible
inference method that controls the null rejection rates well regardless
of whether the data are stationary or not and of the strength of the
serial dependence as verified for some representative data-generating
processes in a simple location model.}{\footnotesize\par}
\end{abstract}
 \indent {\bf{JEL Classification}}: C12, C13, C18, C22, C32, C51\\ 
\noindent {\bf{Keywords}}: Asymptotic expansion, Fixed-$b$, HAC standard errors, HAR inference, Long-run variance, Nonstationarity.  

\onehalfspacing
\thispagestyle{empty}

\pagebreak{}

\section{Introduction}

The construction of standard errors robust to autocorrelation and
heteroskedasticity is important for empirical work because economic
and financial time series exhibit temporal dependence. The early literature
focused on heteroskedasticity and autocorrelation consistent (HAC)
estimators of the asymptotic variance of test statistics (or simply
the long-run variance (LRV) of the relevant series) {[}see, e.g.,
\citeauthor{newey/west:87} (\citeyear{newey/west:87}; \citeyear{newey/west:94}),
\citet{andrews:91}, \citet{andrews/monahan:92}, \citet{hansen:92ecma},
\citet{dejong/davidson:00}{]}. This approach aims at devising a good
estimate of the LRV. Over the last twenty years, the literature has
focused on methods based on fixed-$b$ asymptotics. These involve
an inconsistent estimate of the LRV that keeps the bandwidth at a
fixed fraction of the sample size. This approach was initiated by
\citet{Kiefer/vogelsang/bunzel:00} and \citeauthor{Kiefer/vogelsang:02}
(\citeyear{Kiefer/vogelsang:02}; \citeyear{vogelsang/kiefer:2002ET}).
They developed the analysis assuming stationarity and showed that
valid heteroskedasticity and autocorrelation robust (HAR) inference
is feasible even without a consistent estimator of the LRV. Inconsistency
results in a pivotal nonstandard limiting distribution whose critical
value can be obtained by simulations (e.g., a $t$-statistic on a
coefficient in the linear regression model will not follow asymptotically
a standard normal distribution but a distribution involving a ratio
of Gaussian processes). Theoretical results based on asymptotic expansions
suggested that fixed-$b$ HAR test statistics exhibit an error in
rejection probability (ERP) that is smaller than that associated to
test statistics based on HAC estimators {[}see \citet{jansson:04}
and \citet{sun/phillips/jin:08}{]}. This supported extensive finite-sample
evidence in the literature documenting that the fixed-$b$ approach
leads to HAR test statistics with more accurate null rejection rates
when the data are stationary with strong temporal dependence than
those associated to test statistics based on HAC estimators. Since
then the literature has mostly concentrated on various refinements
of fixed-$b$ HAR inference while maintaining the stationarity assumption,
mostly to have tests having null rejection rates closer to the nominal
level. 

Although stationarity rarely holds in economic and financial time
series, the literature has surprisingly ignored investigating the
theoretical and empirical properties of existing fixed-$b$ HAR inference
when stationarity does not hold. By nonstationary we mean non-constant
moments. As in the literature, we consider processes whose sum of
absolute autocovariances is finite. This rules out processes with
unbounded second moments (e.g., unit root). Nonstationarity can occur
for several reasons: changes in the moments of the relevant time series
induced by changes in the model parameters that govern the data (e.g.,
the Great Moderation with the decline in variance for many macroeconomic
variables, the effects of the financial crisis of 2007\textendash 2008
or of the COVID-19 pandemic); smooth changes in the distributions
governing the data that arise from transitory dynamics from one regime
to another. Unfortunately, the theoretical properties of fixed-$b$
HAR inference change substantially when stationarity does not hold.
The contribution of the paper is to establish such theoretical results
and discuss their relevance for inference in empirical work.

We show that the limiting distribution of HAR test statistics under
fixed-$b$ asymptotics is not pivotal when the data are nonstationary.
It takes the form of a complicated function of Gaussian processes
and depends on the second moments of the relevant series. For example,
in the case of the linear regression model, it depends on the second
moments of the regressors and errors. Hence, fixed-$b$ inference
methods based on stationarity are not theoretically valid in general.
The nuisance parameters entering the fixed-$b$ limiting distribution
can be consistently estimated under small-$b$ asymptotics but only
with nonparametric rate of convergence. We develop asymptotic expansions
under nonstationarity and we show that the ERP is an order of magnitude
larger than that obtained under stationarity by \citet{jansson:04}
and \citet{sun/phillips/jin:08} {[}cf. $O(T^{-\gamma})$ with $\gamma<1/2$
versus $O(T^{-1})$ where $T$ is the sample size{]}. Further, we
show that the ERP of fixed-$b$ HAR tests is also larger than that
of HAR tests based on HAC estimators. It follows from our results
that if one uses fixed-$b$ methods based on the pivotal fixed-$b$
limiting distribution obtained under stationarity but the data are
nonstationary, then the ERP does not even converge to zero as the
sample size increases because that is not the correct limiting distribution.
Hence, our results provide formal support to the claims in \citet{ibragimov/muller:10}
and \citet{mueller:14} who mentioned that the stationarity assumption
required by existing fixed-$b$ methods can be a limitation. 

The pivotal property breaks down because under nonstationarity the
LRV estimator that uses a fixed-bandwidth is not asymptotically proportional
to the LRV. A non-pivotal limiting distribution results in a much
more complex type of inference in practice. The increase in the ERP
from the stationary case arises from fact that the nuisance parameters
have to be estimated. It is the discrepancy between these estimates
and their probability limits that is reflected in the leading term
of the asymptotic expansion.

Our theoretical results reconcile with recent finite-sample evidence
that showed that fixed-$b$ HAR tests can perform poorly when the
data are nonstationary. These  issues have been documented extensively
by \textcolor{MyBlue}{Belotti et al.} \citeyearpar{belotti/casini/catania/grassi/perron_HAC_Sim_Bandws},
\citet{casini_hac}, \citet{casini/perron_PrewhitedHAC} and \citet{casini/perron_Low_Frequency_Contam_Nonstat:2020}
who considered $t$-tests in the linear regression models as well
as HAR tests outside the linear regression model, and a variety of
data-generating processes. They provided  evidence that existing
fixed-$b$ HAR tests can be severely undersized and can exhibit non-monotonic
power. The more nonstationary the data are, the stronger the distortions.
This is especially visible in HAR inference contexts characterized
by a stationary null hypothesis and a nonstationary alternative hypothesis
{[}e.g., tests for structural breaks, tests for regime-switching,
tests for time-varying parameters and threshold effects, and tests
for forecast evaluation{]}. In such cases, the power of fixed-$b$
HAR tests can be zero irrespective of how large the sample size is
and how far the alternative is from the null value. 

In our simulation analysis we focus on HAR inference in the linear
regression model. The empirical results corroborate the predictions
of our ERP results as the existing fixed-$b$ method yields substantial
under-rejections. On the other hand, LRV estimators using small-$b$
bandwidths and standard asymptotic distributions avoid these under-rejection
issues but likely exhibit over-rejections in the context of strongly
persistent (stationary or nonstationary) processes. To address these
problems, we propose a feasible inference method based on the non-pivotal
nonstationary fixed-$b$ limiting distribution that involves replacing
the nuisance parameters by nonparametric estimates and obtaining the
critical values by simulating the limiting distribution. For some
representative data-generating processes in a simple location model
the new method leads to HAR test statistics with accurate null rejection
rates irrespective of whether the data are stationary or not and of
the strength of the serial dependence.

Recent works in HAR inference {[}see, e.g., \citet{sun:14} and \citet{lazarus/lewis/stock/watson:18}{]}
considered the use of small-$b$ asymptotics (i.e., small-bandwidths)
in conjunction with fixed-$b$ critical values. These bandwidths are
typically larger than the MSE-optimal bandwidths used for the HAC
estimators. The idea is that as $b_{T}\rightarrow0$ the fixed-$b$
limiting distribution approximates the standard asymptotic distribution
based on small-$b$ asymptotics. Although our results are obtained
for fixed-bandwidths, they might suggest that using the critical values
from the new fixed-$b$ limiting distribution would improve the finite-sample
performance under nonstationarity. This is an interesting research
question which, however, deserves its own research. 

The remainder of the paper is organized as follows. Section \ref{Section: Statistical Enviromnent}
introduces the statistical problem in the well-known setting of the
linear regression model. In Section \ref{Section: Fixed--Limiting-Distribution}
we study the limiting distribution of $t$- and $F$-type test statistics.
Section \ref{Section ERP} develops the asymptotic expansions and
presents the results on the ERP. Section \ref{Section: Finite-Sample-Effectiveness}
presents Monte Carlo simulations. Section \ref{Section Conclusions}
concludes the paper. The supplemental material {[}cf. \citet{casini_fixed_b_erp_supp}{]}
contains all mathematical proofs. 

\section{HAR Testing in the Linear Regression Model\label{Section: Statistical Enviromnent}}

We consider the linear regression model
\begin{align}
y_{t} & =x'_{t}\beta_{0}+e_{t},\qquad t=1,\,2,\ldots,\,T,\label{Eq. (1) KV (2002)}
\end{align}
 where $\beta_{0}\in\Theta\subset\mathbb{R}^{p}$, $y_{t}$ is an
observation on the dependent variable, $x_{t}$ is a $p$-vector of
regressors and $e_{t}$ is an unobserved disturbance that is autocorrelated
and possibly conditionally heteroskedastic, and $\mathbb{E}(e_{t}|\,x_{t})=0$.
The problem addressed is testing linear hypotheses about $\beta_{0}$.
We consider the ordinary least squares (OLS) estimator $\widehat{\beta}=(\sum_{t=1}^{T}x_{t}x'_{t})^{-1}\sum_{t=1}^{T}x_{t}y_{t}$.
Let $V_{t}=x_{t}e_{t}$. Define $S_{\left\lfloor Tr\right\rfloor }=\sum_{t=1}^{\left\lfloor Tr\right\rfloor }V_{t}$
where $\left\lfloor Tr\right\rfloor $ denotes the integer part of
$Tr$. Using ordinary manipulations,
\begin{align*}
\sqrt{T}\left(\widehat{\beta}-\beta_{0}\right) & =\left(T^{-1}\sum_{t=1}^{T}x_{t}x'_{t}\right)^{-1}T^{-1/2}S_{T}.
\end{align*}
 The variance of $T^{-1/2}S_{T}$ plays an important role for constructing
tests about $\beta_{0}$. Its exact formula depends on the assumptions
about $\left\{ V_{t}\right\} $. We begin with the following notational
conventions.

A function $g\left(\cdot\right):\,\left[0,\,1\right]\mapsto\mathbb{R}$
is said to be piecewise (Lipschitz) continuous if it is (Lipschitz)
continuous except on a set of discontinuity points that has zero Lebesgue
measure. A matrix is said to be piecewise (Lipschitz) continuous if
each of its element is piecewise (Lipschitz) continuous. Let $W_{p}\left(r\right)$
denote a $p$-vector of independent standard Wiener processes where
$r\in\left[0,\,1\right]$. We use $\overset{\mathbb{P}}{\rightarrow},\,\Rightarrow$
and $\overset{d}{\rightarrow}$ to denote convergence in probability,
weak convergence and convergence in distribution, respectively. The
following assumptions are sufficient to establish the asymptotic distribution
of the test statistics. Let $\Omega\left(u\right)$ denote some $p\times p$
positive semidefinite matrix. 
\begin{assumption}
\label{Assumption: Assumption 1 in KV (2002), S_Tr, Nonstationarity}$T^{-1/2}S_{\left\lfloor Tr\right\rfloor }\Rightarrow\int_{0}^{r}\Sigma\left(u\right)dW_{p}\left(u\right)$
where $\Sigma\left(u\right)$ is  given by the Cholesky decomposition
$\Omega\left(u\right)=\Sigma\left(u\right)\Sigma\left(u\right)'$
and is piecewise continuous with  $\sup_{u\in\left[0,\,1\right]}\left\Vert \Sigma\left(u\right)\right\Vert <\infty$.
\end{assumption}
\begin{assumption}
\label{Assumption: Assumption 2 in KV (2002), Qr, Nonstationarity}$T^{-1}\sum_{t=1}^{\left\lfloor Tr\right\rfloor }x_{t}x'_{t}\overset{\mathbb{P}}{\rightarrow}\int_{0}^{r}Q\left(u\right)du$
uniformly in $r$ where $Q\left(u\right)$ is piecewise continuous
with  $\sup_{u\in\left[0,\,1\right]}\left\Vert Q\left(u\right)\right\Vert <\infty$. 
\end{assumption}
Assumption \ref{Assumption: Assumption 1 in KV (2002), S_Tr, Nonstationarity}
states a functional law for nonstationary processes {[}see, e.g.,
\citet{aldous:1978} and \citet{merlevede/peligra/utev:2019}{]}.
If $\left\{ V_{t}\right\} $ is second-order stationary, then $\Sigma\left(u\right)=\Sigma$
for all $u$ and Assumption \ref{Assumption: Assumption 1 in KV (2002), S_Tr, Nonstationarity}
reduces to $T^{-1/2}S_{\left\lfloor Tr\right\rfloor }\Rightarrow\Sigma W_{p}\left(r\right)$.
The fixed-$b$ literature has routinely used the assumption of second-order
stationarity {[}see, e.g., \citet{Kiefer/vogelsang/bunzel:00}, \citet{jansson:04},
\citet{sun/phillips/jin:08} and \citet{lazarus/lewis/stock:17}{]}.
We relax this assumption substantially as we allow for general time-variation
in the second moments of the regressors and errors which encompasses
most of the nonstationary processes used in econometrics and statistics.
For example, it allows for structural breaks, regime-switching, time-varying
parameters and segmented local stationarity in the second moments
of $\{V_{t}\}$. With regards to the temporal dependence, Assumption
\ref{Assumption: Assumption 1 in KV (2002), S_Tr, Nonstationarity}
holds under a variety of regularity conditions. For example, standard
mixing conditions and (time-varying) invertible ARMA processes are
allowed. 

Assumption \ref{Assumption: Assumption 2 in KV (2002), Qr, Nonstationarity}
allows for structural breaks as well as smooth variation in the second
moments of the regressors.\footnote{Assumption \ref{Assumption: Assumption 2 in KV (2002), Qr, Nonstationarity}
also allows for polynomial trending regressors as long as they are
written in the form $(t/T)^{l}$ ($l\geq0$), or more generally, written
as a piecewise continuous function of the time trend, say $g(t/T)$.} The fixed-$b$ literature required $Q\left(u\right)=Q$ for all $u$
in which case Assumption \ref{Assumption: Assumption 2 in KV (2002), Qr, Nonstationarity}
reduces to $T^{-1}\sum_{t=1}^{\left\lfloor Tr\right\rfloor }x_{t}x'_{t}\overset{\mathbb{P}}{\rightarrow}rQ$.
The latter is quite restrictive in practice. The uniform convergence,
boundness and positive definiteness of $Q\left(\cdot\right)$ are
satisfied for a fairly general class of processes. As in previous
works, Assumption \ref{Assumption: Assumption 1 in KV (2002), S_Tr, Nonstationarity}-\ref{Assumption: Assumption 2 in KV (2002), Qr, Nonstationarity}
rule out unit roots and long memory. Let 
\begin{align}
\mathrm{Var}\left(T^{-1/2}S_{T}\right) & =\sum_{k=-T+1}^{T-1}\Gamma_{T,k},\qquad\Gamma_{T,k}=\begin{cases}
T^{-1}\sum_{t=k+1}^{T}\mathbb{E}(V_{t}V'_{t-k}) & \mathrm{for\,}k\geq0\\
T^{-1}\sum_{t=-k+1}^{T}\mathbb{E}(V_{t+k}V'_{t}) & \mathrm{for\,}k<0
\end{cases}.\label{Eq. (2) KV (2002), Var ST}
\end{align}
Under Assumption \ref{Assumption: Assumption 1 in KV (2002), S_Tr, Nonstationarity}-\ref{Assumption: Assumption 2 in KV (2002), Qr, Nonstationarity},
the limit of $\mathrm{Var}(T^{-1/2}S_{T})$ is given by {[}cf. \citet{casini_hac}{]}
\begin{align*}
\lim_{T\rightarrow\infty}\mathrm{Var}\left(T^{-1/2}S_{T}\right)\triangleq\Omega & =\int_{0}^{1}c\left(u,\,0\right)du+\sum_{k=1}^{\infty}\int_{0}^{1}\left(c\left(u,\,k\right)+c\left(u,\,k\right)'\right)du,
\end{align*}
 where $c\left(u,\,k\right)=\mathbb{E}(V_{\left\lfloor Tu\right\rfloor }V_{\left\lfloor Tu\right\rfloor -k})+O(T^{-1})$.
By the Cholesky decomposition $\Omega\left(u\right)=\Sigma\left(u\right)\Sigma\left(u\right)'$
and so $\Omega=\int_{0}^{1}\Omega\left(u\right)du$. Note that $\Omega=2\pi\int_{0}^{1}f\left(u,\,0\right)du$
where $f\left(u,\,0\right)$ is the local spectral density matrix
of $\left\{ V_{t}\right\} $ at rescaled time $u$ and frequency $0$.
For $u$ a continuity point, $f\left(u,\,\omega\right)$ is defined
implicitly by the relation $\mathbb{E}(V_{\left\lfloor Tu\right\rfloor }V_{\left\lfloor Tu\right\rfloor -k})=\int_{-\pi}^{\pi}e^{i\omega k}f\left(u,\,\omega\right)d\omega$;
see \citet{casini_hac} for more details. If $\left\{ V_{t}\right\} $
is second-order stationary, then $\Omega=\Sigma\Sigma'=2\pi f\left(0\right)$
since $f(u,\,0)=f(0)$. 

Under Assumption \ref{Assumption: Assumption 1 in KV (2002), S_Tr, Nonstationarity}-\ref{Assumption: Assumption 2 in KV (2002), Qr, Nonstationarity},
it directly follows, using standard arguments, that 
\begin{align}
\sqrt{T}\left(\widehat{\beta}-\beta_{0}\right) & \overset{d}{\rightarrow}\overline{Q}^{-1}\Omega^{1/2}W_{p}\left(1\right)\sim\mathscr{N}\left(\mathbf{0},\,\overline{Q}^{-1}\Omega\overline{Q}^{-1}\right),\label{Eq. (4) KV (2002)}
\end{align}
where $\Omega^{1/2}$ is the matrix square-root of $\Omega$ and $\overline{Q}\triangleq\int_{0}^{1}Q\left(u\right)du$.
Under second-order stationarity $Q\left(u\right)=Q$, $\Sigma\left(u\right)=\Sigma$
and \eqref{Eq. (4) KV (2002)} reduces to
\begin{align}
\sqrt{T}\left(\widehat{\beta}-\beta_{0}\right) & \overset{d}{\rightarrow}Q^{-1}\Sigma W_{p}\left(1\right)\sim\mathscr{N}\left(\mathbf{0},\,Q^{-1}\Omega Q^{-1}\right).\label{Eq. (4) KV (2002)-1}
\end{align}
 The classical approach to testing hypotheses about $\beta_{0}$ is
based on studentization. Provided that a consistent estimator of
$\overline{Q}^{-1}\Omega\overline{Q}^{-1}$ can be constructed, it
is possible to construct a test statistic whose asymptotic distribution
is free of nuisance parameters. The term $\overline{Q}$ can be consistently
estimated straightforwardly using $\widehat{Q}=T^{-1}\sum_{t=1}^{T}x_{t}x'_{t}$.
Consistent estimators of $\Omega$ are known as HAC estimators {[}see,
e.g., \citet{newey/west:87}, \citet{andrews:91}, \citet{dejong/davidson:00}
and \citet{casini_hac}{]}. HAC estimators take the following general
form, 
\begin{align*}
\widehat{\Omega}_{\mathrm{HAC}}\triangleq\sum_{k=-T+1}^{T-1}K\left(b_{T}k\right)\widehat{\Gamma}\left(k\right) & ,
\end{align*}
where
\begin{align}
\widehat{\Gamma}\left(k\right)=\begin{cases}
T^{-1}\sum_{t=k+1}^{T}\widehat{V}_{t}\widehat{V}'_{t-k} & \mathrm{for\,}k\geq0\\
T^{-1}\sum_{t=-k+1}^{T}\widehat{V}_{t+k}\widehat{V}'_{t} & \mathrm{for\,}k<0
\end{cases} & ,\label{Eq. Definition of Gamma(k)}
\end{align}
 $\widehat{V}_{t}=x_{t}\widehat{e}_{t}$ and $\left\{ \widehat{e}_{t}\right\} $
are the OLS residuals, $K\left(\cdot\right)$ is a  kernel and $b_{T}$
is a bandwidth sequence. Under $b_{T}\rightarrow0$ at an appropriate
rate, we have $\widehat{\Omega}_{\mathrm{HAC}}\overset{\mathbb{P}}{\rightarrow}\Omega.$
An alternative to $\widehat{\Omega}_{\mathrm{HAC}}$ is the double-kernel
HAC (DK-HAC) estimator, say $\widehat{\Omega}_{\mathrm{DK-HAC}}$,
proposed by \citet{casini_hac} to flexibly account for nonstationarity.
$\widehat{\Omega}_{\mathrm{DK-HAC}}$ uses an additional kernel for
smoothing over time; see \citet{casini_hac} for details.  Under
appropriate conditions on the bandwidths, we have $\widehat{\Omega}_{\mathrm{DK-HAC}}\overset{\mathbb{P}}{\rightarrow}\Omega$.
Hence, equipped with either $\widehat{\Omega}_{\mathrm{HAC}}$ or
$\widehat{\Omega}_{\mathrm{DK-HAC}}$, HAR inference is standard because
test statistics follow asymptotically standard distributions. 

An alternative approach to HAR inference relies on inconsistent estimation
of $\Omega.$ \citeauthor{Kiefer/vogelsang:02} (\citeyear{Kiefer/vogelsang:02},
\citeyear{vogelsang/kiefer:2002ET}, \citeyear{kiefer/vogelsang:05})
proposed to use the following ``estimator'',\footnote{As a notational matter, it is useful to remark that the more recent
fixed-$b$ literature does not refer to $\widehat{\Omega}_{\mathrm{fixed}-b}$
as an estimator. This recent literature rather uses the terminology
``fixed-$b$'' to refer to an asymptotic embedding. We may sometime
refer to $\widehat{\Omega}_{\mathrm{fixed}-b}$ as an estimator. This
should not create any confusion since our results are provided for
the case of $b$ fixed which corresponds to the early fixed-$b$ literature.} 
\begin{align}
\widehat{\Omega}_{\mathrm{fixed}-b}\triangleq\sum_{k=-T+1}^{T-1}K\left(\frac{k}{bT}\right)\widehat{\Gamma}\left(k\right) & ,\label{Eq.: Fixed-b Estimator}
\end{align}
where $b\in(0,\,1]$ is fixed. Note that $\widehat{\Omega}_{\mathrm{fixed}-b}$
is equivalent to $\widehat{\Omega}_{\mathrm{HAC}}$ with $b_{T}=\left(bT\right)^{-1}$.
$\widehat{\Omega}_{\mathrm{fixed}-b}$ is inconsistent for $\Omega$.
 \citet{Kiefer/vogelsang/bunzel:00} showed that an asymptotic distribution
theory for HAR tests is possible even with an inconsistent estimate
of $\Omega.$ One first has to derive the limiting distribution of
$\widehat{\Omega}_{\mathrm{fixed}-b}$ under the null hypothesis.
Then, one can use it to obtain the limiting distribution of the test
statistic of interest which typically involves a ratio of Gaussian
processes. Thus, from the inconsistency of $\widehat{\Omega}_{\mathrm{fixed}-b}$,
HAR test statistics will not follow asymptotically standard distributions. 

\section{\label{Section: Fixed--Limiting-Distribution}Fixed-$\boldsymbol{b}$
Limiting Distribution of HAR Tests }

In this section we study the limiting distribution of the HAR tests
for linear hypothesis in the linear regression model under fixed-$b$
asymptotics when the data are nonstationary. Consider testing the
null hypothesis $H_{0}:\,R\beta_{0}=r$ against the alternative hypothesis
$H_{1}:\,R\beta_{0}\neq r$ where $R$ is a $q\times p$ matrix with
rank $q$ and $r$ is a $q\times1$ vector. Using $\widehat{\Omega}_{\mathrm{fixed}-b}$
an $F$-test can be constructed as follows:
\begin{align*}
F_{\mathrm{fixed}-b} & =T\left(R\widehat{\beta}-r\right)'\left[R\widehat{Q}^{-1}\widehat{\Omega}_{\mathrm{fixed}-b}\widehat{Q}^{-1}R'\right]^{-1}\left(R\widehat{\beta}-r\right)/q.
\end{align*}
For testing one restriction, $q=1$, one can use the following $t$-statistic:
\begin{align*}
t_{\mathrm{fixed}-b} & =\frac{T^{1/2}\left(R\widehat{\beta}-r\right)}{\sqrt{R\widehat{Q}^{-1}\widehat{\Omega}_{\mathrm{fixed}-b}\widehat{Q}^{-1}R'}}.
\end{align*}
Let $B_{p}\left(r\right)=W_{p}\left(r\right)-rW_{p}\left(1\right)$
denote the $p\times1$ vector of Brownian bridges. Consider the following
class of kernels, 
\begin{align}
\boldsymbol{K} & =\{K\left(\cdot\right):\,\mathbb{R}\rightarrow\left[-1,\,1\right]:\,K\left(0\right)=1,\,K\left(x\right)=K\left(-x\right),\,\forall x\in\mathbb{R}\label{Eq. (2.6) K1 Kernel class}\\
 & \quad{\textstyle \quad\int\nolimits _{-\infty}^{\infty}}K^{2}\left(x\right)dx<\infty,\,K\left(\cdot\right)\,\mathrm{is\,continuous\,at\,0}\}.\nonumber 
\end{align}
Examples of kernels in $\boldsymbol{K}$ include the Truncated, Bartlett,
Parzen, Quadratic Spectral (QS) and Tukey-Hanning kernels. \citet{kiefer/vogelsang:05}
showed under stationarity that 
\begin{align}
\widehat{\Omega}_{\mathrm{fixed}-b} & \Rightarrow-\Sigma\left(\frac{1}{b^{2}}\int_{0}^{1}\int_{0}^{1}K''\left(\frac{r-s}{b}\right)B_{p}\left(r\right)B_{p}\left(s\right)'drds\right)\Sigma',\label{Eq.: Omega_Fixedb}
\end{align}
for $K\in\boldsymbol{K}$ with $K''\left(x\right)$ assumed to exist
for $x\in\left[-1,\,1\right]$ and to be continuous.\footnote{Note that $K''\left(x\right)$ does not exist for some popular kernels.
This is the case for the  Bartlett kernel for which $K''\left(0\right)$
does not exist. However, \citet{kiefer/vogelsang:05} showed that
for the Bartlett kernel it holds that 
\begin{align*}
\widehat{\Omega}_{\mathrm{fixed}-b} & \Rightarrow\Sigma\biggl(\frac{2}{b}\int_{0}^{1}B_{p}\left(r\right)B_{p}\left(r\right)'dr\\
 & \quad-\frac{1}{b}\int_{0}^{1-b}\left(B_{p}\left(r+b\right)B_{p}\left(r\right)'+B_{p}\left(r+b\right)B_{p}\left(r\right)'\right)dr\biggr)\Sigma'.
\end{align*}
 Recall that the Bartlett kernel is defined as $K_{\mathrm{BT}}\left(x\right)=1-\left|x\right|$
for $\left|x\right|\leq1$ and $K_{\mathrm{BT}}\left(x\right)=0$
otherwise. } A key feature of the result in \eqref{Eq.: Omega_Fixedb} is that
$\widehat{\Omega}_{\mathrm{fixed}-b}$ is asymptotically proportional
to $\Omega$ through $\Sigma\Sigma'$. The null asymptotic distributions
of $F_{\mathrm{fixed}-b}$ and $t_{\mathrm{fixed}-b}$ under stationarity
are given, respectively, by
\begin{align*}
F_{\mathrm{fixed}-b}\Rightarrow & W_{q}\left(1\right)'\left[-\frac{1}{b^{2}}\int_{0}^{1}\int_{0}^{1}K''\left(\frac{r-s}{b}\right)B_{q}\left(r\right)B_{q}\left(s\right)'drds\right]^{-1}W_{q}\left(1\right)/q,
\end{align*}
and
\begin{align*}
t_{\mathrm{fixed}-b} & \Rightarrow\frac{W_{1}\left(1\right)}{\sqrt{-\frac{1}{b^{2}}\int_{0}^{1}\int_{0}^{1}K''\left(\frac{r-s}{b}\right)B_{1}\left(r\right)B_{1}\left(s\right)drds}}.
\end{align*}
Both null distributions are pivotal. Thus, valid testing is possible
without consistent estimation of $\Omega.$ This result crucially
hinges on stationarity. To see this, consider $t_{\mathrm{fixed}-b}$
for the single-regressor case ($p=1)$ and for the null hypothesis
$H_{0}:\,\beta_{0}=0$. Its numerator and denominator are asymptotically
equivalent to, respectively, $Q^{-1}\Sigma W_{1}\left(1\right)$ and
\begin{align*}
Q^{-1}\Sigma & \left(-\frac{1}{b^{2}}\int_{0}^{1}\int_{0}^{1}K''\left(\frac{r-s}{b}\right)B_{1}\left(r\right)B_{1}\left(s\right)drds\right)^{1/2}.
\end{align*}
Since $W_{1}\left(1\right)$ and $B_{1}\left(r\right)$ are independent,
$t_{\mathrm{fixed}-b}$ is a ratio of two independent random variables.
The factor $Q^{-1}\Sigma$ cancels because it appears in both numerator
and denominator. It follows that the asymptotic null distribution
is pivotal. We show that this argument break downs when  stationarity
does not hold. Under nonstationarity the factor in the denominator
corresponding to $Q^{-1}\Sigma$ will depend on the rescaled time
$s$ and $r$, and enter the integrand. Thus, it will not cancel out.

We now present the results about the fixed-$b$ limiting distribution
of the HAR tests. For $r\in\left[0,\,1\right],$ let 
\begin{align*}
\widetilde{B}_{p}\left(r\right)=\widetilde{B}_{p}\left(r,\,\Sigma,\,Q\right) & \triangleq\int_{0}^{r}\Sigma\left(u\right)dW_{p}\left(u\right)-\left(\int_{0}^{r}Q\left(u\right)du\right)\overline{Q}^{-1}\int_{0}^{1}\Sigma\left(u\right)dW_{p}\left(u\right).
\end{align*}

We begin with the following theorem which provides the limiting distribution
of $\widehat{\Omega}_{\mathrm{fixed}-b}$. 
\begin{thm}
\label{Theorem: Limiting Distribution Omega_fixedb}Let Assumption
\ref{Assumption: Assumption 1 in KV (2002), S_Tr, Nonstationarity}-\ref{Assumption: Assumption 2 in KV (2002), Qr, Nonstationarity}
hold and $K\in\boldsymbol{K}$. Then, we have: (i) If $K''\left(x\right)$
exists for $x\in\left[-1,\,1\right]$ and is continuous, then
\begin{align}
\widehat{\Omega}_{\mathrm{fixed}-b} & \Rightarrow-\frac{1}{b^{2}}\int_{0}^{1}\int_{0}^{1}K''\left(\frac{r-s}{b}\right)\widetilde{B}_{p}\left(r,\,\Sigma,\,Q\right)\widetilde{B}_{p}\left(s,\,\Sigma,\,Q\right)'drds\label{Eq. Asymptotic Distribution of Omega fixed_b}\\
 & \triangleq\mathscr{G}_{b}.\nonumber 
\end{align}
 (ii) If  $K\left(x\right)=K_{\mathrm{BT}}\left(x\right)$, then
\begin{align}
\widehat{\Omega}_{\mathrm{fixed}-b} & \Rightarrow\frac{2}{b}\int_{0}^{1}\left(\widetilde{B}_{p}\left(r,\,\Sigma,\,Q\right)\widetilde{B}_{p}\left(r,\,\Sigma,\,Q\right)'\right)dr\label{Eq. Asymptotic Distribution of Omega fixed_b, Barteltt}\\
 & \quad-\frac{1}{b}\int_{0}^{1-b}\left(\widetilde{B}_{p}\left(r+b,\,\Sigma,\,Q\right)\widetilde{B}_{p}\left(r,\,\Sigma,\,Q\right)'+\widetilde{B}_{p}\left(r,\,\Sigma,\,Q\right)\widetilde{B}_{p}\left(r+b,\,\Sigma,\,Q\right)'\right)dr\nonumber \\
 & \triangleq\mathscr{G}_{\mathrm{BT},b}.\nonumber 
\end{align}
\end{thm}
Theorem \ref{Theorem: Limiting Distribution Omega_fixedb} show that,
unlike in the stationary case, $\widehat{\Omega}_{\mathrm{fixed}-b}$
is not asymptotically proportional to $\Omega$. This anticipates
that asymptotically pivotal tests for null hypotheses involving $\beta_{0}$
cannot be constructed. The limiting distribution depends on $K''\left(\cdot\right),$
$b$ and most importantly on $\Sigma\left(\cdot\right)$ and $Q\left(\cdot\right)$
so that it is not free of nuisance parameters. 

We now present the limiting distribution of $F_{\mathrm{fixed}-b}$
and $t_{\mathrm{fixed}-b}$ under $H_{0}.$ 
\begin{thm}
\label{Theorem 2: KV (2002)}Let Assumption \ref{Assumption: Assumption 1 in KV (2002), S_Tr, Nonstationarity}-\ref{Assumption: Assumption 2 in KV (2002), Qr, Nonstationarity}
hold and $K\in\boldsymbol{K}$. Then, we have: (i) If $K''\left(x\right)$
exists for $x\in\left[-1,\,1\right]$ and is continuous, then 
\begin{align*}
F_{\mathrm{fixed}-b} & \Rightarrow\left(R\overline{Q}^{-1}\int_{0}^{1}\Sigma\left(u\right)dW_{p}\left(u\right)\right)'\left(R\overline{Q}^{-1}\mathscr{G}_{b}\overline{Q}^{-1}R'\right)^{-1}R\overline{Q}^{-1}\int_{0}^{1}\Sigma\left(u\right)dW_{p}\left(u\right)/q,
\end{align*}
where $\mathscr{G}_{b}$ is defined in \eqref{Eq. Asymptotic Distribution of Omega fixed_b}.
If $q=1,$ then 
\begin{align*}
t_{\mathrm{fixed}-b} & \Rightarrow\frac{R\overline{Q}^{-1}\int_{0}^{1}\Sigma\left(u\right)dW_{p}\left(u\right)}{\sqrt{R\overline{Q}^{-1}\mathscr{G}_{b}\overline{Q}^{-1}R'}}.
\end{align*}
 (ii) If the Bartlett kernel is used, $K\left(x\right)=K_{\mathrm{BT}}\left(x\right)$,
then 
\begin{align*}
F_{\mathrm{fixed}-b} & \Rightarrow\left(R\overline{Q}^{-1}\int_{0}^{1}\Sigma\left(u\right)dW_{p}\left(u\right)\right)'\left(R\overline{Q}^{-1}\mathscr{G}_{\mathrm{BT},b}\overline{Q}^{-1}R'\right)^{-1}R\overline{Q}^{-1}\int_{0}^{1}\Sigma\left(u\right)dW_{p}\left(u\right)/q,
\end{align*}
where $\mathscr{G}_{\mathrm{BT},b}$ is defined in \eqref{Eq. Asymptotic Distribution of Omega fixed_b, Barteltt}.
If $q=1,$ then 
\begin{align*}
t_{\mathrm{fixed}-b} & \Rightarrow\frac{R\overline{Q}^{-1}\int_{0}^{1}\Sigma\left(u\right)dW_{1}\left(u\right)}{\sqrt{R\overline{Q}^{-1}\mathscr{G}_{\mathrm{BT},b}\overline{Q}^{-1}R'}}.
\end{align*}
\end{thm}
Theorem \ref{Theorem 2: KV (2002)} shows that the asymptotic distribution
of the $F$ and $t$ test statistics under fixed-$b$ asymptotics
under nonstationarity are not pivotal. This contrasts with the stationary
case where the asymptotic distributions depend only on the kernel
and bandwidth. Consequently,  fixed-$b$ inference based on stationarity
is not theoretically valid under nonstationarity. The limiting distributions
of $F_{\mathrm{fixed}-b}$ and $t_{\mathrm{fixed}-b}$ depend on nuisance
parameters such as the time-varying autocovariance function of $\left\{ V_{t}\right\} $
through $\Sigma\left(\cdot\right)$ and the second moments of the
regressors through $Q\left(\cdot\right)$. An inspection of the proof
shows that it is practically impossible to make $F_{\mathrm{fixed}-b}$
and $t_{\mathrm{fixed}-b}$ pivotal by studentization based on any
sequence of inconsistent covariance matrix estimates.  This follows
because the LRV is time-varying and, as noted above, this break downs
the property that both numerator and denominator are asymptotically
proportional to $\Omega$ so that the nuisance parameters cancel out.
On the other hand, the property that under nonstationarity the fixed-$b$
asymptotic framework yields non-pivotal asymptotic distributions that
depend on the underlying second-order properties of $\left\{ V_{t}\right\} $
may suggest that reliable HAR inference is more challenging.

Theorem \ref{Theorem 2: KV (2002)} suggests that valid inference
under fixed-$b$ asymptotics is going to be more complex in terms
of practical implementation relative to when the data are stationary.
In the literature, complexity in the implementation has been recognized
as a strong disadvantage for the success of a given method in empirical
work {[}see, e.g., \citet{lazarus/lewis/stock:17}{]}. The simplest
way to use Theorem \ref{Theorem 2: KV (2002)} for conducting inference
is to replace the nuisance parameters by consistent estimates. This
means constructing estimates of $\Sigma\left(u\right)$, $Q\left(u\right)$
and $\overline{Q}.$ For $\overline{Q}$ the argument is straightforward.
As under stationarity, one can use $\widehat{Q}=T^{-1}\sum_{t=1}^{T}x_{t}x'_{t}$
since $\widehat{Q}-\overline{Q}\overset{\mathbb{P}}{\rightarrow}0$
also under nonstationarity. More complex is the case for $\Sigma\left(u\right)$
and $Q\left(u\right)$. Nonparametric estimators  for $\Sigma\left(u\right)$
and $Q\left(u\right)$ can be constructed. This requires introducing
bandwidths and kernels as well as a criterion for their choice. Then,
one plugs-in these estimates into the limit distribution and the critical
value can be obtained by simulations. However, since $\Sigma\left(u\right)$,
$Q\left(u\right)$ and $\overline{Q}$ depend on the data, the critical
values need to be obtained on a case-by-case basis. We consider this
approach in our simulation analysis in Section \ref{Section: Finite-Sample-Effectiveness}
below.

It is interesting to briefly discuss the properties of fixed-$b$
HAR inference when $\left\{ V_{t}\right\} $ follows more general
forms of nonstationary, i.e., $\left\{ V_{t}\right\} $ does not satisfy
Assumption \ref{Assumption: Assumption 1 in KV (2002), S_Tr, Nonstationarity}-\ref{Assumption: Assumption 2 in KV (2002), Qr, Nonstationarity}.
Assumption \ref{Assumption: Assumption 1 in KV (2002), S_Tr, Nonstationarity}-\ref{Assumption: Assumption 2 in KV (2002), Qr, Nonstationarity}
are satisfied if $\left\{ V_{t}\right\} $ is, e.g., segmented locally
stationary, locally stationary and, of course, stationary. However,
if $\left\{ V_{t}\right\} $ is a sequence of unconditionally heteroskedastic
random variables such that $Q\left(s\right)$ and $\Sigma\left(s\right)$
do not satisfy the smoothness restrictions in Assumption \ref{Assumption: Assumption 1 in KV (2002), S_Tr, Nonstationarity}-\ref{Assumption: Assumption 2 in KV (2002), Qr, Nonstationarity}
then Theorem \ref{Theorem: Limiting Distribution Omega_fixedb}-\ref{Theorem 2: KV (2002)}
do not hold. For example, consider $V_{t}=\rho_{t}V_{t-1}+u_{t}$
where $u_{t}\sim i.i.d.\,\mathscr{N}\left(0,\,1\right)$ and $\rho_{t}\in\left(-1,\,1\right)$
for all $t$. Segmented local stationarity corresponds to $\rho_{t}$
being piecewise continuous, local stationarity corresponds to $\rho_{t}$
being continuous and stationarity corresponds to $\rho_{t}$ being
constant. If $\rho_{t}$ does not satisfy any of these restrictions,
Assumption \ref{Assumption: Assumption 1 in KV (2002), S_Tr, Nonstationarity}-\ref{Assumption: Assumption 2 in KV (2002), Qr, Nonstationarity}
do not hold. For unconditionally heteroskedastic random variables
the asymptotic distributions of $F_{\mathrm{fixed}-b}$ and $t_{\mathrm{fixed}-b}$
remain unknown since they cannot be characterized. Thus, for general
nonstationary random variables fixed-$b$ inference based on the asymptotic
distribution is infeasible. This highlights one major difference from
HAR inference based on consistent estimation of $\Omega.$ HAC and
DK-HAC estimators are consistent for $\Omega$ also for general nonstationary
random variables so that HAR test statistics follow asymptotically
the usual standard distributions. For example, a $t$-statistic studentized
by a HAC or DK-HAC estimator will follow asymptotically a standard
normal. 

Before concluding this section, we note that a few papers explored
fixed-$b$ asymptotics in settings that involve other forms of nonstationarity
that do not fall within the standard HAR inference problem with $\sqrt{T}$-asymptotically
normal OLS estimator. \citet{bunzel/vogelsang:2005} allowed for deterministic
trends and integrated of order one (I(1)) errors. \citet{vogelsang/wagner:2013}
considered fixed-$b$ asymptotics for unit root tests while \citet{vogelsang/wagner:2014}
focused on cointegrating regression. \citet{xu:2012} and \citet{demetrescu/hanck/kruse-becher:2023}
considered multivariate trend tests and hypothesis tests in a GMM
framework, respectively, allowing for time-varying volatility and
no serial correlation.

\section{\label{Section ERP}Error in Rejection Probability in a Gaussian
Location Model }

We develop high-order asymptotic expansions and obtain the ERP of
fixed-$b$ HAR tests. The results in \citet{velasco/robinson:01},
\citet{jansson:04} and \citet{sun/phillips/jin:08} suggest that
under stationarity the ERP of $F_{\mathrm{fixed}-b}$ and of $t_{\mathrm{fixed}-b}$
are smaller than those of the  conventional HAC-based HAR tests.
We show that the opposite is true when stationarity does not hold.

Consider the location model $y_{t}=\beta_{0}+e_{t}$ $(t=1,\ldots,\,T)$.
We have $V_{t}=e_{t}$. Under the assumption that $V_{t}$ is stationary
and Gaussian, \citet{velasco/robinson:01} developed second-order
Edgeworth expansions and showed that
\begin{align}
\mathbb{P}\left(t_{\mathrm{HAC}}\leq z\right)-\varPhi\left(z\right) & =d\left(z\right)\left(Tb_{T}\right)^{-1/2}+o\left(\left(Tb_{T}\right)^{-1/2}\right),\label{Eq. ERP HAC}
\end{align}
 for any $z\in\mathbb{R}$ where 
\begin{align*}
t_{\mathrm{HAC}} & =\frac{\sqrt{T}\left(\widehat{\beta}-\beta_{0}\right)}{\sqrt{\widehat{\Omega}_{\mathrm{HAC}}}},
\end{align*}
$b_{T}\rightarrow0$, $\varPhi\left(\cdot\right)$ is the distribution
function of the standard normal and $d\left(\cdot\right)$ is an odd
function. The ERP is the leading term of the right-hand side of \eqref{Eq. ERP HAC}.
Since $b_{T}=O(T^{-\eta})$ with $0<\eta<1$, the ERP of $t_{\mathrm{HAC}}$
is $O(T^{-\gamma})$ with $\gamma<1/2$. It follows that the leading
term of $\mathbb{P}(F_{\mathrm{HAC}}\leq c)$ where $F_{\mathrm{HAC}}=T(\widehat{\beta}-\beta_{0})^{2}/\widehat{\Omega}_{\mathrm{HAC}}$
is of the form $2d(\sqrt{c})T^{-\gamma}=O(T^{-\gamma})$ for any $c>0.$

\citet{jansson:04} and \citet{sun/phillips/jin:08} showed that\footnote{Actually \citet{jansson:04} showed that the bound was $O(T^{-1}\log T)$.
Using a different proof strategy, \citet{sun/phillips/jin:08} showed
that the $\log T$ term can be dropped. }
\begin{align}
\mathbb{P}\left(F_{\mathrm{fixed}-b}\leq c\right)-\mathbb{P}\left(\frac{W_{1}\left(1\right)^{2}}{\int_{0}^{1}\int_{0}^{1}-K''\left(r-s\right)B_{1}\left(r\right)B_{1}\left(s\right)'drds}\leq c\right) & =O\left(T^{-1}\right).\label{Eq. ERP Fixed-b Stationarity}
\end{align}
Thus, $F_{\mathrm{fixed}-b}$ has a smaller ERP than $F_{\mathrm{HAC}}$
{[}cf. $O(T^{-1})$ versus $O(T^{-\gamma})${]}. This implies that
the rate of convergence of $F_{\mathrm{fixed}-b}$ to its (nonstandard)
limiting null distribution is faster than the rate of convergence
of $F_{\mathrm{HAC}}$ to a $\chi_{1}^{2}$. These results reconciled
with finite-sample evidence in the literature showing that the null
rejection rates of $F_{\mathrm{fixed}-b}$ and $t_{\mathrm{fixed}-b}$
are more accurate than those of $F_{\mathrm{HAC}}$ and $t_{\mathrm{HAC}}$,
respectively, when that data are stationary.

We now address the question of whether these results extend to nonstationarity.
It turns out that the answer is negative. This provides an analytical
explanation for the Monte Carlo experiments that have appeared recently
in \citet{casini_hac}, \citet{casini/perron_Low_Frequency_Contam_Nonstat:2020}
and \citet{casini/perron_PrewhitedHAC} who found serious distortions
in the rejection rates of fixed-$b$ HAR tests under the null and
alternative hypotheses when the data are nonstationary. These distortions
being often much larger than those corresponding to the conventional
HAC-based HAR tests. 

Theorem \ref{Theorem 2: KV (2002)} showed that the fixed-$b$ HAR
tests are not pivotal. Thus, a natural way to conduct inference based
on the fixed-$b$ asymptotic distribution is to construct consistent
estimates of its nuisance parameters. We introduce a general nonparametric
estimator of $\Sigma\left(u\right)$. Let 
\begin{align*}
\boldsymbol{K}_{2} & =\{K_{2}\left(\cdot\right):\,\mathbb{R}\rightarrow\left[0,\,\infty\right]:\,K_{2}\left(x\right)=K_{2}\left(1-x\right),\,{\textstyle \int}K_{2}\left(x\right)dx=1,\\
 & \qquad\qquad K_{2}\left(x\right)=0,\,\mathrm{for\,}\,x\notin\left[0,\,1\right],\,K_{2}\left(\cdot\right)\,\mathrm{is\,continuous}\},
\end{align*}
and
\begin{align}
\widehat{\Omega}\left(u\right)=\widehat{\Sigma}^{2}\left(u\right) & =\sum_{k=-T+1}^{T-1}K_{h_{1}}\left(h_{1}k\right)\widehat{c}_{T,h_{2}}\left(u,\,k\right),\label{Eq. Sigma_hat(u)}
\end{align}
where $K_{h_{1}}\left(\cdot\right)\in\boldsymbol{K}$, $h_{1}$ is
a bandwidth sequence satisfying $h_{1}\rightarrow0$,
\begin{align*}
\widehat{c}_{T,h_{2}}\left(u,\,k\right) & =\left(Th_{2}\right)^{-1}\sum_{s=|k|+1}^{T}K_{h_{2}}\left(\frac{\left(\left\lfloor Tu\right\rfloor -\left(s-|k|/2\right)\right)/T}{h_{2}}\right)\widehat{V}_{s}\widehat{V}{}_{s-|k|},
\end{align*}
 with $K_{h_{2}}\left(\cdot\right)\in\boldsymbol{K}_{2}$ and $h_{2}$
is a bandwidth sequence satisfying $h_{2}\rightarrow0$. Since $x_{t}=1$
for all $t$, we have $Q\left(u\right)=1$ for all $u\in\left[0,\,1\right]$
in \eqref{Eq. Asymptotic Distribution of Omega fixed_b}-\eqref{Eq. Asymptotic Distribution of Omega fixed_b, Barteltt}.
Thus, we set $\widehat{Q}\left(u\right)=1$ for all $u\in\left[0,\,1\right]$.
For arbitrary $x_{t},$ one can take 
\begin{align*}
\widehat{Q}\left(u\right) & =\left(Th_{2}\right)^{-1}\sum_{s=1}^{T}K_{h_{2}}\left(\frac{\left(\left\lfloor Tu\right\rfloor -s\right)/T}{h_{2}}\right)x_{s}^{2}.
\end{align*}
 As in the literature, we focus on the simple location model with
Gaussian errors. The Gaussianity assumption can be relaxed by considering
distributions with, for example, Gram-Charlier representations at
the expenses of more complex derivations {[}see, e.g., \citet{phillips:1980}{]}.
The following assumption on $V_{t}$ facilitates the development of
the higher order expansions and is weaker than the one used by \citet{sun/phillips/jin:08}
since they also imposed second-order stationarity.
\begin{assumption}
\label{Assumption 3 in Sun et al.}$\{V_{t}\}$ is a mean-zero Gaussian
process with $\sup_{1\leq t\leq T}\sum_{k=-\infty}^{\infty}k^{2}|\mathbb{E}(V_{t}V_{t-k})|<\infty$. 
\end{assumption}
In order to develop the asymptotic expansions we use the following
 conditions on the kernel which were also used by \citet{andrews:91}
and \citet{sun/phillips/jin:08}. 
\begin{assumption}
\label{Assumption: Assumption 2 in SPJ (2008)}(i) $K\left(x\right):\,\mathbb{R}\rightarrow\left[0,\,1\right]$
is symmetric  and satisfies $K\left(0\right)=1$, $\int_{0}^{\infty}xK\left(x\right)dx<\infty$
and $|K\left(x\right)-K\left(y\right)|\leq C_{1}\left|x-y\right|$
for all $x,\,y\in\mathbb{R}$ and some $C_{1}<\infty$. 

(ii) $q_{0}\geq1$ where $q_{0}$ is the Parzen characteristic exponent
defined by
\begin{align*}
q_{0} & =\max\left\{ \widetilde{q}:\,\widetilde{q}\in\mathbb{Z}_{+},\,\overline{K}_{\widetilde{q}}=\lim_{x\rightarrow0}\frac{1-K\left(x\right)}{\left|x\right|^{\widetilde{q}}}<\infty\right\} .
\end{align*}

(iii) $K\left(x\right)$ is positive semidefinite, i.e., for any square
integrable function $g\left(x\right)$, $\int_{0}^{\infty}\int_{0}^{\infty}K\left(s-t\right)$
$g\left(s\right)g\left(t\right)dsdt\geq0$. 
\end{assumption}
All of the commonly used kernels with the exception of the truncated
kernel satisfy Assumption \ref{Assumption: Assumption 2 in SPJ (2008)}-(i,
ii). \citet{sun/phillips/jin:08} required piecewise smoothness on
$K\left(\cdot\right)$ instead of the Lipschitz condition. Part (iii)
ensures that the associated LRV estimator is positive semidefinite.
The commonly used kernels that satisfy part (i, iii) are the Bartlett,
Parzen and quadratic spectral (QS) kernels. For the Bartlett kernel,
$q_{0}=1$, while for the Parzen and QS kernels, $q_{0}=2.$ 

As in \citet{sun/phillips/jin:08}, we present the asymptotic expansion
for the test statistic studentized by $\widehat{\Omega}_{\mathrm{fixed}-b}$
defined in \eqref{Eq.: Fixed-b Estimator}. Let $K_{b}=K\left(\cdot/b\right)$.
Lemma \ref{Lemma: Theorem 1 in PSJ (2007)} in the supplement extends
Theorem \ref{Theorem: Limiting Distribution Omega_fixedb} to $\widehat{\Omega}_{\mathrm{fixed}-b}$
using the kernels that satisfy Assumption \ref{Assumption: Assumption 2 in SPJ (2008)}.
Under Assumption \ref{Assumption: Assumption 1 in KV (2002), S_Tr, Nonstationarity}
and \ref{Assumption 3 in Sun et al.}-\ref{Assumption: Assumption 2 in SPJ (2008)},
Lemma \ref{Lemma: Theorem 1 in PSJ (2007)} shows that $\widehat{\Omega}_{\mathrm{fixed}-b}\Rightarrow\mathscr{G}_{b}$
where 
\begin{align*}
\mathscr{G}_{b}= & \int_{0}^{1}\int_{0}^{1}K_{b}\left(r-s\right)d\widetilde{B}_{1}\left(r\right)d\widetilde{B}_{1}\left(s\right),
\end{align*}
 and 
\begin{align*}
\widetilde{B}_{1}\left(r\right) & =\int_{0}^{r}\left(\Sigma\left(u\right)dW_{1}\left(u\right)-r\left(\int_{0}^{1}\Sigma\left(u\right)dW_{1}\left(u\right)\right)\right).
\end{align*}
 Let $C_{\Omega}=\sup_{s\in\left[0,\,1\right]}\Omega\left(s\right)$,
$C_{2,\Omega}=\max\{C_{\Omega},\,1\}$, 
\begin{align*}
t_{\mathrm{fixed}-b} & \triangleq\frac{\sqrt{T}\left(\widehat{\beta}-\beta_{0}\right)}{\sqrt{\widehat{\Omega}_{\mathrm{fixed}-b}}},
\end{align*}
and 
\begin{align*}
\mathscr{\widehat{G}}_{b} & =\int_{0}^{1}\int_{0}^{1}K_{b}\left(r-s\right)\left(\int_{0}^{r}\widehat{\Sigma}\left(u\right)dW_{1}\left(u\right)-r\int_{0}^{1}\widehat{\Sigma}\left(u\right)dW_{1}\left(u\right)\right)\\
 & \quad\times\left(\int_{0}^{s}\widehat{\Sigma}\left(u\right)dW_{1}\left(u\right)-s\int_{0}^{1}\widehat{\Sigma}\left(u\right)dW_{1}\left(u\right)\right)drds.
\end{align*}

\begin{thm}
\label{Theorem ERP Fixedb}Let Assumption \ref{Assumption: Assumption 1 in KV (2002), S_Tr, Nonstationarity},
\ref{Assumption 3 in Sun et al.}-\ref{Assumption: Assumption 2 in SPJ (2008)},
$h_{1}\rightarrow0$, $h_{2}\rightarrow0$, $Th_{1}h_{2}\rightarrow\infty$
and $\sqrt{Th_{1}h_{2}}\left(h_{1}^{2}+h_{2}^{2}\right)\rightarrow0$
hold. Provided that $b$ is fixed such that $b<1/(16C_{2,\Omega}$
$\int_{-\infty}^{\infty}|K\left(x\right)|dx)$, we have 
\begin{align*}
\sup_{z\in\mathbb{R}_{+}}\left|\mathbb{P}\left(\left|t_{\mathrm{fixed}-b}\right|\leq z\right)-\mathbb{P}\left(\left|\frac{\int_{0}^{1}\widehat{\Sigma}\left(u\right)dW_{1}\left(u\right)}{\sqrt{\mathscr{\widehat{G}}_{b}}}\right|\leq z\right)\right| & =O\left(\left(Th_{1}h_{2}\right)^{-1/2}\right).
\end{align*}
  
\end{thm}
Theorem \ref{Theorem ERP Fixedb} shows that the ERP associated to
$t_{\mathrm{fixed}-b}$ is $O((Th_{1}h_{2})^{-1/2})$. This is an
order of magnitude larger than the ERP associated to $t_{\mathrm{fixed}-b}$
under stationarity, $O(T^{-1})$, where the latter was established
by \citet{jansson:04} and \citet{sun/phillips/jin:08}. The increase
in the ERP of $t_{\mathrm{fixed}-b}$ is the price one has to pay
for not having a pivotal distribution under nonstationarity. This
is intuitive. Without a pivotal distribution, one has to obtain estimates
of the nuisance parameters. However, the nuisance parameters can be
consistently estimated only under small-$b$ asymptotics. The latter
estimates enjoy a nonparametric rate of convergence which then results
in a larger ERP since it is the discrepancy between these estimates
and their probability limits that is reflected in the leading term
of the asymptotic expansion.

The conventional fixed-$b$ methods use a fixed bandwidth and the
critical value from the pivotal fixed-$b$ limiting distribution obtained
under the assumption of stationarity. Our results suggest that the
ERP associated to such fixed-$b$ HAR tests is $O\left(1\right)$.
This follows because that critical value is not theoretically valid,
i.e., it is from the pivotal fixed-$b$ limiting distribution which,
however, is different from the non-pivotal fixed-$b$ limiting distribution
under nonstationarity. Thus, as $T\rightarrow\infty$ the ERP does
not converge to zero, implying large distortions in the null rejection
rates even for unbounded sample sizes. 

Theorem \ref{Theorem ERP Fixedb} implies that the theoretical properties
of fixed-$b$ inference changes substantially depending on whether
the data are stationarity or not.  In particular, it suggests that
the approximations based on fixed-$b$ asymptotics obtained under
stationarity in the literature are not  valid and do not provide
a good approximation when stationarity does not hold. This contrasts
to HAR inference tests based on consistent long-run variance estimators
which are valid also under nonstationarity and have the same asymptotic
distribution regardless of whether the data are stationary or not.\footnote{It is useful to remind that even though they are generally valid their
finite-sample performance can be poor if there is strong dependence
under either stationarity or nonstationarity as documented in the
literature.} Theorem \ref{Theorem ERP Fixedb} also provides formal support to
the arguments in \citet{ibragimov/muller:10} and \citet{mueller:14}
who mentioned that the stationarity assumption used by fixed-$b$
methods is a disadvantage relative to conventional methods including
the $t$-statistic approach. 

Additional comments: 1. It is useful to compare Theorem \ref{Theorem ERP Fixedb}
with the results for the ERP associated to $t_{\mathrm{HAC}}$. Under
stationarity \citet{velasco/robinson:01} showed that the ERP associated
to $t_{\mathrm{HAC}}$ is $O((Tb_{T})^{-1/2})$ where $b_{T}\rightarrow0.$
\citet{casini/perron_Low_Frequency_Contam_Nonstat:2020} showed that
under nonstationarity the ERP associated to $t_{\mathrm{HAC}}$ has
the same order as under stationarity, i.e., $O((Tb_{T})^{-1/2})$.
Thus, it is sufficient to compare $O((Th_{1}h_{2})^{-1/2})$ and $O((Tb_{T})^{-1/2})$.
Since $h_{1}$ and $b_{T}$ are the bandwidths used for smoothing
over lagged autocovariances, they may have a similar order. It follows
that the ERP associated to $t_{\mathrm{fixed}-b}$ may be larger than
that associated to $t_{\mathrm{HAC}}$. In addition, the ERP associated
to $t_{\mathrm{fixed}-b}$ based on conventional fixed-$b$ methods
that rely on stationarity is much larger than the ERP associated to
$t_{\mathrm{HAC}}$ since the former is $O\left(1\right)$.

2. A more recent development in the  literature {[}see, e.g., \citet{sun:14}
and \citet{lazarus/lewis/stock/watson:18}{]} considered the use of
small-$b$ asymptotics (i.e., small-bandwidths) and fixed-$b$ critical
values. These bandwidths are typically larger than the MSE-optimal
bandwidths used for $t_{\mathrm{HAC}}$ {[}see eq. (3) in \citet{lazarus/lewis/stock/watson:18},
and the equation for $b^{*}$ on p. 666 of \citet{sun:14} and the
related discussion there{]}. As $b_{T}\rightarrow0$ the fixed-$b$
limiting distribution approximates the standard asymptotic distribution
based on small-$b$ asymptotics. Thus, the fixed-$b$ critical values
converge to the standard normal critical values for the case of a
$t$-test. In the limit the ERP of these HAR tests should be the same
as that of $t_{\mathrm{HAC}}$. However, recent simulation results
in the literature show that these HAR tests have different finite-sample
rejection probabilities from those of $t_{\mathrm{HAC}}$. Hence,
although the result in Theorem \ref{Theorem ERP Fixedb} only speaks
for fixed-bandwidths, it might suggest that using the fixed-$b$ critical
values from the new fixed-$b$ limiting distribution may improve the
finite-sample performance of these recent fixed-$b$ methods under
nonstationarity. 

3. Overall, the theoretical results contrast with what the early
fixed-$b$ literature showed under stationarity {[}see \citet{jansson:04}{]},
namely that the original fixed-$b$ HAR inference is theoretically
superior to HAR inference based on consistent long-run variance estimators. 

4. Our theoretical results complement the recent finite-sample evidence
in \textcolor{MyBlue}{Belotti et al.} \citeyearpar{belotti/casini/catania/grassi/perron_HAC_Sim_Bandws},
\citet{casini_hac}, \citet{casini/perron_PrewhitedHAC} and \citet{casini/perron_Low_Frequency_Contam_Nonstat:2020}.
Their simulation results showed that existing fixed-$b$ HAR inference
tests perform poorly in terms of the accuracy of the null rejection
rates and of power when stationarity does not hold. They considered
$t$-tests in the linear regression models and HAR tests outside the
linear regression model, and a variety of data-generating processes.
They provided  evidence that fixed-$b$ HAR tests can be severely
undersized and can exhibit non-monotonic power. Some of these issues
are generated by the low frequency contamination induced by nonstationarity
which biases upward each sample autocovariance $\widehat{\Gamma}\left(k\right)$.
Since $\widehat{\Omega}_{\mathrm{fixed}-b}$ uses many lagged autocovariances
as $b$ is fixed, it is inflated which then results in  size distortions
and lower power. 

5. The fixed-$b$ limiting distribution under nonstationarity is complex
to use in practice as it depends in a complicated way on nuisance
parameters. The procedure in Theorem \ref{Theorem ERP Fixedb} replaces
the nuisance parameter $\Sigma\left(\cdot\right)$ by a consistent
estimate. This procedure represents a natural starting point to study
the properties of fixed-$b$ inference under nonstationarity and so
the corresponding ERP results may provide general guidance. There
are certainly other procedures that could be used. It is beyond the
scope of the paper to investigate how best to use the non-pivotal
fixed-$b$ limiting distribution. If one wants to consider other
procedures such as finding a conservative upper bound for the critical
value that holds under all possible values of the nuisance parameter,
bootstrap-based autocorrelation robust tests, modification of the
test statistic\footnote{\citet{hwang/sun:2017} proposed a modification to the trinity of
test statistics in the two-step GMM setting and showed that the modified
test statistics are asymptotically $F$ distributed under fixed-$b$
asymptotics.}, etc., then one has to face the challenge that these methods are
not as simple as HAC-based inference which can be a disadvantage as
recently argued by \citet{lazarus/lewis/stock:17}. Future work should
investigate on possible alternative fixed-$b$ procedures that exploit
the results in Section \ref{Section: Fixed--Limiting-Distribution}. 

6. The requirement $Th_{1}h_{2}\rightarrow\infty$ is a standard condition
for consistency of nonparametric estimators such as $\widehat{\Omega}\left(u\right)$.
The requirement $b<1/(16C_{2,\Omega}\int_{-\infty}^{\infty}|K\left(x\right)|dx)$
is similar to the one used by \citet{sun/phillips/jin:08}. It can
be relaxed at the expenses of more complex derivations.

7. As remarked at the end of Section \ref{Section: Fixed--Limiting-Distribution},
for unconditionally heteroskedastic random variables,  standard fixed-$b$
HAR inference is infeasible. Thus, the associated ERP does not convergence
to zero. In contrast, HAR inference based on consistent long-run variance
estimator is valid and the associated ERP is again $O((Tb_{T})^{-1/2})$
with $b_{T}\rightarrow0.$ 

\section{\label{Section: Finite-Sample-Effectiveness}Finite-Sample Effectiveness
of the Limit Theory}

In this section we conduct a Monte Carlo analysis to evaluate the
effectiveness of the theoretical results of Section \ref{Section: Fixed--Limiting-Distribution}-\ref{Section ERP}.
We consider the empirical null rejection rates and local power of
the $t$-statistic in a simple location model: 
\begin{align*}
y_{t} & =\beta_{0}+e_{t},\qquad t=1,\ldots,\,T.
\end{align*}
 We consider the following data-generating processes for $e_{t}$.
In model M1 we specify $e_{t}$ as an AR(1) with a break in the autoregressive
coefficient, 
\begin{align*}
e_{t} & =\begin{cases}
0.8e_{t-1}+u_{t}, & t\leq0.2T\\
0.3e_{t-1}+u_{t}, & t>0.2T
\end{cases},
\end{align*}
where $u_{t}\sim\mathrm{i.i.d.\,}\mathscr{N}\left(0,\,1\right)$ and
$e_{0}\sim\mathscr{N}\left(0,\,1\right)$. Further, the initial condition
of $e_{t}$ in the second regime is not the realized $e_{0.2T}$ but
we set $e_{0.2T}\sim\mathscr{N}\left(0,\,1\right)$ so that the two
regimes are independent.\footnote{The results are unchanged when we use the realized $e_{0.2T}$ as
the initial condition for the second regime.} Model M2 involves a locally stationary AR(1): 
\begin{align*}
e_{t} & =\rho_{t}e_{t-1}+u_{t},\qquad\qquad\rho_{t}=0.85\cos\left(1.5t/T\right),
\end{align*}
 where $u_{t}\sim\mathrm{i.i.d.\,}\mathscr{N}\left(0,\,1\right)$
and $e_{0}\sim\mathscr{N}\left(0,\,1\right)$. Note that $\rho_{t}$
varies between 0.055 to 0.850. In model M3 we consider a stationary
AR(1) $e_{t}=0.9e_{t-1}+u_{t}$ where $u_{t}\sim\mathrm{i.i.d.\,}\mathscr{N}\left(0,\,1\right)$
and $e_{0}\sim\mathscr{N}\left(0,\,1\right)$. We consider the following
test statistic:
\begin{align*}
t_{b} & =\frac{\sqrt{T}\left(\widehat{\beta}-\beta_{0}\right)}{\sqrt{\widehat{\Omega}_{b}}},
\end{align*}
 where $\widehat{\beta}=\overline{y}=T^{-1}\sum_{t=1}^{T}y_{t}$ and
$\widehat{\Omega}_{b}=\widehat{\Omega}_{\mathrm{fixed-}b}$ as defined
in \eqref{Eq.: Fixed-b Estimator} with the Bartlett kernel and $\widehat{V}_{t}=y_{t}-\overline{y}$.
We set $\beta_{0}=0$ and consider the null hypothesis that $\beta_{0}\leq0$
against the alternative that $\beta_{0}>0$. The results for a two-sided
version of the test are qualitatively similar. We report the results
for the sample sizes $T=250,\,500$ and 5,000 replications were used
in all cases. To illustrate how the reference distribution works as
the bandwidth $b$ varies in finite-sample, we compute the rejection
probabilities for $t_{b}$ implemented using $b=0.02,\,0.04,\,0.06,\ldots,\,0.96,\,0.98,\,1$.
We set the asymptotic significance level to 0.05 and consider the
following critical values. The usual standard normal critical value
of 1.645 for all values of $b$; the critical value from the stationary
fixed-$b$ distribution in \citet{kiefer/vogelsang:05} for the corresponding
$b$; the critical values from the infeasible and feasible nonstationary
fixed-$b$ distribution for the corresponding $b$. The infeasible
version is obtained by simulating the distribution in Theorem \ref{Theorem 2: KV (2002)}
with the true value of $\Sigma\left(u\right)$ and $Q\left(u\right)$.
Note that in our setting $Q\left(u\right)=1$ for all $u.$ The feasible
version is obtained by simulating the distribution in Theorem \ref{Theorem 2: KV (2002)}
where $\Sigma\left(u\right)$ is replaced by $\widehat{\Sigma}\left(u\right)$
defined in \eqref{Eq. Sigma_hat(u)} with $h_{1}=(Th_{2})^{-4/5}$
and $h_{2}=T{}^{-1/3}$, $K_{h_{1}}$ is Bartlett kernel and $K_{h_{2}}$
is the rectangular kernel.  Other choices for $h_{1}$ and $h_{2}$
that satisfy the conditions of Theorem \ref{Theorem ERP Fixedb} are
possible. However, we verified in unreported simulations that the
empirical rejection probabilities are only very marginally sensitive
to the choice of $h_{1}$ and $h_{2}$ (i.e., within the $\pm1\%$
range about the ones reported below). In particular, any combination
of $\left(h_{1},\,h_{2}\right)$ leads to empirical rejection rates
that are more accurate than those from the stationary fixed-$b$ distribution
of \citet{kiefer/vogelsang:05}. Similarly, different choices for
the kernels $K_{h_{1}}$ and $K_{h_{2}}$ yield quantitative similar
results. We do not report them for brevity.

\begin{center}
\begin{figure}[H]
\includegraphics[width=16cm,height=7cm]{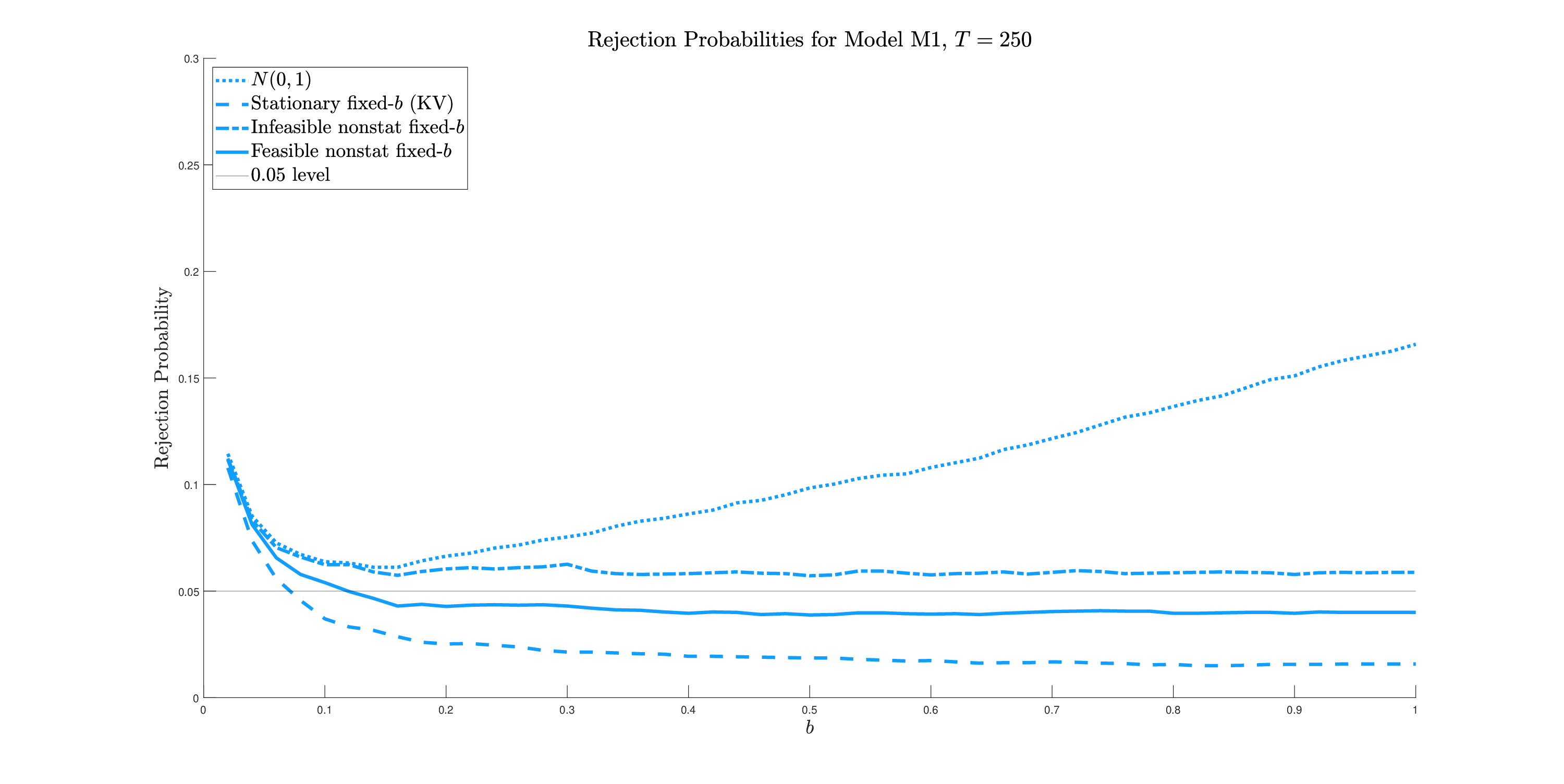}

{\footnotesize{}\caption{{\scriptsize{}\label{Fig_M1_Size_T250}}Small-sample null rejection
rates for model M1. The sample size is $T=250.$}
}{\footnotesize\par}
\end{figure}
\end{center}

\begin{center}
\begin{figure}[H]
\includegraphics[width=16cm,height=7cm]{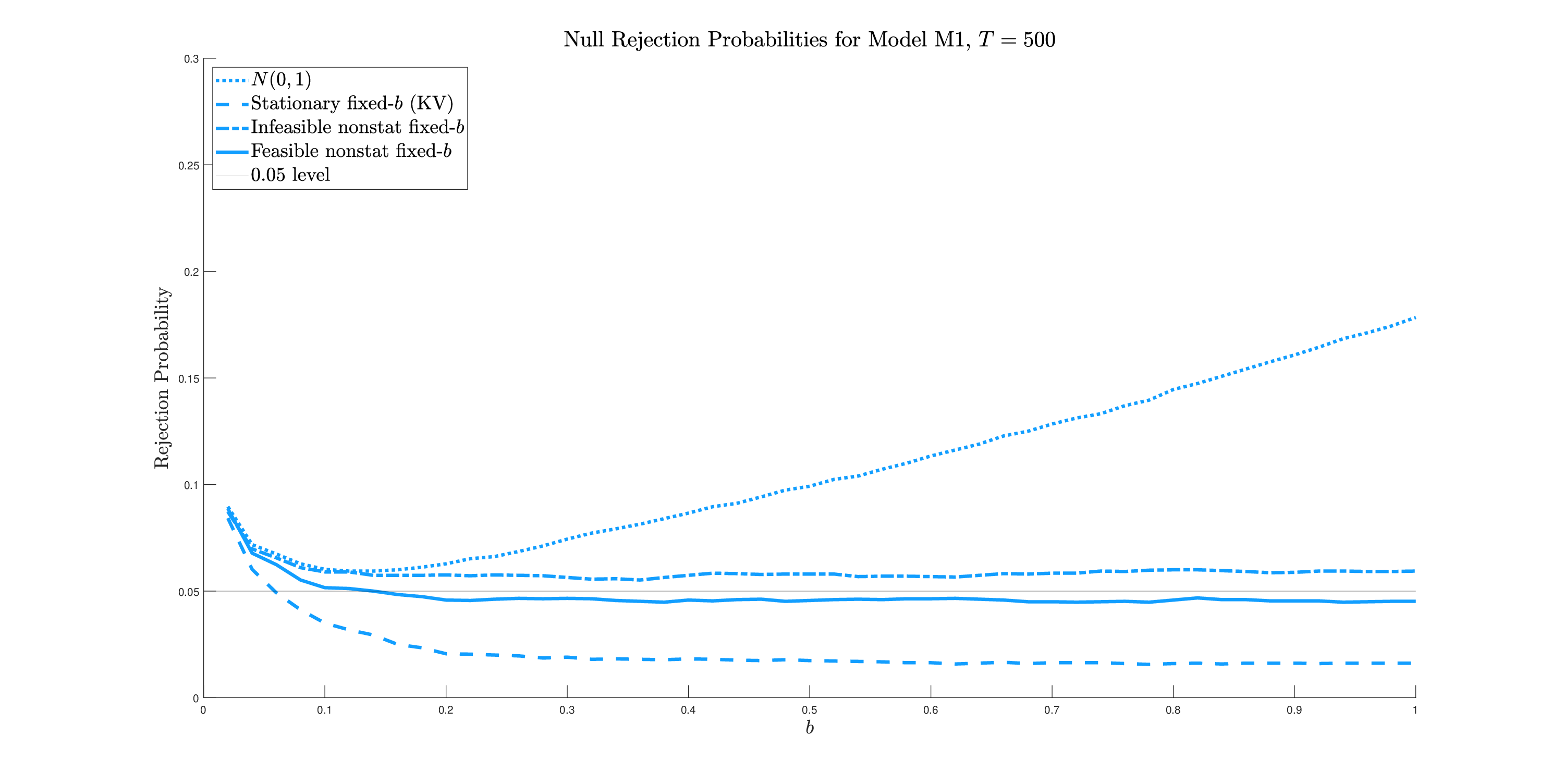}

{\footnotesize{}\caption{{\scriptsize{}\label{Fig_M1_Size_T500}}Small-sample null rejection
rates for model M1. The sample size is $T=500.$}
}{\footnotesize\par}
\end{figure}
\end{center}

\begin{center}
\begin{figure}[H]
\includegraphics[width=16cm,height=7cm]{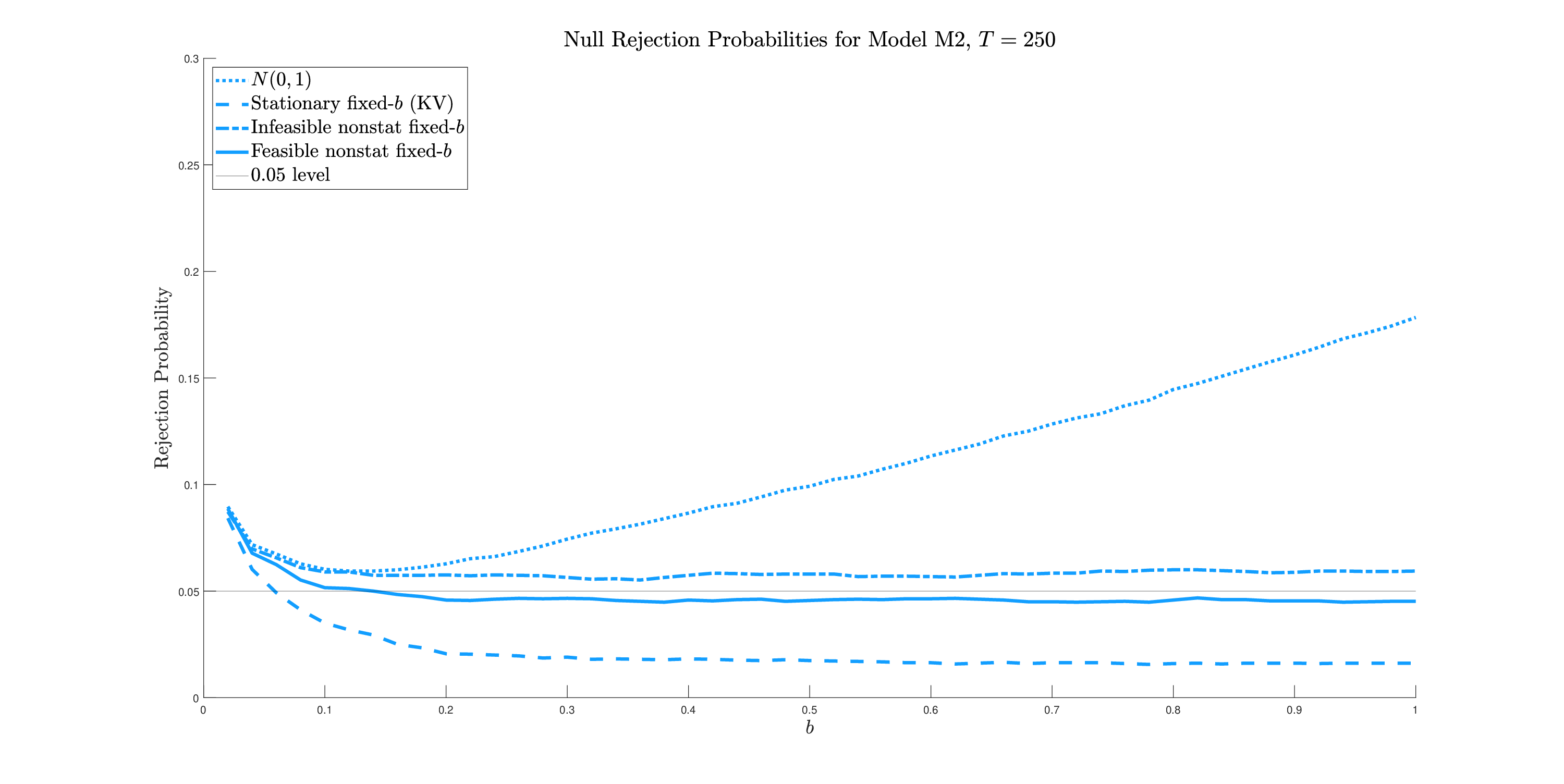}

{\footnotesize{}\caption{{\scriptsize{}\label{Fig_M2_Size_T250}}Small-sample null rejection
rates for model M2. The sample size is $T=250.$}
}{\footnotesize\par}
\end{figure}
\end{center}

\begin{center}
\begin{figure}[H]
\includegraphics[width=16cm,height=7cm]{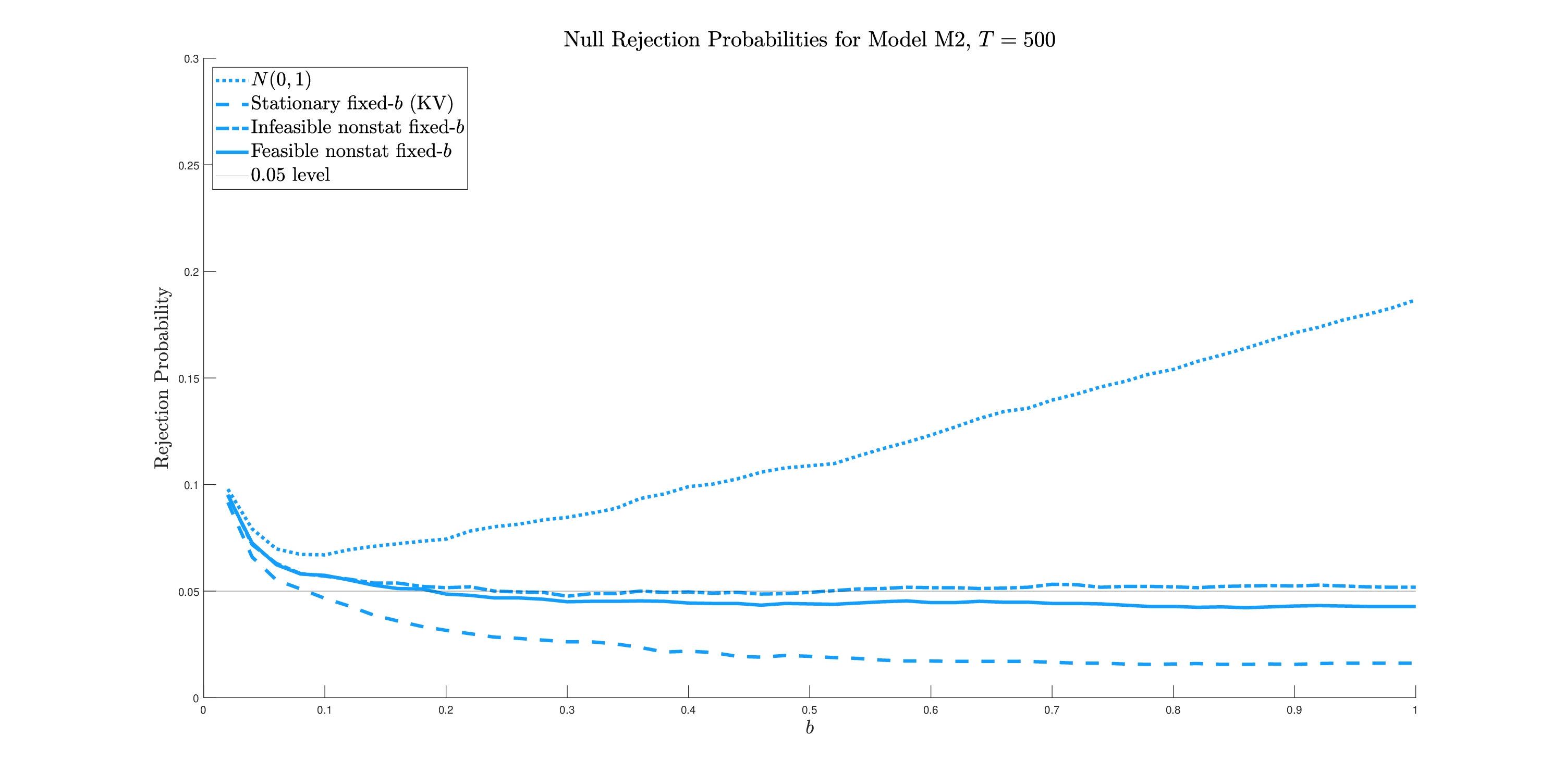}

{\footnotesize{}\caption{{\scriptsize{}\label{Fig_M2_Size_T500}}Small-sample null rejection
rates for model M2. The sample size is $T=500.$}
}{\footnotesize\par}
\end{figure}
\end{center}

\begin{center}
\begin{figure}[H]
\includegraphics[width=16cm,height=7cm]{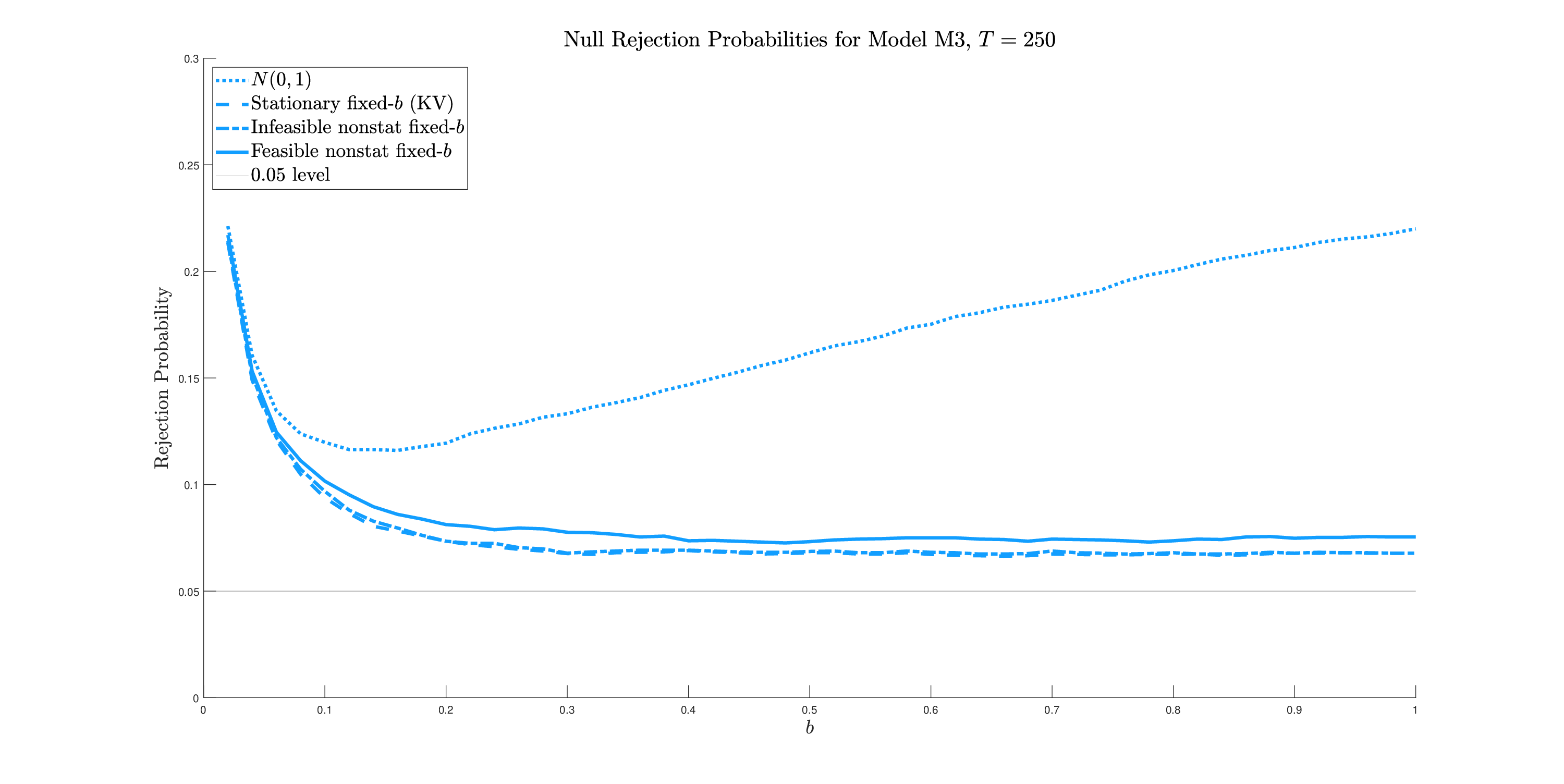}

{\footnotesize{}\caption{{\scriptsize{}\label{Fig_M3_Size_T250}}Small-sample null rejection
rates for model M3. The sample size is $T=250.$}
}{\footnotesize\par}
\end{figure}
\end{center}

\begin{center}
\begin{figure}[H]
\includegraphics[width=16cm,height=7cm]{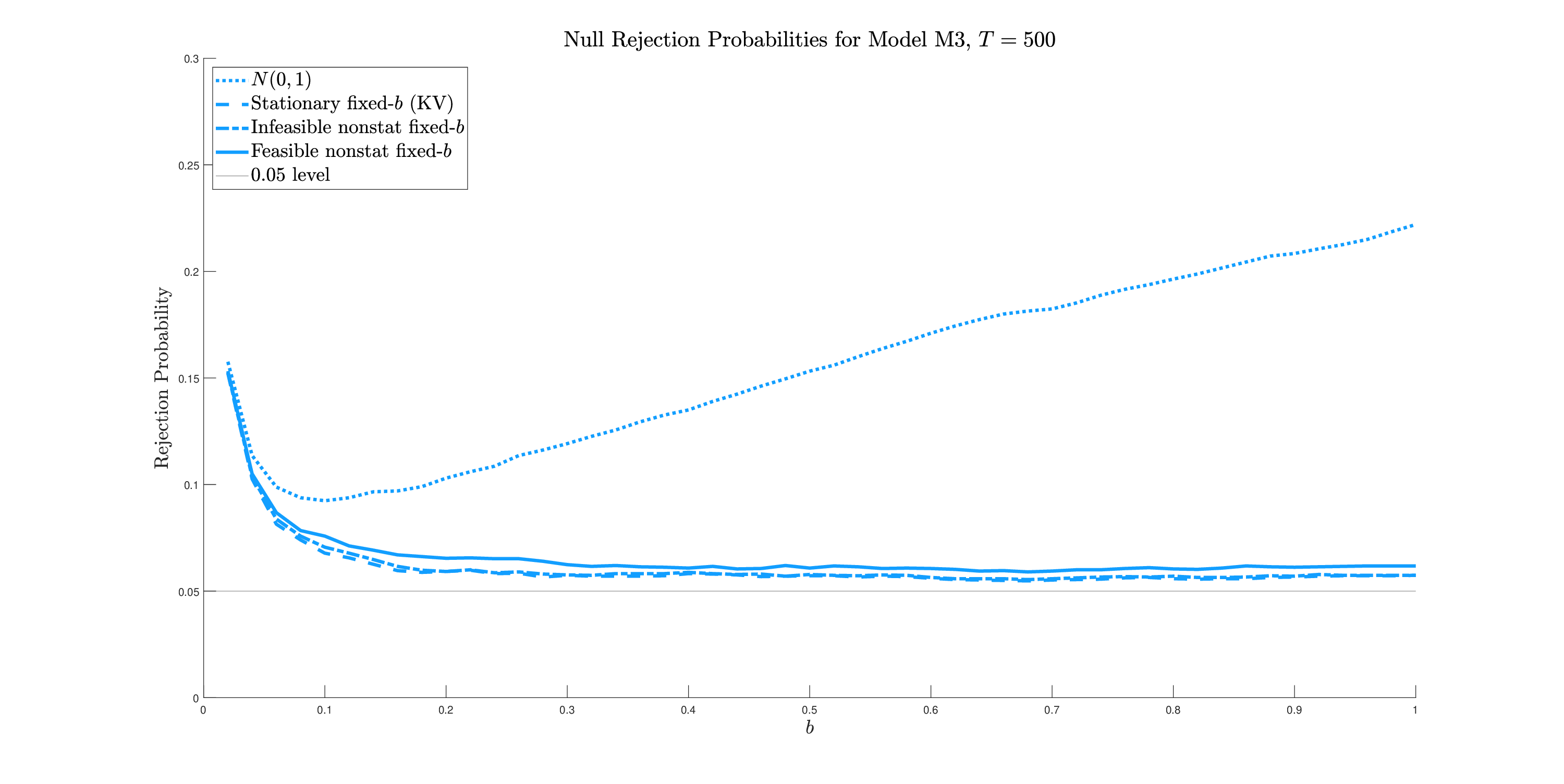}

{\footnotesize{}\caption{{\scriptsize{}\label{Fig_M3_Size_T500}}Small-sample null rejection
rates for model M3. The sample size is $T=500.$}
}{\footnotesize\par}
\end{figure}
\end{center}

The results about the null rejection rates are depicted in Figures
\ref{Fig_M1_Size_T250}-\ref{Fig_M3_Size_T500}. To facilitate the
reading of the figures, we also report the null rejection rates for
each method for $b=0.25,\,0.5,\,0.75,\,1$ in Table \ref{Table 1}.
In each figure, the line with the label, $N\left(0,\,1\right)$, plots
the rejection probabilities when the critical value 1.645 is used.
The figures also depict plots of the rejection probabilities using
the stationary fixed-$b$ asymptotic critical values and the infeasible
and feasible nonstationary fixed-$b$ asymptotic critical values.
The results are striking. In the nonstationary models M1 and M2 the
stationary fixed-$b$ method yields rejection rates that are substantially
below the significance level for all values of $b$ except for $b$
very small.\footnote{This is consistent with the empirical results in \citet{casini/perron_Low_Frequency_Contam_Nonstat:2020}.}
The latter feature is expected since for small $b$ the fixed-$b$
asymptotic theory reduces to the small-$b$ asymptotics, or the fixed-$b$
distribution reduces to the standard normal distribution. Given that
a small $b$ means that a small number of sample autocovariances are
used in $\widehat{\Omega}_{b}$, we expect the test statistic to over-reject.\footnote{This feature also appeared in the simulations of \citet{kiefer/vogelsang:05}.}
In contrast, both the infeasible and feasible nonstationary fixed-$b$
distributions lead to empirical rejection rates that are very close
to 0.05 for all values of $b$ except for $b$ very small. Interestingly,
for $T=500$ the stationary fixed-$b$ critical values yield rejection
rates that sometimes are worse than for $T=250$, i.e., the under-rejection
becomes more pronounced with a larger the sample size. This does not
happen when the nonstationary fixed-$b$ critical values are used,
in fact they yield more accurate rejection rates as the sample size
increases. Hence, these results corroborate the relevance of the nonstationary
fixed-$b$ distribution theory relative to the stationary fixed-$b$
distribution theory when the data are nonstationary. 

The pattern of the rejection rates when the standard normal critical
value is used is similar to that found by \citet{kiefer/vogelsang:05}.
When $b$ is small, there are substantial over-rejections. The rejections
fall as $b$ increases but then rise again as $b$ increases further.
For all values of $b$ the rejection rates are beyond 0.05. This confirms
the size distortions documented in the literature and that the accuracy
of the small-$b$ asymptotics depends on $b$.

It is useful to analyze the performance of the nonstationary fixed-$b$
distribution when the data are indeed stationary with strong dependence.
The results are reported in Figure \ref{Fig_M3_Size_T250}-\ref{Fig_M3_Size_T500}.
First note that Theorem \ref{Theorem 2: KV (2002)} implies that when
the data are stationary the nonstationary fixed-$b$ distribution
reduces to the stationary fixed-$b$ distribution obtained by \citet{kiefer/vogelsang:05}.
In fact, the figures show that the rejection rates corresponding to
the infeasible nonstationary fixed-$b$ critical values coincide with
those corresponding to the stationary fixed-$b$ critical values.
As it is well-known, the standard normal critical value leads to large
over-rejections for all $b.$ In contrast, the empirical rejection
rates corresponding to either fixed-$b$ critical values are much
more accurate. The feasible nonstationary fixed-$b$ critical values
yield rejection rates that are essentially the same as the ones from
the stationary fixed-$b$ distribution. For $b\in\left[0.2,\,1\right]$
the rejection rates of either fixed-$b$ method are stable and therefore
equally accurate. For small values of $b$ also the fixed-$b$ critical
values yield over-rejections. This is obvious since a small $b$ does
not correspond to a fixed-$b$ asymptotic theory with $b$ required
to be fixed. Rather, the fixed-$b$ distribution approximates the
standard normal distribution as $b\rightarrow0$ and so no gain is
expected from using the fixed-$b$ critical values for values of $b$
that are too small.

Let us comment on the difference between the infeasible and feasible
nonstationary fixed-$b$ inference. The proof of Theorem \ref{Theorem ERP Fixedb}
suggests that the infeasible fixed-$b$ distribution enjoys a smaller
ERP than its feasible counterpart. The feasible fixed-$b$ method
depends on the nonparametric estimators of $\Sigma\left(u\right)$
and $Q\left(u\right)$ which in turn depend on the choice of $h_{1}$
and $h_{2}$. However, as noted above, the rejection rates are not
much sensitive to the choice of $h_{1}$ and $h_{2}$. For the reported
results, the performance of the feasible inference is not very different
from that of the infeasible inference. The feasible inference is slightly
more accurate in model M1-M2 and slightly less in model M3. Our unreported
simulations involving other choices of $h_{1}$ and $h_{2}$ showed
that the feasible inference can often be slightly worse than the infeasible
inference as the theory would suggest.

\begin{center}
\begin{figure}[H]
\includegraphics[width=16cm,height=8cm]{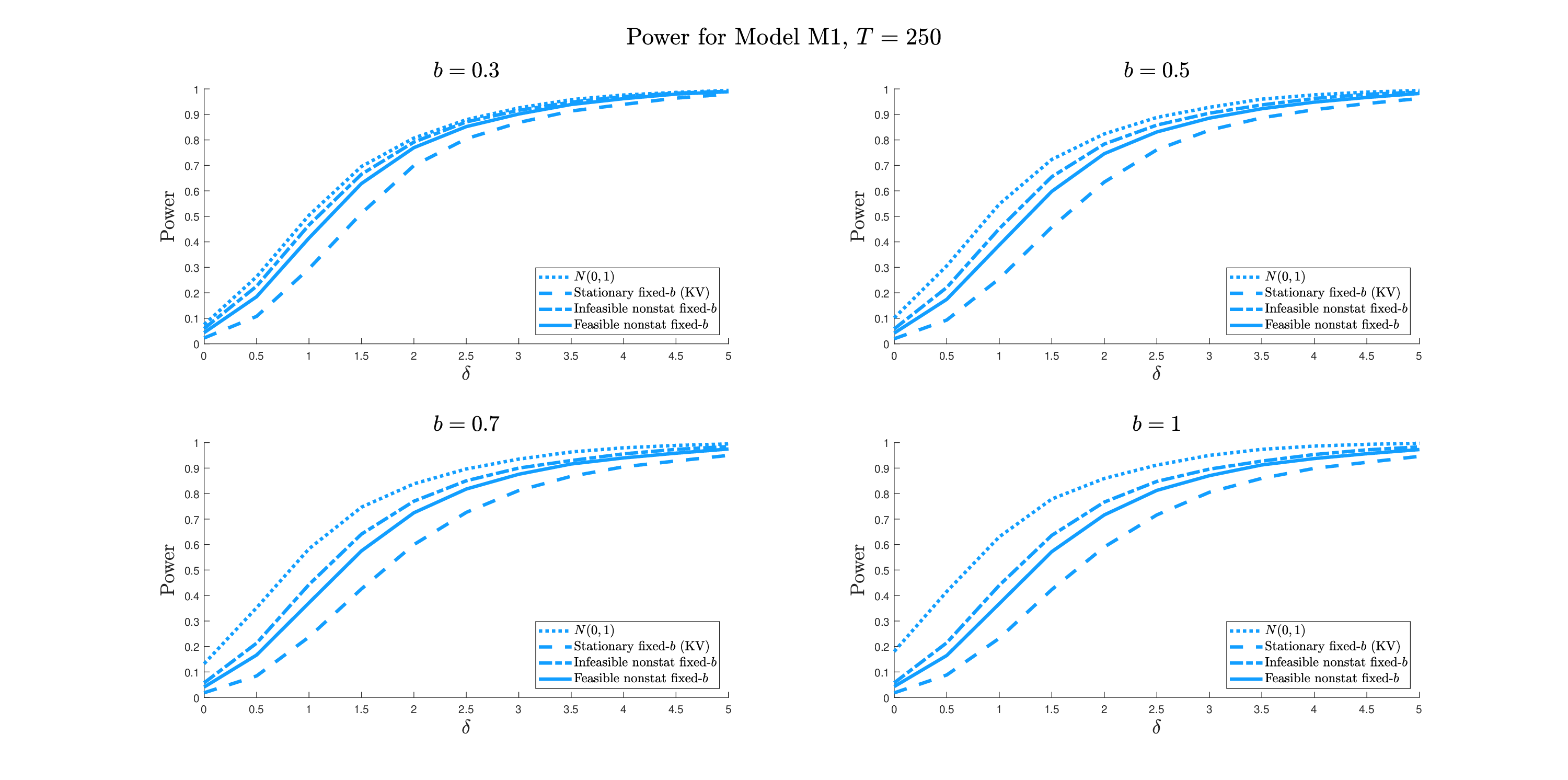}

{\footnotesize{}\caption{{\scriptsize{}\label{Fig_M1_Power_T250_bs}}Small-sample size-unadjusted
power for model M1. The sample size is $T=250.$}
}{\footnotesize\par}
\end{figure}
\end{center}

\begin{center}
\begin{figure}[H]
\includegraphics[width=16cm,height=8cm]{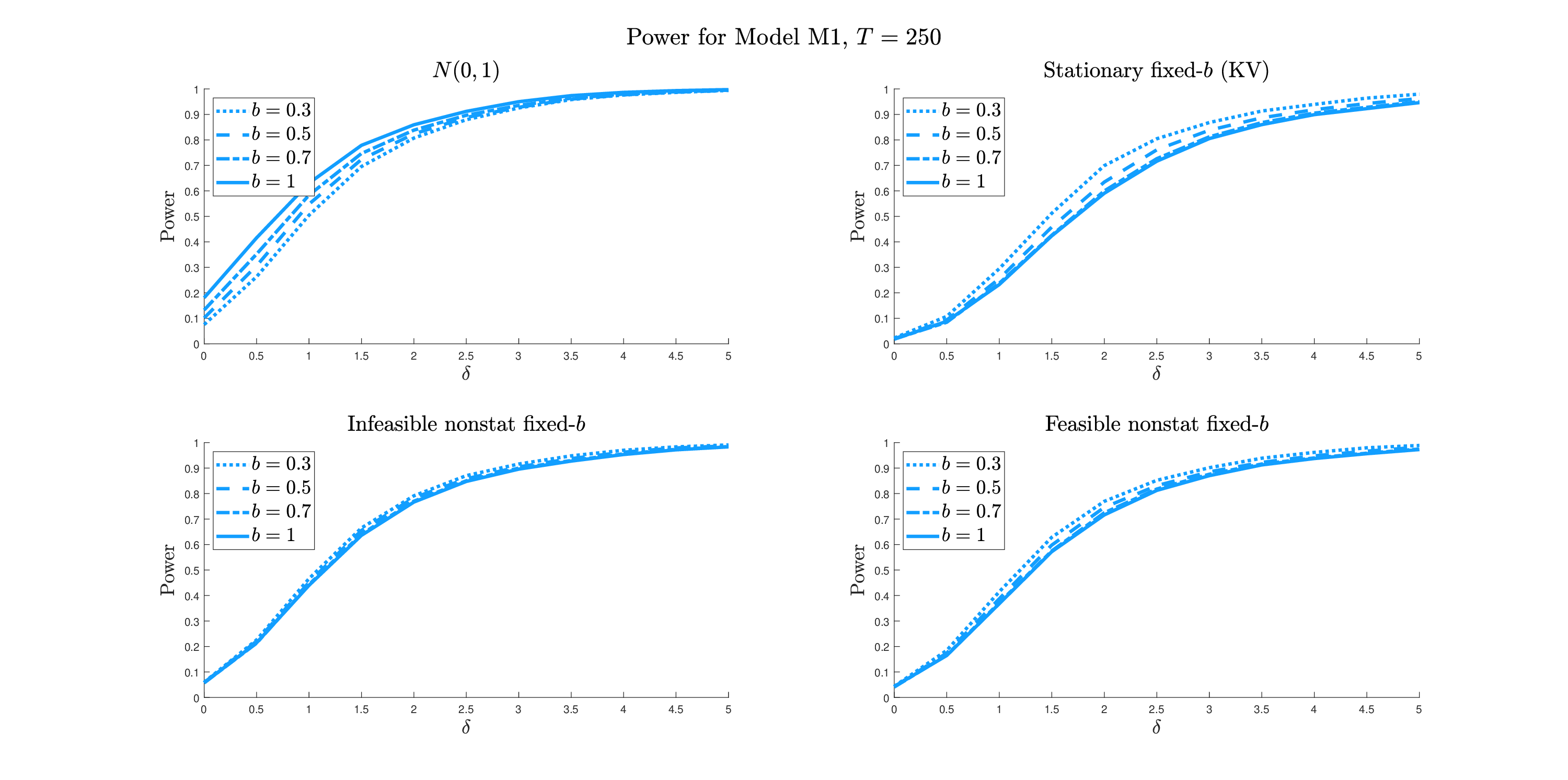}

{\footnotesize{}\caption{{\scriptsize{}\label{Fig_M1_Power_T25_methods}}Small-sample size-unadjusted
power for model M1. The sample size is $T=250.$}
}{\footnotesize\par}
\end{figure}
\end{center}

\begin{center}
\begin{figure}[H]
\includegraphics[width=16cm,height=8cm]{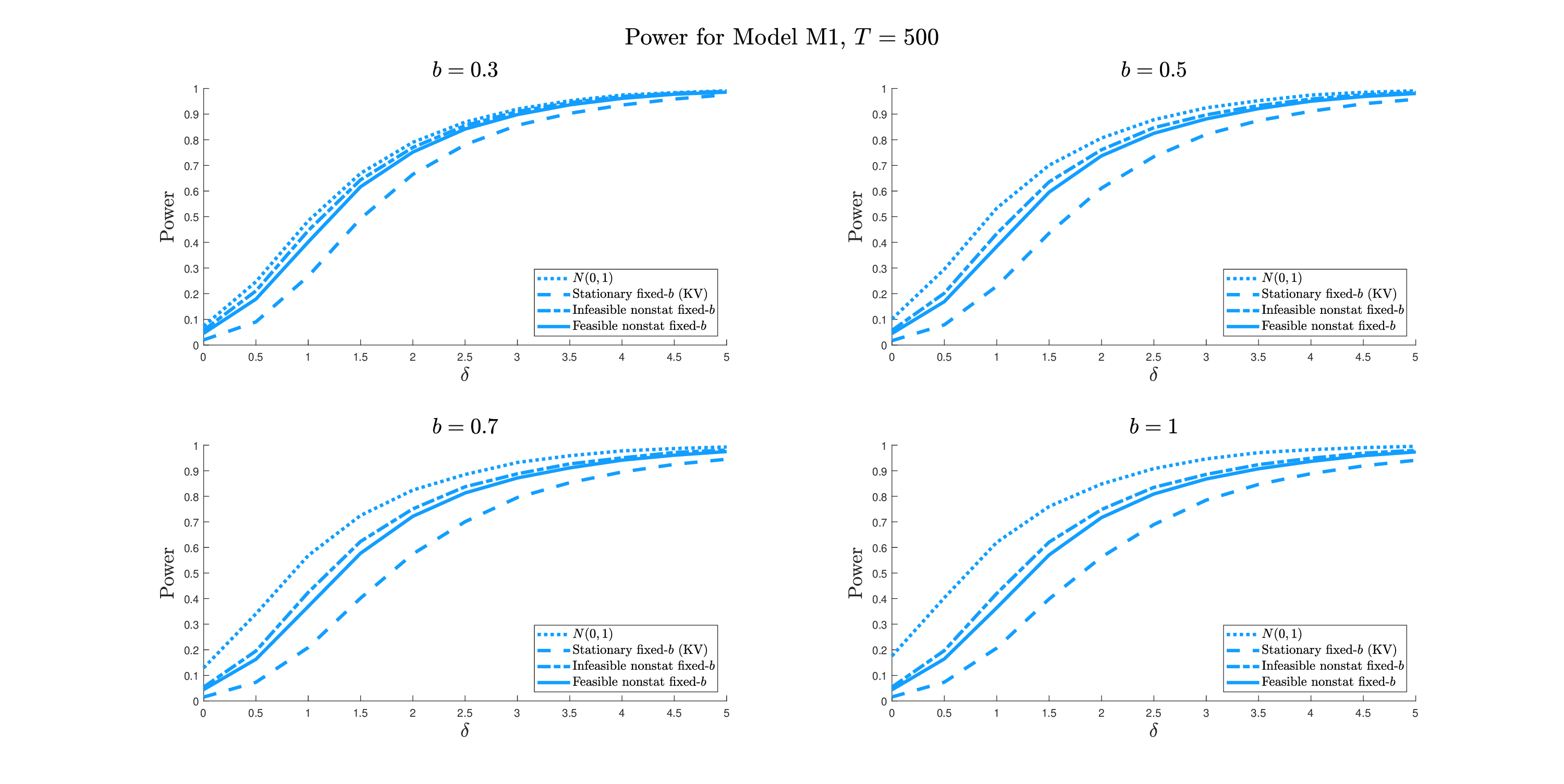}

{\footnotesize{}\caption{{\scriptsize{}\label{Fig_M1_Power_T500_methods}}Small-sample size-unadjusted
power for model M1. The sample size is $T=500.$}
}{\footnotesize\par}
\end{figure}
\end{center}

We now move to discuss the local power results. We consider the following
null and alternative hypotheses 
\begin{align*}
H_{0}:\,\beta_{0}\leq0 & \qquad\mathrm{vs.}\qquad H_{1}:\,\beta_{0}=cT^{-1/2},
\end{align*}
where $c=\delta\sqrt{\Omega}>0$ is a constant. The local power is
computed for $\delta=0,\,0.2,\,0.4,\ldots,\,4.8,\,5$ using 5\% asymptotic
null critical values. What we report is the size-unadjusted power.
We only report the results for model M1 since the results for model
M2-M3 are qualitative similar. Figure \ref{Fig_M1_Power_T250_bs}
and \ref{Fig_M1_Power_T500_methods} plot the power across methods
for a given value of $b$. Figure \ref{Fig_M1_Power_T25_methods}
plots the power across values of $b$ for a given method. The power
corresponding to the standard normal critical value is much higher
than that associated to the fixed-$b$ methods. This is intuitive
given that the standard normal critical value leads to oversized tests.
Hence, it is fair to focus on the power comparison between the stationary
fixed-$b$ method and the infeasible and feasible nonstationary fixed-$b$
methods since they are not oversized. The figures show that the power
gain from using the nonstationary fixed-$b$ critical values is roughly
10\% for $T=250$ and roughly 15\% for $T=500.$ All tests have monotonic
power and as $\delta$ increases the power differences become smaller
and smaller. These features continue to hold for model M2. Thus, the
under-rejections of the stationary fixed-$b$ inference lead to power
losses relative to the nonstationary fixed-$b$ inference. In model
M3 the power functions associated to the stationary and nonstationary
fixed-$b$ methods are essentially the same. Thus, there is no loss
in using the nonstationary fixed-$b$ inference when the data are
stationary. 

To sum up, the theoretical results of Section \ref{Section: Fixed--Limiting-Distribution}-\ref{Section ERP}
provide useful predictions about the finite-sample accuracy of the
stationary and nonstationary fixed-$b$ distributions. The nonstationary
fixed-$b$ distribution yields empirical null rejection rates that
are accurate for both stationary and nonstationary data. The stationary
fixed-$b$ distribution yields null rejection rates that are substantially
below the significance level when the data are nonstationary. It follows
that its corresponding power is lower than that associated to the
nonstationary fixed-$b$ distribution.

Finally, recent works by \citet{casini_hac} and \citet{casini/perron_Low_Frequency_Contam_Nonstat:2020}
showed that in the context of nonstationary alternative hypotheses
the stationary fixed-$b$ method as well as the traditional small-$b$
HAC methods exhibit non-monotonic power. This involves testing problems
outside the linear regression (e.g., tests for structural breaks,
time-varying parameters and regime switching, and tests for forecast
evaluation). By construction the ERP results in Section \ref{Section ERP}
are only relevant for the size properties of the tests and thus are
not suitable for nonstationary alternative hypotheses. We verified
via simulations that also the nonstationary fixed-$b$ method can
suffer from non-monotonic power in those contexts, though by a smaller
extent. The only methods that have monotonic power are those based
on the DK-HAC estimators of \citet{casini_hac}. Hence, it would be
interesting to combine the DK-HAC estimation with the nonstationary
fixed-$b$ asymptotics in future research. 

\section{\label{Section Conclusions}Conclusions}

This paper has shown that the theoretical properties of fixed-$b$
HAR inference change depending on whether the data are stationary
or not. Under nonstationarity we established that fixed-$b$ HAR test
statistics have a limiting distribution that is not pivotal and that
their ERP is an order of magnitude larger than that under stationarity
and can be larger than that of HAR tests based on traditional HAC
estimators. These theoretical results reconcile with recent finite-sample
evidence showing that fixed-$b$ HAR test statistics can perform poorly
when the data are nonstationary, both in terms of distortions in the
null rejection rates and of non-monotonic power. Overall, the results
highlight a new facet of the trade-off between size and power in HAR
inference, i.e., the methods that achieve better null rejection rates
under stationarity are the ones that suffer more from under-rejection
under nonstationarity (i.e., time-varying autocovariance structure),
and vice-versa. A new inference method based on the nonstationary
fixed-$b$ distribution is proposed and it is shown to provide accurate
null rejection rates for hypothesis testing in the linear regression
model irrespective of whether the data are stationary or not and of
the strength of the dependence as verified for some representative
data-generating processes in a simple location model.

\subsection*{Supplemental Materials}

The supplement for online publication {[}cf. \citet{casini_fixed_b_erp_supp}{]}
contains the proofs of the results. 

\section{Appendix}

\subsection{Figures}

\begin{singlespace}

\end{singlespace}

\subsection{Table}

\begin{table}[H]
\caption{\label{Table 1}Small-sample null rejection rates for model M1-M3}

\bigskip{}

\centering{}%
\begin{tabular}{ccccccccc}
Model M1 & \multicolumn{4}{c}{$T=250$} & \multicolumn{4}{c}{$T=500$}\tabularnewline
Critical value\textbackslash$b${} & 0.25 & 0.5 & 0.75 & 1 & 0.25 & 0.5 & 0.75 & 1\tabularnewline
\hline 
\hline 
$N\left(0,\,1\right)$ & 0.071 & 0.098 & 0.128 & 0.166 & 0.066 & 0.099 & 0.133 & 0.178\tabularnewline
Stationary fixed-$b$ (KV) & 0.025 & 0.019 & 0.016 & 0.015 & 0.020 & 0.017 & 0.016 & 0.015\tabularnewline
Infeasible nonstat fixed-$b$ & 0.060 & 0.057 & 0.059 & 0.059 & 0.057 & 0.058 & 0.059 & 0.059\tabularnewline
Feasible nonstat fixed-$b$ & 0.044 & 0.039 & 0.041 & 0.040 & 0.046 & 0.046 & 0.045 & 0.045\tabularnewline
\hline 
\multicolumn{9}{c}{}\tabularnewline
Model M2 & \multicolumn{4}{c}{$T=250$} & \multicolumn{4}{c}{$T=500$}\tabularnewline
Critical value\textbackslash$b${} & 0.25 & 0.5 & 0.75 & 1 & 0.25 & 0.5 & 0.75 & 1\tabularnewline
\hline 
\hline 
$N\left(0,\,1\right)$ & 0.066 & 0.099 & 0.133 & 0.178 & 0.080 & 0.109 & 0.146 & 0.187\tabularnewline
Stationary fixed-$b$ (KV) & 0.020 & 0.017 & 0.016 & 0.016 & 0.028 & 0.019 & 0.016 & 0.016\tabularnewline
Infeasible nonstat fixed-$b$ & 0.058 & 0.058 & 0.059 & 0.059 & 0.050 & 0.049 & 0.052 & 0.051\tabularnewline
Feasible nonstat fixed-$b$ & 0.046 & 0.045 & 0.045 & 0.045 & 0.047 & 0.044 & 0.044 & 0.043\tabularnewline
\hline 
\multicolumn{9}{c}{}\tabularnewline
Model M3 & \multicolumn{4}{c}{$T=250$} & \multicolumn{4}{c}{$T=500$}\tabularnewline
Critical value\textbackslash$b${} & 0.25 & 0.5 & 0.75 & 1 & 0.25 & 0.5 & 0.75 & 1\tabularnewline
\hline 
\hline 
$N\left(0,\,1\right)$ & 0.126 & 0.162 & 0.191 & 0.220 & 0.109 & 0.153 & 0.189 & 0.222\tabularnewline
Stationary fixed-$b$ (KV) & 0.071 & 0.071 & 0.068 & 0.068 & 0.058 & 0.057 & 0.056 & 0.057\tabularnewline
Infeasible nonstat fixed-$b$ & 0.072 & 0.069 & 0.068 & 0.068 & 0.059 & 0.058 & 0.057 & 0.057\tabularnewline
Feasible nonstat fixed-$b$ & 0.079 & 0.073 & 0.074 & 0.075 & 0.065 & 0.061 & 0.060 & 0.062\tabularnewline
\hline 
\end{tabular}
\end{table}

\newpage{}

\newpage{}

\bibliographystyle{elsarticle-harv}
\bibliography{References_JoE}
\addcontentsline{toc}{section}{References}

\newpage{}

\newpage{}

\clearpage 
\pagenumbering{arabic}
\renewcommand*{\thepage}{A-\arabic{page}}
\appendix

\pagebreak{}

\section*{}
\addcontentsline{toc}{part}{Supplemental Material}

\begin{center}
\title{\textbf{\Large{Supplement to ``The Fixed-$\boldsymbol{b}$ Limiting Distribution and the ERP of HAR Tests Under Nonstationarity"}}} 
\maketitle
\end{center}
\medskip{} 
\medskip{} 
\medskip{} 
\thispagestyle{empty}

\begin{center}
\author{\textsc{\textcolor{MyBlue}{Alessandro Casini}}}\\ 
\medskip{}
\medskip{} 
\medskip{} 

\small{{Department of Economics and Finance}}\\
\small{{University of Rome Tor Vergata}}\\
\medskip{}
\medskip{} 
\medskip{} 
\medskip{} 
\date{\small{\today}} 
\medskip{} 
\medskip{} 
\medskip{} 
\end{center}
\begin{abstract}
{\footnotesize{}This supplemental material is for online publication.
It contains the proofs of the results.}{\footnotesize\par}
\end{abstract}
\setcounter{page}{0}
\setcounter{section}{0}
\renewcommand*{\theHsection}{\the\value{section}}

\newpage{}

\begin{singlespace} 
\noindent 
\small

\allowdisplaybreaks


\renewcommand{\thepage}{S-\arabic{page}}   
\renewcommand{\thesection}{S.\Alph{section}}   
\renewcommand{\theequation}{S.\arabic{equation}}




\section{\label{Section Mathematical-Appendix}Mathematical Proofs}

\subsection{Proof of Theorem \ref{Theorem: Limiting Distribution Omega_fixedb}}

Define
\begin{align*}
Q_{T}\left(r\right)=T^{-1}\sum_{t=1}^{\left\lfloor Tr\right\rfloor }x_{t}x'_{t}, & \qquad X_{T}\left(r\right)=T^{-1/2}S_{\left\lfloor Tr\right\rfloor }.
\end{align*}
 Let $K_{b}\left(\cdot\right)=K\left(\cdot/b\right)$ and 
\begin{align*}
D_{b,T}\left(r\right) & =T^{2}\left[\left(K_{b}\left(\frac{\left\lfloor Tr\right\rfloor +1}{T}\right)-K_{b}\left(\frac{\left\lfloor Tr\right\rfloor }{T}\right)\right)-\left(K_{b}\left(\frac{\left\lfloor Tr\right\rfloor }{T}\right)-K_{b}\left(\frac{\left\lfloor Tr\right\rfloor -1}{T}\right)\right)\right].
\end{align*}
By symmetry of $K\left(\cdot\right)$, it follows the symmetry of
$D_{b,T}\left(\cdot\right)$.  If $K''\left(r\right)$ is assumed
to exist, then $\lim_{T\rightarrow\infty}D_{b,T}\left(r\right)=K_{b}''\left(r\right)$.
The convergence is uniform in $r$ if $K''\left(r\right)$ is continuous.
From Assumption \ref{Assumption: Assumption 1 in KV (2002), S_Tr, Nonstationarity}-\ref{Assumption: Assumption 2 in KV (2002), Qr, Nonstationarity}
it follows that $(Q_{T}\left(r\right),\,X_{T}\left(r\right)',\,D_{b,T}\left(r\right))\Rightarrow(\int_{0}^{r}Q\left(u\right)du,\,(\int_{0}^{r}\Sigma\left(u\right)dW_{p}\left(u\right))',\,K_{b}''\left(r\right))$
jointly.  

Define $K_{i,j}=\left(\left(i-j\right)/\left(bT\right)\right)$. We
have
\begin{align*}
\widehat{\Omega}_{\mathrm{fixed}-b} & =T^{-1}\sum_{i=1}^{T}\sum_{j=1}^{T}K_{i,j}\widehat{V}_{i}\widehat{V}'_{j}=T^{-1}\sum_{i=1}^{T}\widehat{V}_{i}\left(\sum_{j=1}^{T}K_{i,j}\widehat{V}'_{j}\right).
\end{align*}
Note that
\begin{align}
T^{2} & \left(\left(K_{i,j}-K_{i,j+1}\right)-\left(K_{i+1,j}-K_{i+1,j+1}\right)\right)\label{Eq. (A.1) in KV (2002)}\\
 & =-T^{2}\left[\left(K_{b}\left(\frac{i-j+1}{T}\right)-K_{b}\left(\frac{i-j}{T}\right)\right)-\left(K_{b}\left(\frac{i-j}{T}\right)-K_{b}\left(\frac{i-j-1}{T}\right)\right)\right]\nonumber \\
 & =D_{b,T}\left(\left(i-j\right)/T\right)\nonumber 
\end{align}
Define $\widehat{S}_{t}=\sum_{j=1}^{t}\widehat{V}_{j}$. Note that
$\widehat{S}_{T}=\mathbf{0}$ by the normal equations for OLS. We
have 
\begin{align}
T^{-1/2}\widehat{S}_{\left\lfloor Tr\right\rfloor } & =T^{-1/2}S_{\left\lfloor Tr\right\rfloor }-T^{-1}\sum_{t=1}^{\left\lfloor Tr\right\rfloor }x_{t}x'_{t}\left(T^{-1}\sum_{t=1}^{T}x_{t}x'_{t}\right)^{-1}T^{-1/2}S_{T}\label{Eq. S_hat_Tr}\\
 & =X_{T}\left(r\right)-Q_{T}\left(r\right)Q_{T}\left(1\right)^{-1}X_{T}\left(1\right).\nonumber 
\end{align}
Using the identity
\begin{align}
\sum_{l=1}^{T}a_{l}b_{l} & =\sum_{l=1}^{T-1}\left(\left(a_{l}-a_{l+1}\right)\sum_{j=1}^{l}b_{j}\right)+a_{T}\sum_{j=1}^{T}b_{j},\label{Eq. (Identity)}
\end{align}
 first applied to $\sum_{j=1}^{T}K_{i,j}\widehat{V}'_{j}$ and then
again to the sum over $i$, \citeReferencesSupp{vogelsang/kiefer:2002ET}
showed that
\begin{align}
\widehat{\Omega}_{\mathrm{fixed}-b} & =T^{-1}\sum_{i=1}^{T-1}T^{-1}\sum_{j=1}^{T-1}T^{2}\left(\left(K_{i,j}-K_{i,j+1}\right)-\left(K_{i+1,j}-K_{i+1,j+1}\right)\right)T^{-1/2}\widehat{S}_{i}T^{-1/2}\widehat{S}'_{j}.\label{Eq. 16 KV05}
\end{align}
We first consider part (i). Using \eqref{Eq. (A.1) in KV (2002)}-\eqref{Eq. S_hat_Tr}
in \eqref{Eq. 16 KV05} we have  
\begin{align*}
\widehat{\Omega}_{\mathrm{fixed}-b} & =\int_{0}^{1}\int_{0}^{1}-D_{b,T}\left(r-s\right)\left[X_{T}\left(r\right)-Q_{T}\left(r\right)Q_{T}\left(1\right)^{-1}X_{T}\left(1\right)\right]\left[X_{T}\left(s\right)-Q_{T}\left(s\right)Q_{T}\left(1\right)^{-1}X_{T}\left(1\right)\right]'drds\\
 & \Rightarrow-\int_{0}^{1}\int_{0}^{1}K_{b}''\left(r-s\right)\left(\int_{0}^{r}\Sigma\left(u\right)dW_{p}\left(u\right)-\left(\int_{0}^{r}Q\left(u\right)du\right)\overline{Q}^{-1}\int_{0}^{1}\Sigma\left(u\right)dW_{p}\left(u\right)\right)\\
 & \quad\times\left(\int_{0}^{s}\Sigma\left(u\right)dW_{p}\left(u\right)-\left(\int_{0}^{s}Q\left(u\right)du\right)\overline{Q}^{-1}\int_{0}^{1}\Sigma\left(u\right)dW_{p}\left(u\right)\right)'drds\\
 & =-\frac{1}{b^{2}}\int_{0}^{1}\int_{0}^{1}K''\left(\frac{r-s}{b}\right)\widetilde{B}_{p}\left(r,\,\Sigma,\,Q\right)\widetilde{B}_{p}\left(s,\,\Sigma,\,Q\right)'drds\\
 & =\mathscr{G}_{b},
\end{align*}
where we have used Assumption \ref{Assumption: Assumption 1 in KV (2002), S_Tr, Nonstationarity}-\ref{Assumption: Assumption 2 in KV (2002), Qr, Nonstationarity},
the continuous mapping theorem since $\widehat{\Omega}_{\mathrm{fixed}-b}$
is a continuous function of $(Q_{T}\left(r\right),$ $X_{T}\left(r\right)',\,D_{b.T}\left(r\right))$
and $K_{b}''\left(x\right)=b^{-2}K''\left(x/b\right)$.

We now move to part (ii). Suppose that the Bartlett kernel $K_{\mathrm{BT}}$
is used. Let 
\begin{align*}
\Delta^{2}K_{ij} & \triangleq\left(K_{i,j}-K_{i,j+1}\right)-\left(K_{i+1,j}-K_{i+1,j+1}\right).
\end{align*}
 Note that $\Delta^{2}K_{i,j}=2/(bT)$ for $\left|i-j\right|=0$,
$\Delta^{2}K_{i,j}=-1/\left(bT\right)+1-\left\lfloor bT\right\rfloor /\left(bT\right)$
for $\left|i-j\right|=\left\lfloor bT\right\rfloor $, $\Delta^{2}K_{i,j}=-\left(1-\left\lfloor bT\right\rfloor /bT\right)$
for $\left|i-j\right|=\left\lfloor bT\right\rfloor +1$ and $\Delta^{2}K_{i,j}=0$
otherwise. Using this into \eqref{Eq. 16 KV05} we obtain 
\begin{align*}
\widehat{\Omega}_{\mathrm{fixed}-b} & =T^{-1}\sum_{i=1}^{T-1}T^{-1}\sum_{j=1}^{T-1}T^{2}\Delta^{2}K_{i,j}T^{-1/2}\widehat{S}_{i}T^{-1/2}\widehat{S}'_{j}\\
 & =\frac{2}{bT}\sum_{i=1}^{T-1}T^{-1/2}\widehat{S}_{i}T^{-1/2}\widehat{S}'_{i}\\
 & \quad+T\left[-\frac{1}{bT}+1-\frac{\left\lfloor bT\right\rfloor }{bT}\right]T^{-1}\sum_{i=1}^{T-\left\lfloor bT\right\rfloor -1}\left(T^{-1/2}\widehat{S}_{i+\left\lfloor bT\right\rfloor }T^{-1/2}\widehat{S}'_{i}+T^{-1/2}\widehat{S}{}_{i}T^{-1/2}\widehat{S}'_{i+\left\lfloor bT\right\rfloor }\right)\\
 & \quad-\left(1-\frac{\left\lfloor bT\right\rfloor }{bT}\right)\sum_{i=1}^{T-\left[bT\right]-2}\left(T^{-1/2}\widehat{S}_{i+\left\lfloor bT\right\rfloor +1}T^{-1/2}\widehat{S}'_{i}+T^{-1/2}\widehat{S}{}_{i}T^{-1/2}\widehat{S}'_{i+\left\lfloor bT\right\rfloor +1}\right)\\
 & =\frac{2}{bT}\sum_{i=1}^{T-1}T^{-1/2}\widehat{S}_{i}T^{-1/2}\widehat{S}'_{i}\\
 & \quad+T\left[-\frac{1}{bT}+1-\frac{\left\lfloor bT\right\rfloor }{bT}\right]T^{-1}\sum_{i=1}^{T-\left\lfloor bT\right\rfloor -1}\left(T^{-1/2}\widehat{S}_{i+\left\lfloor bT\right\rfloor }T^{-1/2}\widehat{S}'_{i}+T^{-1/2}\widehat{S}{}_{i}T^{-1/2}\widehat{S}'_{i+\left\lfloor bT\right\rfloor }\right)\\
 & \quad-\left(1-\frac{\left\lfloor bT\right\rfloor }{bT}\right)\sum_{i=1}^{T-\left[bT\right]-1}\left(T^{-1/2}\widehat{S}_{i+\left\lfloor bT\right\rfloor +1}T^{-1/2}\widehat{S}'_{i}+T^{-1/2}\widehat{S}{}_{i}T^{-1/2}\widehat{S}'_{i+\left\lfloor bT\right\rfloor +1}\right),
\end{align*}
where we have used that $\widehat{S}_{T}\widehat{S}'_{T-\left\lfloor bT\right\rfloor -1}=\mathbf{0}$
and $\widehat{S}{}_{T-\left\lfloor bT\right\rfloor -1}\widehat{S}'_{T}=\mathbf{0}$.
Note that
\begin{align*}
\left(1-\frac{\left\lfloor bT\right\rfloor }{bT}\right) & \sum_{i=1}^{T-\left\lfloor bT\right\rfloor -1}T^{-1/2}\widehat{S}_{i+\left[bT\right]+1}T^{-1/2}\widehat{S}'_{i}\\
 & =\left(1-\frac{\left\lfloor bT\right\rfloor }{bT}\right)\sum_{i=1}^{T-\left\lfloor bT\right\rfloor -1}\left(T^{-1/2}\widehat{S}_{i+\left\lfloor bT\right\rfloor }T^{-1/2}\widehat{S}'_{i}+T^{-1/2}\widehat{V}_{i+\left\lfloor bT\right\rfloor +1}T^{-1/2}\widehat{S}'_{i}\right)\\
 & =\left(1-\frac{\left\lfloor bT\right\rfloor }{bT}\right)\sum_{i=1}^{T-\left\lfloor bT\right\rfloor -1}T^{-1/2}\widehat{S}_{i+\left\lfloor bT\right\rfloor }T^{-1/2}\widehat{S}'_{i}+\left(1-\frac{\left\lfloor bT\right\rfloor }{bT}\right)T^{-1}\sum_{i=1}^{T-\left\lfloor bT\right\rfloor -1}\widehat{V}_{i+\left\lfloor bT\right\rfloor +1}\widehat{S}'_{i}\\
 & =\left(1-\frac{\left\lfloor bT\right\rfloor }{bT}\right)\sum_{i=1}^{T-\left\lfloor bT\right\rfloor -1}T^{-1/2}\widehat{S}_{i+\left\lfloor bT\right\rfloor }T^{-1/2}\widehat{S}'_{i}+o_{\mathbb{P}}\left(1\right),
\end{align*}
 where the $o_{\mathbb{P}}\left(1\right)$ term follows from $\lim_{T\rightarrow\infty}\left(1-\frac{\left\lfloor bT\right\rfloor }{bT}\right)=0$
and $T^{-1}\sum_{i=1}^{T-\left\lfloor bT\right\rfloor -1}\widehat{V}_{i+\left\lfloor bT\right\rfloor +1}\widehat{S}'_{i}=O_{\mathbb{P}}\left(1\right).$
It follows that
\begin{align*}
\widehat{\Omega}_{\mathrm{fixed}-b} & =\frac{2}{bT}\sum_{i=1}^{T-1}T^{-1/2}\widehat{S}_{i}T^{-1/2}\widehat{S}'_{i}\\
 & \quad-\frac{1}{bT}\sum_{i=1}^{T-\left\lfloor bT\right\rfloor -1}\left(T^{-1/2}\widehat{S}_{i+\left\lfloor bT\right\rfloor }T^{-1/2}\widehat{S}'_{i}+T^{-1/2}\widehat{S}{}_{i}T^{-1/2}\widehat{S}'_{i+\left\lfloor bT\right\rfloor }\right)+o_{\mathbb{P}}\left(1\right).
\end{align*}
 Using \eqref{Eq. S_hat_Tr} and Assumption \ref{Assumption: Assumption 1 in KV (2002), S_Tr, Nonstationarity}-\ref{Assumption: Assumption 2 in KV (2002), Qr, Nonstationarity}
we yield, 
\begin{align*}
\widehat{\Omega}_{\mathrm{fixed}-b} & \Rightarrow\frac{2}{b}\int_{0}^{1}\left(\int_{0}^{r}\Sigma\left(u\right)dW_{p}\left(u\right)-\left(\int_{0}^{r}Q\left(u\right)du\right)\overline{Q}^{-1}\int_{0}^{1}\Sigma\left(u\right)dW_{p}\left(u\right)\right)\\
 & \quad\times\left(\int_{0}^{r}\Sigma\left(u\right)dW_{p}\left(u\right)-\left(\int_{0}^{r}Q\left(u\right)du\right)\overline{Q}^{-1}\int_{0}^{1}\Sigma\left(u\right)dW_{p}\left(u\right)\right)'dr\\
 & \quad-\frac{1}{b}\int_{0}^{1-b}\Biggl(\left(\int_{0}^{r+b}\Sigma\left(u\right)dW_{p}\left(u\right)-\left(\int_{0}^{r+b}Q\left(u\right)du\right)\overline{Q}^{-1}\int_{0}^{1}\Sigma\left(u\right)dW_{p}\left(u\right)\right)\\
 & \quad\times\left(\int_{0}^{r}\Sigma\left(u\right)dW_{p}\left(u\right)-\left(\int_{0}^{r}Q\left(u\right)du\right)\overline{Q}^{-1}\int_{0}^{1}\Sigma\left(u\right)dW_{p}\left(u\right)\right)'\\
 & \quad+\left(\int_{0}^{r}\Sigma\left(u\right)dW_{p}\left(u\right)-\left(\int_{0}^{r}Q\left(u\right)du\right)\overline{Q}^{-1}\int_{0}^{1}\Sigma\left(u\right)dW_{p}\left(u\right)\right)\\
 & \quad\times\left(\int_{0}^{r+b}\Sigma\left(u\right)dW_{p}\left(u\right)-\left(\int_{0}^{r+b}Q\left(u\right)du\right)\overline{Q}^{-1}\int_{0}^{1}\Sigma\left(u\right)dW_{p}\left(u\right)\right)'\Biggr)dr\\
 & =\frac{2}{b}\int_{0}^{1}\widetilde{B}_{p}\left(r,\,\Sigma,\,Q\right)\widetilde{B}_{p}\left(r,\,\Sigma,\,Q\right)'dr\\
 & \quad-\frac{1}{b}\int_{0}^{1-b}\left(\widetilde{B}_{p}\left(r+b,\,\Sigma,\,Q\right)\widetilde{B}_{p}\left(r,\,\Sigma,\,Q\right)'+\widetilde{B}_{p}\left(r,\,\Sigma,\,Q\right)\widetilde{B}_{p}\left(r+b,\,\Sigma,\,Q\right)'\right)dr,
\end{align*}
which concludes the proof. $\square$ 

\subsection{Proof of Theorem \ref{Theorem 2: KV (2002)}}

We begin with part (i). Using Theorem \ref{Theorem: Limiting Distribution Omega_fixedb}
we have 

\begin{align*}
F_{\mathrm{fixed}-b} & =\left(RQ_{T}\left(1\right)^{-1}X_{T}\left(1\right)\right)'\left(RQ_{T}\left(1\right)^{-1}\widehat{\Omega}_{\mathrm{fixed}-b}Q_{T}\left(1\right)^{-1}R'\right)^{-1}RQ_{T}\left(1\right)^{-1}X_{T}\left(1\right)/q\\
 & \Rightarrow\left(R\overline{Q}^{-1}\int_{0}^{1}\Sigma\left(u\right)dW_{p}\left(u\right)\right)'\left(R\overline{Q}^{-1}\mathscr{G}_{b}\overline{Q}^{-1}R'\right)^{-1}R\overline{Q}^{-1}\int_{0}^{1}\Sigma\left(u\right)dW_{p}\left(u\right)/q
\end{align*}
where $\mathscr{G}_{b}$ is defined in \eqref{Eq. Asymptotic Distribution of Omega fixed_b}.
If $q=1,$ then 
\begin{align*}
t_{\mathrm{fixed}-b} & =\frac{RQ_{T}\left(1\right)^{-1}X_{T}\left(1\right)}{\sqrt{RQ_{T}\left(1\right)^{-1}\widehat{\Omega}_{\mathrm{fixed}-b}Q_{T}\left(1\right)^{-1}R'}}\\
 & \Rightarrow\frac{R\overline{Q}^{-1}\int_{0}^{1}\Sigma\left(u\right)dW_{p}\left(u\right)}{\sqrt{R\overline{Q}^{-1}\mathscr{G}_{b}\overline{Q}^{-1}R'}}.
\end{align*}
 Part (ii) can be proved in a similar manner. $\square$ 

\subsection{Proof of Theorem \ref{Theorem ERP Fixedb}}

Throughout the proof, let $\widehat{\Omega}_{b}=\widehat{\Omega}_{\mathrm{fixed}-b}$
and 
\begin{align*}
Z_{T,0}\left(z\right)\triangleq\mathbb{P}\left(\left|\frac{\sqrt{T}\left(\widehat{\beta}-\beta_{0}\right)}{\sqrt{\widehat{\Omega}_{b}}}\right|\leq z\right),\quad & \qquad\qquad Z_{0}\left(z\right)\triangleq\mathbb{P}\left(\left|\frac{\int_{0}^{1}\Sigma\left(u\right)dW_{1}\left(u\right)}{\sqrt{\mathscr{G}_{b}}}\right|\leq z\right),\\
\widehat{Z}_{0}\left(z\right)\triangleq\mathbb{P} & \left(\left|\frac{\int_{0}^{1}\widehat{\Sigma}\left(u\right)dW_{1}\left(u\right)}{\sqrt{\mathscr{\widehat{G}}_{b}}}\right|\leq z\right).
\end{align*}
We have 
\begin{align*}
\sup_{z\in\mathbb{R}_{+}}\left|Z_{T,0}\left(z\right)-\widehat{Z}_{0}\left(z\right)\right| & \leq\sup_{z\in\mathbb{R}_{+}}\left|Z_{T,0}\left(z\right)-Z_{0}\left(z\right)\right|+\sup_{z\in\mathbb{R}_{+}}\left|Z_{0}\left(z\right)-\widehat{Z}_{0}\left(z\right)\right|\\
 & \triangleq D_{1}+D_{2}.
\end{align*}
 We show that $D_{1}=O(T^{-1})$ and $D_{2}=O((Th_{1}h_{2})^{-1/2})$.
  We begin with some preliminary lemmas. The first lemma below
generalizes Theorem \ref{Theorem: Limiting Distribution Omega_fixedb}
to allow for general kernels satisfying Assumption \ref{Assumption: Assumption 2 in SPJ (2008)}
and for a $p$-vector $\widehat{V}_{t}$. Let 
\begin{align*}
\widetilde{B}_{p}\left(r\right) & =\int_{0}^{r}\Sigma\left(u\right)dW_{p}\left(u\right)-r\left(\int_{0}^{1}\Sigma\left(u\right)dW_{p}\left(u\right)\right),
\end{align*}
 and
\begin{align*}
K_{b}^{*}\left(r,\,s\right) & \triangleq K_{b}\left(r-s\right)-\int_{0}^{1}K_{b}\left(r-t\right)dt-\int_{0}^{1}K_{b}\left(\tau-s\right)d\tau+\int_{0}^{1}\int_{0}^{1}K_{b}\left(t-\tau\right)dtd\tau.
\end{align*}

\begin{lem}
\label{Lemma: Theorem 1 in PSJ (2007)} Let Assumption \ref{Assumption: Assumption 1 in KV (2002), S_Tr, Nonstationarity}
and \ref{Assumption: Assumption 2 in SPJ (2008)} hold. We have:

(i) $\widehat{\Omega}_{b}\Rightarrow\mathscr{G}_{b}$ where $\mathscr{G}_{b}=\int_{0}^{1}\int_{0}^{1}K_{b}\left(r-s\right)d\widetilde{B}_{p}\left(r\right)d\widetilde{B}_{p}\left(s\right)'$;

(ii) $\mu_{b}=\mathbb{E}\left(\mathscr{G}_{b}\right)=\int_{0}^{1}K_{b}^{*}\left(s,\,s\right)\Omega\left(s\right)ds$.
\end{lem}
\noindent\textit{Proof of Lemma} \ref{Lemma: Theorem 1 in PSJ (2007)}.
We begin with part (i). Since $K\left(\cdot\right)$ is positive semidefinite,
Mercer's Theorem implies that 
\begin{align}
K\left(r-s\right) & =\sum_{n=1}^{\infty}\frac{1}{\lambda_{n}}g_{n}\left(r\right)g_{n}\left(s\right),\label{Eq. (13) in PSJ (2007), Mercer's Theorem}
\end{align}
 where $\lambda_{n}^{-1}>0$ are the eigenvalues of $K\left(\cdot\right)$
and $g_{n}\left(\cdot\right)$ are the corresponding eigenfunctions,
i.e., $g_{n}\left(s\right)=\lambda_{n}\int_{0}^{1}K\left(r-s\right)g_{n}\left(r\right)dr$.
The convergence of the right-hand side over $\left(r,\,s\right)\in\left[0,\,1\right]\times\left[0,\,1\right]$
is uniform. 

Using Assumption \ref{Assumption: Assumption 1 in KV (2002), S_Tr, Nonstationarity}
and \eqref{Eq. (13) in PSJ (2007), Mercer's Theorem} we have 
\begin{align*}
\widehat{\Omega}_{b} & =\frac{1}{T}\sum_{t=1}^{T}\sum_{s=1}^{T}K_{b}\left(\frac{t-s}{T}\right)\widehat{V}_{t}\widehat{V}'_{s}\\
 & =\sum_{n=1}^{\infty}\frac{1}{\lambda_{n}}\frac{1}{T}\sum_{t=1}^{T}\sum_{s=1}^{T}g_{n}\left(t/(bT)\right)g_{n}\left(s/(bT)\right)\widehat{V}_{t}\widehat{V}'_{s}\\
 & =\sum_{n=1}^{\infty}\frac{1}{\lambda_{n}}\left(\frac{1}{\sqrt{T}}\sum_{t=1}^{T}\widehat{V}_{t}g_{n}\left(t/(bT)\right)\right)\left(\frac{1}{\sqrt{T}}\sum_{s=1}^{T}g_{n}\left(s/(bT)\right)\widehat{V}'_{s}\right)\\
 & \Rightarrow\sum_{n=1}^{\infty}\frac{1}{\lambda_{n}}\left(\int_{0}^{1}g_{n}\left(r/b\right)d\widetilde{B}_{p}\left(r\right)\right)\left(\int_{0}^{1}g_{n}\left(s/b\right)d\widetilde{B}_{p}\left(s\right)\right)'\\
 & =\int_{0}^{1}\int_{0}^{1}\sum_{n=1}^{\infty}\frac{1}{\lambda_{n}}g_{n}\left(r/b\right)g_{n}\left(s/b\right)d\widetilde{B}_{p}\left(r\right)d\widetilde{B}_{p}\left(s\right)'\\
 & =\int_{0}^{1}\int_{0}^{1}K_{b}\left(r-s\right)d\widetilde{B}_{p}\left(r\right)d\widetilde{B}_{p}\left(s\right)'.\\
 & =\mathscr{G}_{b}
\end{align*}
For part (ii), after some algebraic manipulations we can write 
\begin{align*}
\mathscr{G}_{b} & =\int_{0}^{1}\int_{0}^{1}K_{b}^{*}\left(r,\,s\right)\Sigma\left(r\right)dW_{p}\left(r\right)\left(\Sigma\left(s\right)dW_{p}\left(s\right)\right)'.
\end{align*}
 It follows that 
\begin{align*}
\mathbb{E}\left(\mathscr{G}_{b}\right) & =\mathbb{E}\left(\int_{0}^{1}\int_{0}^{1}K_{b}^{*}\left(r,\,s\right)\Sigma\left(r\right)dW_{p}\left(r\right)\left(\Sigma\left(s\right)dW_{p}\left(s\right)\right)'\right)\\
 & =\int_{0}^{1}K_{b}^{*}\left(s,\,s\right)\Sigma\left(s\right)\Sigma\left(s\right)'ds\\
 & =\int_{0}^{1}K_{b}^{*}\left(s,\,s\right)\Omega\left(s\right)ds,
\end{align*}
 which concludes the proof. $\square$ 

\medskip{}

Let $p=1$. \citeReferencesSupp{sun/phillips/jin:08} showed that
$K_{b}^{*}\left(r,\,s\right)$ is positive semidefinite. Thus, using
Mercer's theorem, we have 
\begin{align}
K_{b}^{*}\left(r,\,s\right) & =\sum_{n=1}^{\infty}\lambda_{n}^{*}g_{n}^{*}\left(r\right)g_{n}^{*}\left(s\right),\label{Eq. (25) SPJ (2008)}
\end{align}
where $\lambda_{n}^{*}>0$ are the eigenvalues of $K_{b}^{*}\left(\cdot,\,\cdot\right)$
and $g_{n}^{*}\left(r\right)$ are the corresponding eigenfunctions,
i.e., $\lambda_{n}^{*}g_{n}^{*}\left(s\right)=\int_{0}^{\infty}K_{b}^{*}\left(r,\,s\right)g_{n}^{*}\left(r\right)dr$.
 Since $\Sigma\left(s\right)>0$, we can write 
\begin{align*}
K_{b}^{*}\left(r,\,s\right)\Sigma\left(r\right)\Sigma\left(s\right) & =\sum_{n=1}^{\infty}\lambda_{n}^{*}g_{n}^{**}\left(r\right)g_{n}^{**}\left(s\right),
\end{align*}
where $g_{n}^{**}\left(r\right)=\Sigma\left(r\right)g_{n}^{*}\left(r\right)$.
Then, $\lambda_{n}^{*}g_{n}^{**}\left(s\right)=\int_{0}^{\infty}K_{b}^{*}\left(r,\,s\right)\Sigma\left(r\right)\Sigma\left(s\right)g_{n}^{**}\left(r\right)dr$.
It follows that  $\mathscr{G}_{b}=\sum_{n=1}^{\infty}\lambda_{n}^{*}Z_{n}^{2}$
where $Z_{n}\sim i.i.d.\,\mathscr{N}(0,\,1)$. Thus, the characteristic
function of $\Omega^{-1}(\mathscr{G}_{b}-\mu_{b})$ is given by
\begin{align}
\psi\left(t\right) & =\mathbb{E}\left(e^{it\Omega^{-1}\left(\mathscr{G}_{b}-\mu_{b}\right)}\right)=\prod_{n=1}^{\infty}\left(1-2i\lambda_{n}^{*}\Omega^{-1}t\right)^{-1/2}\left(e^{-it\Omega^{-1}\mu_{b}}\right),\label{Eq. (26) in SPJ (2008)}
\end{align}
 and the cumulant generating function is
\begin{align}
\log\psi\left(t\right) & =\sum_{m=2}^{\infty}\left(2^{m-1}\left(m-1\right)!\sum_{n=1}^{\infty}\left(\lambda_{n}^{*}\Omega^{-1}\right)^{m}\right)\frac{\left(it\right)^{m}}{m!}.\label{Eq. (27) SPJ (2008)}
\end{align}
 Let $\kappa_{j}$ $\left(j=1,\ldots\right)$ be the $j$th cumulant
of $\Omega^{-1}(\mathscr{G}_{b}-\mu_{b})$. Then 
\begin{align}
\kappa_{1}=0\quad\mathrm{and}\quad & \kappa_{m}=2^{m-1}\left(m-1\right)!\sum_{n=1}^{\infty}\left(\lambda_{n}^{*}\Omega^{-1}\right)^{m}\qquad\mathrm{for}\,m\geq2.\label{Eq. (28) SPJ (2008)}
\end{align}
 Let $\tau_{m+1}=\tau_{1}$. For $m\geq2$, some algebraic manipulations
show that
\begin{align}
\kappa_{m} & =2^{m-1}\left(m-1\right)!\Omega^{-m}\sum_{n=1}^{\infty}\left(\left(g_{n}^{**}\left(s\right)\right)^{-1}\int_{0}^{\infty}K_{b}^{*}\left(r,\,s\right)\Sigma\left(r\right)\Sigma\left(s\right)g_{n}^{**}\left(r\right)dr\right)^{m}\label{Eq. (29) SPJ (2008)}\\
 & =2^{m-1}\left(m-1\right)!\Omega^{-m}\int_{0}^{1}\cdots\int_{0}^{1}\left(\prod_{j=1}^{m}\Omega\left(\tau_{j}\right)K_{b}^{*}\left(\tau_{j},\,\tau_{j+1}\right)\right)d\tau_{1}\cdots d\tau_{m}.\nonumber 
\end{align}
 Let $\Xi_{m}=\Omega^{-m}\mathbb{E}((\mathscr{G}_{b}-\mu_{b})^{m})$
for $m\geq1$. 
\begin{lem}
\label{Lemma: Lemma 1 SPJ (2008)}Let $\overline{C}_{1}=4\int_{-\infty}^{\infty}|K\left(v\right)|dv,$
$C_{\Omega}=\sup_{s\in\left[0,\,1\right]}\Omega\left(s\right),$ $D_{m}>0$
be a constant depending on $m$ and Assumption \ref{Assumption: Assumption 2 in SPJ (2008)}
hold. Then, for $m\geq1$, we have 
\begin{align}
\left|\kappa_{m}\right| & \leq2^{m}\left(m-1\right)!\Omega^{-m}C_{\Omega}^{m}\left(\overline{C}_{1}b\right)^{m-1},\label{Eq. (A.1) SPJ (2008)}
\end{align}
 and 
\begin{align}
\left|\Xi_{m}\right| & \leq D_{m}2^{2m}m!\Omega^{-m}C_{\Omega}^{m}\left(\overline{C}_{1}b\right)^{m-1}.\label{Eq. (A.2) SPJ (2008)}
\end{align}
\end{lem}
\noindent\textit{Proof of Lemma} \ref{Lemma: Lemma 1 SPJ (2008)}.
From eq. (A.3) in \citeReferencesSupp{sun/phillips/jin:08},\footnote{Any reference to equations in \citeReferencesSupp{sun/phillips/jin:08}
corresponds to the long version of the working paper available in
Sun's webpage.} 
\begin{align}
\left|\int_{0}^{1}\cdots\int_{0}^{1}\left(\prod_{j=1}^{m}K_{b}^{*}\left(\tau_{j},\,\tau_{j+1}\right)\right)d\tau_{1}\cdots d\tau_{m}\right| & \leq2\left(\sup_{s}\int_{0}^{1}\left|K_{b}^{*}\left(r,\,s\right)\right|dr\right)^{m-1}.\label{Eq. (A.3) in SPJ (2008)}
\end{align}
We have 
\begin{align}
\sup_{s}\int_{0}^{1}\left|K_{b}^{*}\left(r,\,s\right)\right|dr\leq & b\overline{C}_{1},\label{Eq. (A.5) in SPJ (2008)}
\end{align}
from which it follows that 
\begin{align}
\left|\int_{0}^{1}\cdots\int_{0}^{1}\left(\prod_{j=1}^{m}K_{b}^{*}\left(\tau_{j},\,\tau_{j+1}\right)\right)d\tau_{1}\cdots d\tau_{m}\right| & \leq2\left(b\overline{C}_{1}\right)^{m-1}.\label{Eq. (A.6) SPJ (2008)}
\end{align}
 Using \eqref{Eq. (A.6) SPJ (2008)} and some algebraic manipulations
we have for $m\geq2$, 
\begin{align}
\kappa_{m} & =2^{m-1}\left(m-1\right)!\Omega^{-m}\sum_{n=1}^{\infty}\left(\lambda_{n}^{*}\right)^{m}\label{Eq. (A.7) SPJ (2008)}\\
 & =2^{m-1}\left(m-1\right)!\Omega^{-m}\int_{0}^{1}\cdots\int_{0}^{1}\left(\prod_{j=1}^{m}\Omega\left(\tau_{j}\right)K_{b}^{*}\left(\tau_{j},\,\tau_{j+1}\right)\right)d\tau_{1}\cdots d\tau_{m}\nonumber \\
 & \leq2^{m-1}\left(m-1\right)!\Omega^{-m}C_{\Omega}^{m}\int_{0}^{1}\cdots\int_{0}^{1}\left|\prod_{j=1}^{m}K_{b}^{*}\left(\tau_{j},\,\tau_{j+1}\right)\right|d\tau_{1}\cdots d\tau_{m}\nonumber \\
 & \leq2^{m}\left(m-1\right)!\Omega^{-m}C_{\Omega}^{m}\left(b\overline{C}_{1}\right)^{m-1},\nonumber 
\end{align}
where $C_{\Omega}=\sup_{s\in\left[0,\,1\right]}\Omega\left(s\right)$.
The moments $\left\{ \Xi_{j}\right\} $ and cumulants $\left\{ \kappa_{j}\right\} $
are related by the following
\begin{align}
\Xi_{m} & =\sum_{\pi_{p}}\frac{m!}{\left(j_{1}!\right)^{m_{1}}\left(j_{2}!\right)^{m_{2}}\cdots\left(j_{l}!\right)^{m_{l}}}\frac{1}{m_{1}!m_{2}!\cdots m_{l}!}\prod_{j=\pi_{p}}\kappa_{j},\label{Eq. (A.8) in SPJ (2008)}
\end{align}
 where the sum is taken over all partitions $\pi_{p}\in\Pi$ such
that
\begin{align}
\pi_{p} & =\left[\underset{m_{1}\,\mathrm{times}}{\underbrace{j_{1},\cdots j_{1}}},\,\underset{m_{2}\,\mathrm{times}}{\underbrace{j_{2},\ldots,\,j_{2}}},\ldots,\,\underset{m_{l}\,\mathrm{times}}{\underbrace{j_{l},\ldots,\,j_{l}}}\right],\label{Eq. (A.9) in SPJ (2008)}
\end{align}
 for some integer $l$, sequence $\left\{ j_{i}\right\} _{i=1}^{l}$
such that $j_{1}>j_{2}>\cdots>j_{l}$ and $m=\sum_{i=1}^{l}m_{i}j_{i}$. 

Using \eqref{Eq. (A.7) SPJ (2008)}-\eqref{Eq. (A.9) in SPJ (2008)}
yield
\begin{align}
\left|\Xi_{m}\right| & <2^{m}m!\Omega^{-m}C_{\Omega}^{m}\left(\overline{C}_{1}b\right)^{m-1}\sum_{\pi}\frac{\left(j_{1}\right)^{-m_{1}}\left(j_{2}\right)^{-m_{2}}\cdots\left(j_{l}\right)^{-m_{l}}}{m_{1}!m_{2}!\cdots m_{l}!}j_{1}^{m}\nonumber \\
 & \leq D_{m}2^{2m}m!\Omega^{-m}C_{\Omega}^{m}\left(\overline{C}_{1}b\right)^{m-1},\label{Eq. (A.10) SPJ (2008)}
\end{align}
where the last inequality follows from
\begin{align}
\sum_{\pi_{p}}\frac{\left(j_{1}\right)^{-m_{1}}\left(j_{2}\right)^{-m_{2}}\cdots\left(j_{l}\right)^{-m_{l}}}{m_{1}!m_{2}!\cdots m_{l}!} & \leq\sum_{\pi_{p}}\frac{1}{m_{1}!m_{2}!\cdots m_{l}!}<2^{m},\label{Eq. (A.11) SPJ (2008)}
\end{align}
and $D_{m}=\sup_{\pi_{p}\in\Pi}(j_{1}\in\pi_{p})$. $\square$ 

\medskip{}

We now develop an asymptotic expansion of $Z_{T,0}=\mathbb{P}(\sqrt{T}(\widehat{\beta}-\beta_{0})/\sqrt{\widehat{\Omega}_{b}}\leq z)$
for $\beta=\beta_{0}+d/\sqrt{T}$. When $d=0$ (resp., $d\neq0$)
the expansion can be used to approximate the size (resp., power) of
the $t$-statistic. Since $V_{t}$ is autocorrelated, $\widehat{\beta}$
and $\widehat{\Omega}_{b}$ are statistically dependent. Thus, we
decompose $\widehat{\beta}$ and $\widehat{\Omega}_{b}$ into statistically
independent components. Let $V=(V_{1},\ldots,\,V_{T})'$, $y=(y_{1},\ldots,\,y_{T})'$,
$l_{T}=\left(1,\ldots,\,1\right)'$ and $\Upsilon_{T}=\mathrm{Var}\left(V\right)$.
The GLS estimator of $\beta$ is $\widetilde{\beta}=(l'_{T}\Upsilon_{T}^{-1}l_{T})^{-1}l'_{T}\Upsilon_{T}^{-1}y$.
Then, 
\begin{align}
\widehat{\beta}-\beta & =\widetilde{\beta}-\beta+\left(l'_{T}l_{T}\right)^{-1}l'_{T}\widetilde{V},\label{Eq. (42) PSJ (2008)}
\end{align}
 where $\widetilde{V}=(I-l_{T}(l'_{T}\Upsilon_{T}^{-1}l_{T})^{-1}l'_{T}\Upsilon_{T}^{-1})V$,
which is statistically independent of $\widetilde{\beta}-\beta$.
Since $\widehat{\Omega}_{b}$ can be written as a quadratic form in
$\widetilde{V}$, it is also statistically independent of $\widetilde{\beta}-\beta$.
From \citeReferencesSupp{casini_hac}
\begin{align}
\Omega_{T} & \triangleq\mathrm{Var}\left(\sqrt{T}\left(\widehat{\beta}-\beta\right)\right)=T^{-1}l'_{T}\Upsilon_{T}l_{T}=\Omega+O\left(T^{-1}\right),\label{Eq. (43) SPJ (2007)}
\end{align}
where $\Omega=2\pi\int_{0}^{1}f\left(u,\,0\right)du$. Similarly,
one can show that 
\begin{align}
\widetilde{\Omega}_{T} & \triangleq\mathrm{Var}\left(\sqrt{T}\left(\widetilde{\beta}-\beta\right)\right)=T\left(l'_{T}\Upsilon_{T}^{-1}l_{T}\right)^{-1}=\Omega+O\left(T^{-1}\right).\label{Eq. (44) SPJ (2007)}
\end{align}
Therefore $T^{-1/2}l'_{T}\widetilde{V}=\mathscr{N}(0,\,O(T^{-1}))$.
As in eq. (45) in \citeReferencesSupp{sun/phillips/jin:08}, using
the independence of $\widetilde{\beta}$ and $(\widetilde{V},\,\widehat{\Omega}_{b})$,
we have  
\begin{align}
\mathbb{P} & \left(\sqrt{T}\left(\widehat{\beta}-\beta_{0}\right)/\sqrt{\widehat{\Omega}_{b}}\leq z\right)\label{Eq. (45) SPJ (2008)}\\
 & =\mathbb{P}\left(\sqrt{T}\left(\widetilde{\beta}-\beta\right)/\sqrt{\widetilde{\Omega}_{T}}+d/\sqrt{\widetilde{\Omega}_{T}}\leq z\sqrt{\widehat{\Omega}_{b}/\widetilde{\Omega}_{T}}\right)+O\left(T^{-1}\right),\nonumber 
\end{align}
uniformly over $z\in\mathbb{R}$ where $\Phi$ and $\varphi$ are
the cdf and pdf of the standard normal distribution, respectively.

Similarly, uniformly over $z\in\mathbb{R}$, we have
\begin{align*}
\mathbb{P}\left(\sqrt{T}\left(\widehat{\beta}-\beta_{0}\right)/\sqrt{\widehat{\Omega}_{b}}\leq-z\right) & =\mathbb{P}\left(\sqrt{T}\left(\widetilde{\beta}-\beta\right)/\sqrt{\widetilde{\Omega}_{T}}+c/\sqrt{\widetilde{\Omega}_{T}}\leq-z\sqrt{\widehat{\Omega}_{b}/\widetilde{\Omega}_{T}}\right)+O\left(T^{-1}\right).
\end{align*}
It follows that 
\begin{align}
Z_{T,d}\left(z\right) & \mathbb{=P}\left(\left|\sqrt{T}\left(\widehat{\beta}-\beta_{0}\right)/\sqrt{\widehat{\Omega}_{b}}\right|\leq z\right)\label{Eq. (46) in SPJ (2008)}\\
 & =\mathbb{P}\left(\left(\sqrt{T}\left(\widetilde{\beta}-\beta\right)/\sqrt{\widetilde{\Omega}_{T}}+d/\sqrt{\widetilde{\Omega}_{T}}\right)^{2}\leq z^{2}\widehat{\Omega}_{b}/\widetilde{\Omega}_{T}\right)+O\left(T^{-1}\right)\nonumber \\
 & =\mathbb{E}\left(G_{\widetilde{d}}\left(z^{2}\widehat{\Omega}_{b}/\widetilde{\Omega}_{T}\right)\right)=\mathbb{\mathbb{E}}\left(G_{\widetilde{d}}\left(z^{2}\zeta_{b,T}\right)\right)+O\left(T^{-1}\right),\nonumber 
\end{align}
uniformly over $z\in\mathbb{R}_{+},$ where $G_{\widetilde{d}}\left(z\right)=G(z;\,\widetilde{d})$
is the cdf of a non-central chi-squared $\chi_{1}(\widetilde{d}^{2})$
with noncentrality parameter $\widetilde{d}^{2}=d^{2}/\Omega_{T}$
and $\zeta_{b,T}=\widehat{\Omega}_{b}/\Omega_{T}$. Note that $\zeta_{b,T}\Rightarrow\mathscr{G}_{b}/\Omega$.
  Let $\mu_{b,T}=\mathbb{E}(\zeta_{b,T})$ and consider a fourth-order
Taylor expansion of $\zeta_{b,T}$ around its mean, 
\begin{align}
Z_{T,d}\left(z\right) & =G_{\widetilde{d}}\left(\mu_{b,T}z^{2}\right)+\frac{1}{2}G_{\widetilde{d}}''\left(\mu_{b,T}z^{2}\right)\mathbb{E}\left(\zeta_{b,T}-\mu_{b,T}\right)^{2}z^{4}\label{Eq. (47) SPJ (2008)}\\
 & \quad+\frac{1}{6}G_{\widetilde{d}}'''\left(\mu_{b,T}z^{2}\right)\mathbb{E}\left(\zeta_{b,T}-\mu_{b,T}\right)^{3}z^{6}+O\left(\mathbb{E}\left(\zeta_{b,T}-\mu_{b,T}\right)^{4}\right)+O\left(T^{-1}\right),\nonumber 
\end{align}
where the $O\left(\cdot\right)$ term holds uniformly over $z\in\mathbb{R}_{+}$. 

Using \eqref{Eq. (46) in SPJ (2008)} we have 
\begin{align}
Z_{T,0}\left(z\right)-Z_{0}\left(z\right) & =\mathbb{P}\left(\left|\frac{\sqrt{T}\left(\widehat{\beta}-\beta_{0}\right)}{\sqrt{\widehat{\Omega}_{b}}}\right|\leq z\right)-\mathbb{P}\left(\left|\frac{\int_{0}^{1}\Sigma\left(u\right)dW_{1}\left(u\right)}{\sqrt{\mathscr{G}_{b}}}\right|\leq z\right)\label{Eq. (58) SPJ (2008)}\\
 & =\mathbb{E}\left(F_{\chi}\left(z^{2}\zeta_{b,T}\right)\right)-\mathbb{E}\left(F_{\chi}\left(z^{2}\mathscr{G}_{b}/\Omega\right)\right)+O\left(T^{-1}\right),\nonumber 
\end{align}
where $F_{\chi}\left(\cdot\right)=G_{0}\left(\cdot\right)$ is the
cdf of the $\chi_{1}^{2}$ distribution. Next, we compute the cumulants
of both $\zeta_{b,T}-\mu_{b,T}$ and $\Omega^{-1}\left(\mathscr{G}_{b}-\mu_{b}\right)$
where $\mu_{b}=\mathbb{E}(\mathscr{G}_{b})$. Note that $\zeta_{b,T}$
is a quadratic form in a Gaussian vector since $\widehat{\Omega}_{b}=T^{-1}\widehat{V}'W_{b}\widehat{V}=T^{-1}V'A_{T}W_{b}A_{T}V$,
where $W_{b}$ is $T\times T$ with $\left(j,\,s\right)$-th element
$K_{b}\left(\left(j-s\right)/T\right)$ and $A_{T}=I_{T}-l_{T}l'_{T}/T$.
The characteristic function of $\zeta_{b,T}-\mu_{b,T}$ is given by
 
\begin{align}
\psi_{b,T}\left(t\right) & =\left|I-2it\frac{\Upsilon_{T}A_{T}W_{b}A_{T}}{T\Omega_{T}}\right|^{-1/2}\exp\left(-it\mu_{b,T}\right),\label{Eq. (59) in SPJ (2008)}
\end{align}
 where $\Upsilon_{T}=\mathbb{E}(uu')$ and the cumulant generating
function is 
\begin{align}
\log\left(\psi_{b,T}\left(t\right)\right) & =-\frac{1}{2}\log\left|I-2it\frac{\Upsilon_{T}A_{T}W_{b}A_{T}}{T\Omega_{T}}\right|-it\mu_{b,T}=\sum_{m=1}^{\infty}\kappa_{m,T}\frac{\left(it\right)^{m}}{m!},\label{Eq. (60) in SPJ (2008)}
\end{align}
 where $\kappa_{m,T}$ is the $m$th cumulant of $\zeta_{b,T}-\mu_{b,T}$.
Note that $\kappa_{1,T}=0$ and 
\begin{align}
\kappa_{m,T} & =2^{m-1}\left(m-1\right)!T^{-m}\left(\Omega_{T}\right)^{-m}\mathrm{Tr}\left(\left(\Upsilon_{T}A_{T}W_{b}A_{T}\right)^{m}\right),\qquad m\geq2.\label{Eq. (61) SPJ (2008)}
\end{align}

\begin{lem}
\label{Lemma 2 in SPJ (2008)}Let Assumption \ref{Assumption 3 in Sun et al.}-\ref{Assumption: Assumption 2 in SPJ (2008)}
hold.   We have: (i) $\mu_{b,T}=\Omega^{-1}\mu_{b}+O\left(T^{-1}\right);$
(ii) $\kappa_{m,T}=\kappa_{m}+O(m!2^{m}T^{-2}(\overline{C}_{1}b)^{m-2})$
uniformly over $m\geq1;$ (iii) $\Xi_{m,T}=\mathbb{E}(\zeta_{b,T}-\mu_{b,T})^{m}=\Xi_{m}+O(m!2^{2m}T^{-2}(\overline{C}_{1}b)^{m-2})$. 
\end{lem}
\noindent\textit{Proof of Lemma} \ref{Lemma 2 in SPJ (2008)}. Note
that $\mu_{b,T}=(T\Omega_{T})^{-1}\mathrm{Tr}(\Upsilon_{T}A_{T}W_{b}A_{T})$.
Let $\widetilde{W}_{b}=A_{T}W_{b}A_{T}$, where its $\left(j,\,s\right)$-th
element is given by 
\begin{align}
\widetilde{K}_{b}\left(\frac{j}{T},\,\frac{s}{T}\right) & =K_{b}\left(\frac{j-s}{T}\right)-\frac{1}{T}\sum_{p=1}^{T}K_{b}\left(\frac{j-p}{T}\right)\label{Eq. (A.15) in SPJ (2008)}\\
 & \quad-\frac{1}{T}\sum_{q=1}^{T}K_{b}\left(\frac{q-s}{T}\right)+\frac{1}{T^{2}}\sum_{p=1}^{T}\sum_{q=1}^{T}K_{b}\left(\frac{p-q}{T}\right).\nonumber 
\end{align}
 Let $\Gamma_{r_{1}/T}\left(r_{1}-r_{2}\right)=\mathbb{E}(V_{r_{1}}V_{r_{2}}).$
 We have 
\begin{align}
\mathrm{Tr} & \left(\Upsilon_{T}\widetilde{W}_{b}\right)\label{Eq. (A.16) SPJ (2008)}\\
 & =\sum_{1\leq r_{1},\,r_{2}\leq T}\mathbb{E}\left(V_{r_{1}}V_{r_{2}}\right)\widetilde{K}_{b}\left(\frac{r_{1}}{T},\,\frac{r_{2}}{T}\right)\nonumber \\
 & =\sum_{r_{2}=1}^{T}\sum_{h=1-r_{2}}^{T-r_{2}}\Gamma_{r_{2}/T}\left(-h\right)\widetilde{K}_{b}\left(\frac{r_{2}+h}{T},\,\frac{r_{2}}{T}\right)\nonumber \\
 & =\left(\sum_{h=1}^{T-1}\sum_{r_{2}=1}^{T-h}+\sum_{h=1-T}^{0}\sum_{r_{2}=1-h}^{T}\right)\Gamma_{r_{2}/T}\left(-h\right)\widetilde{K}_{b}\left(\frac{r_{2}+h}{T},\,\frac{r_{2}}{T}\right).\nonumber 
\end{align}
 Using the Lipschitz property of $K\left(\cdot\right)$ and the fact
that $\sup_{r_{2},\,h}|\Gamma_{r_{2}/T}\left(-h\right)|<\infty$ which
follows from Assumption \ref{Assumption 3 in Sun et al.}, some algebra
shows that 
\begin{align}
\sum_{r_{2}=1}^{T-h} & \Gamma_{r_{2}/T}\left(-h\right)\widetilde{K}_{b}\left(\frac{r_{2}+h}{T},\,\frac{r_{2}}{T}\right)\label{Eq. (A.17) SPJ (2008)}\\
 & =\sum_{r_{2}=1}^{T-h}\Gamma_{r_{2}/T}\left(-h\right)K_{b}\left(\frac{h}{T}\right)-\frac{1}{T}\sum_{r_{2}=1}^{T-h}\sum_{p=1}^{T}\Gamma_{r_{2}/T}\left(-h\right)K_{b}\left(\frac{r_{2}+h-p}{T}\right)\nonumber \\
 & \quad-\frac{1}{T}\sum_{r_{2}=1}^{T-h_{1}}\sum_{q=1}^{T}\Gamma_{r_{2}/T}\left(-h\right)K_{b}\left(\frac{q-r_{2}}{T}\right)+\sum_{r_{2}=1}^{T-h_{1}}\frac{1}{T^{2}}\sum_{p=1}^{T}\sum_{q=1}^{T}\Gamma_{r_{2}/T}\left(-h\right)K_{b}\left(\frac{p-q}{T}\right)\nonumber \\
 & =-\frac{1}{T}\sum_{r_{2}=1}^{T}\sum_{p=1}^{T}\Gamma_{r_{2}/T}\left(-h\right)K_{b}\left(\frac{r_{2}-p}{T}\right)-\frac{1}{T}\sum_{r_{2}=1}^{T}\sum_{q=1}^{T}\Gamma_{r_{2}/T}\left(-h\right)K_{b}\left(\frac{q-r_{2}}{T}\right)\nonumber \\
 & \quad+\sum_{r_{2}=1}^{T}\frac{1}{T^{2}}\sum_{p=1}^{T}\sum_{q=1}^{T}\Gamma_{r_{2}/T}\left(-h\right)K_{b}\left(\frac{p-q}{T}\right)+\sum_{r_{2}=1}^{T-h}\Gamma_{r_{2}/T}\left(-h\right)K_{b}\left(\frac{h}{T}\right)+O\left(|h|\right)\nonumber \\
 & =-\frac{1}{T}\sum_{r_{2}=1}^{T}\sum_{p=1}^{T}\Gamma_{r_{2}/T}\left(-h\right)K_{b}\left(\frac{r_{2}-p}{T}\right)+\sum_{r_{2}=1}^{T-h}\Gamma_{r_{2}/T}\left(-h\right)K_{b}\left(\frac{h}{T}\right)+O\left(|h|\right)+o\left(1\right)\nonumber \\
 & =\sum_{r_{2}=1}^{T}\Gamma_{r_{2}/T}\left(-h\right)K_{b}\left(0\right)-\frac{1}{T}\sum_{r_{2}=1}^{T}\sum_{p=1}^{T}\Gamma_{r_{2}/T}\left(-h\right)K_{b}\left(\frac{r_{2}-p}{T}\right)\nonumber \\
 & \quad+\sum_{r_{2}=1}^{T}\Gamma_{r_{2}/T}\left(-h\right)\left(K_{b}\left(\frac{h}{T}\right)-K_{b}\left(0\right)\right)+O\left(|h|\right)+o\left(1\right)\nonumber \\
 & =\sum_{r_{2}=1}^{T}\Gamma_{r_{2}/T}\left(-h\right)\widetilde{K}_{b}\left(\frac{r_{2}}{T},\,\frac{r_{2}}{T}\right)+\sum_{r_{2}=1}^{T}\Gamma_{r_{2}/T}\left(-h\right)\left(K_{b}\left(\frac{h}{T}\right)-K_{b}\left(0\right)\right)+O\left(|h|\right)+o\left(1\right).\nonumber 
\end{align}
The same arguments yield 
\begin{align}
\sum_{r_{2}=1-h}^{T} & \Gamma_{r_{2}/T}\left(-h\right)\widetilde{K}_{b}\left(\frac{r_{2}+h}{T},\,\frac{r_{2}}{T}\right)\label{Eq. (A.18) in SPJ (2008)}\\
 & =\sum_{r_{2}=1}^{T}\Gamma_{r_{2}/T}\left(-h\right)\widetilde{K}_{b}\left(\frac{r_{2}}{T},\,\frac{r_{2}}{T}\right)+\sum_{r_{2}=1}^{T}\Gamma_{r_{2}/T}\left(-h\right)\left(K_{b}\left(\frac{h}{T}\right)-K_{b}\left(0\right)\right)+O\left(|h|\right)+o\left(1\right).\nonumber 
\end{align}
Using \eqref{Eq. (A.17) SPJ (2008)}-\eqref{Eq. (A.18) in SPJ (2008)}
into \eqref{Eq. (A.16) SPJ (2008)}, we have 
\begin{align}
\mathrm{Tr} & \left(\Upsilon_{T}\widetilde{W}_{b}\right)\label{Eq. (A.19) in PSJ (2008)}\\
 & =\sum_{h=-\infty}^{\infty}\sum_{r_{2}=1}^{T}\Gamma_{r_{2}/T}\left(-h\right)\widetilde{K}_{b}\left(\frac{r_{2}}{T},\,\frac{r_{2}}{T}\right)+\sum_{h=-\infty}^{\infty}\sum_{r_{2}=1}^{T}\Gamma_{r_{2}/T}\left(-h\right)\left(K_{b}\left(\frac{h}{T}\right)-K_{b}\left(0\right)\right)+O\left(1\right)\nonumber \\
 & =\sum_{h=-\infty}^{\infty}\sum_{r_{2}=1}^{T}\Gamma_{r_{2}/T}\left(-h\right)\widetilde{K}_{b}\left(\frac{r_{2}}{T},\,\frac{r_{2}}{T}\right)+\left(Tb\right)^{-q}\sum_{r_{2}=1}^{T}\sum_{h=-\infty}^{\infty}\left|h\right|^{q}\Gamma_{r_{2}/T}\left(-h\right)\left(\frac{K\left(h/Tb\right)-K\left(0\right)}{\left|h/\left(Tb\right)\right|^{q}}\right)+O\left(1\right)\nonumber \\
 & =\sum_{h=-\infty}^{\infty}\sum_{r_{2}=1}^{T}\Gamma_{r_{2}/T}\left(-h\right)\widetilde{K}_{b}\left(\frac{r_{2}}{T},\,\frac{r_{2}}{T}\right)+\left(Tb\right)^{-q}q_{0}\sum_{r_{2}=1}^{T}\sum_{h=-\infty}^{\infty}\left|h\right|^{q}\Gamma_{r_{2}/T}\left(-h\right)\left(1+o\left(1\right)\right)+O\left(1\right).\nonumber 
\end{align}
By Theorem 2.1 in \citeReferencesSupp{casini_hac} and Lemma 4.1
in \citeReferencesSupp{casini/perron_Low_Frequency_Contam_Nonstat:2020},
\begin{align}
\sum_{h=-\infty}^{\infty}\Gamma_{s/T}\left(-h\right) & =2\pi f\left(s/T,\,0\right)\left(1+O\left(\frac{1}{T}\right)\right),\label{Eq. (A.20) in SPJ (2008)}
\end{align}
and 
\begin{align}
\frac{1}{T}\sum_{h=-\infty}^{\infty}\sum_{s=1}^{T}\Gamma_{s/T}\left(-h\right)\widetilde{K}_{b}\left(\frac{s}{T},\,\frac{s}{T}\right) & =\int_{0}^{1}\Omega\left(u\right)K_{b}^{*}\left(u,\,u\right)du+O\left(\frac{1}{T}\right),\label{Eq. (A.21) SPJ (2008)}
\end{align}
where we have used $2\pi f\left(u,\,0\right)=\Omega\left(u\right).$
Using Assumption \ref{Assumption 3 in Sun et al.} and \eqref{Eq. (A.21) SPJ (2008)},
we yield
\begin{align}
\mu_{b,T} & =\Omega^{-1}\int_{0}^{1}\Omega\left(s\right)K_{b}^{*}\left(s,\,s\right)ds\label{Eq. (A.22) in SPJ (2008)}\\
 & +\left(Tb\right)^{-q}q_{0}\left(\Omega_{T}^{-1}T^{-1}\sum_{r_{2}=1}^{T}\sum_{h=-\infty}^{\infty}\left|h\right|^{q}\Gamma_{r_{2}/T}\left(-h\right)\right)\left(1+o\left(1\right)\right)+O\left(\frac{1}{T}\right).\nonumber 
\end{align}
Since $\mu_{b}=\Omega^{-1}\mathbb{E}(\mathscr{G}_{b})=\Omega^{-1}\int_{0}^{1}K_{b}^{*}\left(s,\,s\right)\Omega\left(s\right)ds$,
$b$ is fixed and $q\geq1$, we have $\mu_{b,T}=\mu_{b}+O(T^{-1})$. 

Next, we consider part (ii). For $m>1$, let $r_{2m+1}=r_{1}$, $r_{2m+2}=r_{2}$
and $h_{m+1}=h_{1}$. Using the same argument used for the case $m=1$
in \eqref{Eq. (A.17) SPJ (2008)} and using eq. (A.26) in \citeReferencesSupp{sun/phillips/jin:08},
we have
\begin{align}
\mathrm{Tr}\left(\left(\Upsilon_{T}\widetilde{W}_{b}\right)^{m}\right) & =\sum_{r_{1},\,r_{2},\ldots,\,r_{2m+1}=1}^{T}\prod_{j=1}^{m}\mathbb{E}\left(V_{r_{2j-1}}V_{r_{2j}}\right)\widetilde{K}_{b}\left(\frac{r_{2j}}{T},\,\frac{r_{2j+1}}{T}\right)\label{Eq. (A.23) SPJ (2008)}\\
 & =\sum_{r_{1},\,r_{2},\ldots,\,r_{2m+1}=1}^{T}\sum_{h_{1}=1-r_{2}}^{T-r_{2}}\sum_{h_{2}=1-r_{4}}^{T-r_{4}}\cdots\sum_{h_{m}=1-r_{2m}}^{T-r_{2m}}\prod_{j=1}^{m}\Gamma_{r_{2j}/T}\left(-h_{j}\right)\widetilde{K}_{b}\left(\frac{r_{2j}}{T},\,\frac{r_{2j+1}+h_{j+1}}{T}\right)\nonumber \\
 & =\left(\sum_{h_{1}=1}^{T-1}\sum_{r_{2}=1}^{T-h_{1}}+\sum_{h_{1}=1-T}^{0}\sum_{r_{2}=1-h_{1}}^{T}\right)\cdots\left(\sum_{h_{m}=1}^{T-1}\sum_{r_{2m}=1}^{T-h_{m}}+\sum_{h_{m}=1-T}^{0}\sum_{r_{2m}=1-h_{m}}^{T}\right)\nonumber \\
 & \quad\times\prod_{j=1}^{m}\Gamma_{r_{2j}/T}\left(-h_{j}\right)\widetilde{K}_{b}\left(\frac{r_{2j}}{T},\,\frac{r_{2j+1}+h_{j+1}}{T}\right)\nonumber \\
 & =L_{1}+L_{2},\nonumber 
\end{align}
 where
\begin{align}
L_{1} & =\left(\sum_{h_{1}=1}^{T-1}\sum_{r_{2}=1}^{T-h_{1}}+\sum_{h_{1}=1-T}^{0}\sum_{r_{2}=1-h_{1}}^{T}\right)\cdots\left(\sum_{h_{m}=1}^{T-1}\sum_{r_{2m}=1}^{T-h_{m}}+\sum_{h_{m}=1-T}^{0}\sum_{r_{2m}=1-h_{m}}^{T}\right)\label{Eq. (A.24) SPJ (2008)}\\
 & \quad\prod_{j=1}^{m}\Gamma_{r_{2j}/T}\left(-h_{j}\right)\widetilde{K}_{b}\left(\frac{r_{2j}}{T},\,\frac{r_{2j+1}+h_{j+1}}{T}\right),\nonumber 
\end{align}
 and 
\begin{align}
L_{2} & =O\left(\left(\sum_{h_{1}=1}^{T-1}\sum_{r_{2}=1}^{T-h_{1}}+\sum_{h_{1}=1-T}^{0}\sum_{r_{2}=1-h_{1}}^{T}\right)\right.\cdots\left(\sum_{h_{m}=1}^{T-1}\sum_{r_{2m}=1}^{T-h_{m}}+\sum_{h_{m}=1-T}^{0}\sum_{r_{2m}=1-h_{m}}^{T}\right)\label{Eq. (A.25) in SPJ (2008)}\\
 & \quad\prod_{j=1}^{m}\Gamma_{r_{2j}/T}\left(-h_{j}\right)\left(\frac{\left|h_{j+1}\right|}{Tb}\right)\Biggr).\nonumber 
\end{align}
Using the same arguments as in eq. (A.30)-(A.31) in \citeReferencesSupp{sun/phillips/jin:08},
we can show that 
\begin{align*}
L_{1} & =\sum_{h=-\infty}^{\infty}\sum_{r}\prod_{j=1}^{m}\Gamma_{r_{2j}/T}\left(-h_{j}\right)\left(\widetilde{K}_{b}\left(\frac{r_{2j}}{T},\,\frac{r_{2j+2}}{T}\right)\right)+O\left(2mT^{m-2}\left(\overline{C}_{1}b\right)^{m-2}\right),
\end{align*}
where $O(2mT^{m-2}((\overline{C}_{1}b))^{m-2})$ follows from 
\begin{align*}
\sum_{h_{1}=-\infty}^{\infty} & \sum_{r_{2}=1}^{T}\cdots\sum_{h_{a}=-\infty}^{\infty}\sum_{r_{2a}=1}^{-h_{a}}\cdots\sum_{h_{m}=-\infty}^{\infty}\sum_{r_{2m}=1}^{T}\prod_{j=1}^{m}\left(\sup_{r_{2j}}\left|\Gamma_{r_{2j}/T}\left(-h_{j}\right)\right|\right)\left|h_{a}\right|\prod_{j\neq a}\left|\widetilde{K}_{b}\left(\frac{r_{2j}}{T},\,\frac{r_{2j+2}}{T}\right)\right|\\
 & \leq\left[\sup_{t}\sum_{s=1}^{T}\widetilde{K}_{b}\left(\frac{s}{T},\,\frac{t}{T}\right)\right]^{m-2}\left(\sum_{h=-\infty}^{\infty}\sup_{s}\left|\Gamma_{s/T}\left(-h\right)\right|\right)^{m-1}\left(\sum_{h_{a}=-\infty}^{\infty}\sup_{s}\left|\Gamma_{s/T}\left(-h_{a}\right)\right|\left|h_{a}\right|\right)\\
 & \leq O\left(2mT^{m-2}\left(\overline{C}_{1}b\right)^{m-2}\right),
\end{align*}
 uniformly over $m$ for some integer $a$ such that $1\leq a\leq m$.
A similar argument yields that $L_{2}=o(2mT^{m-2}(\overline{C}_{1}b)^{m-2})$
uniformly over $m.$ Thus, 
\begin{align*}
\mathrm{Tr}\left(\left(\Upsilon_{T}\widetilde{W}_{b}\right)^{m}\right) & =\sum_{h=-\infty}^{\infty}\sum_{r}\prod_{j=1}^{m}\Gamma_{r_{2j}/T}\left(-h_{j}\right)\left(\widetilde{K}_{b}\left(\frac{r_{2j}}{T},\,\frac{r_{2j+2}}{T}\right)\right)+O\left(2mT^{m-2}\left(\overline{C}_{1}b\right)^{m-2}\right),
\end{align*}
 and using $\tau_{1}=\tau_{m+1}$ we yield
\begin{align}
\kappa_{m,T} & =2^{m-1}\left(m-1\right)!T^{-m}\Omega_{T}^{-m}\mathrm{Tr}\left(\left(\Upsilon_{T}A_{T}W_{b}A_{T}\right)^{m}\right)\label{Eq. (A.33) SPJ (2008)}\\
 & =2^{m-1}\left(m-1\right)!\Omega_{T}^{-m}\left(T^{-m}\sum_{r}\Omega\left(r_{2j}/T\right)\widetilde{K}_{b}\left(\frac{r_{2j}}{T},\,\frac{r_{2j+2}}{T}\right)+O\left(2mT^{-2}\left(\overline{C}_{1}b\right)^{m-2}\right)\right)\nonumber \\
 & =2^{m-1}\left(m-1\right)!\Omega^{-m}\left(\int_{0}^{1}\cdots\int_{0}^{1}\left(\prod_{j=1}^{m}\Omega\left(\tau_{j}\right)K_{b}^{*}\left(\tau_{j},\,\tau_{j+1}\right)\right)d\tau_{1}\cdots d\tau_{m}+O\left(2mT^{-2}\left(\overline{C}_{1}b\right)^{m-2}\right)\right)\nonumber \\
 & =\kappa_{m}+O\left(2mT^{-2}\left(\overline{C}_{1}b\right)^{m-2}\right),\nonumber 
\end{align}
uniformly over $m$. 

Next, we consider part (iii). From \eqref{Eq. (A.8) in SPJ (2008)}
and part (ii), we have uniformly over $m$,
\begin{align}
\Xi_{m,T} & =\Xi_{m}+O\left(\frac{m!2^{m}}{T^{2}}\left(\overline{C}_{1}b\right)^{m-2}\sum_{\pi}\frac{m!}{m_{1}!m_{2}!\cdots m_{k}!}\right)\label{Eq. (A.35) in SPJ (2008)}\\
 & =\Xi_{m}+O\left(\frac{m!2^{2m}}{T^{2}}\left(\overline{C}_{1}b\right)^{m-2}\right),\nonumber 
\end{align}
 where we have used $\sum_{\pi}\frac{m!}{m_{1}!m_{2}!\cdots m_{k}!}<2^{m}$.
$\square$ \medskip{}

\noindent\textit{Proof of Theorem }\ref{Theorem ERP Fixedb}. Let
$F_{\chi}^{\left(m\right)}\left(\cdot\right)$ denote the $m$th derivative
of $F_{\chi}\left(\cdot\right)$. Since $F_{\chi}\left(\cdot\right)$
is a bounded function, we can write  
\begin{align}
\mathbb{P}\left(\left|\frac{\int_{0}^{1}\Sigma\left(u\right)dW_{1}\left(u\right)}{\sqrt{\mathscr{G}_{b}}}\right|\leq z\right) & =\lim_{C\rightarrow\infty}\mathbb{E}\left(F_{\chi}\left(z^{2}\mathscr{G}_{b}/\Omega\right)\mathbf{1}\left(\left|\mathscr{G}_{b}-\mu_{b}\right|\leq\Omega C\right)\right)\label{Eq. (A.93) in SPJ (2008)}\\
 & =\lim_{C\rightarrow\infty}\mathbb{E}\sum_{m=1}^{\infty}\frac{1}{m!}F_{\chi}^{\left(m\right)}\left(\mu_{b}z^{2}/\Omega\right)\Omega^{-m}\left(\mathscr{G}_{b}-\mu_{b}\right)^{m}z^{2m}\mathbf{1}\left\{ \left|\mathscr{G}_{b}-\mu_{b}\right|\leq\Omega C\right\} \nonumber \\
 & =\lim_{C\rightarrow\infty}\sum_{m=1}^{\infty}\frac{1}{m!}F_{\chi}^{\left(m\right)}\left(\mu_{b}z^{2}/\Omega\right)\Xi_{m}z^{2m}\mathbf{1}\left\{ \left|\mathscr{G}_{b}-\mu_{b}\right|\leq\Omega C\right\} ,\nonumber 
\end{align}
where $\Xi_{m}=\Omega^{-m}\mathbb{E}((\mathscr{G}_{b}-\mu_{b})^{m})$.
Since $F_{\chi}\left(z^{2}\right)$ decays exponentially as $z\rightarrow\infty$,
there exists a constant $C_{2}>0$ such that $|F_{\chi}^{\left(m\right)}(\mu_{b}z^{2}/\Omega)z^{2m}|<C_{2}$
for all $m$.  Using Lemma \ref{Lemma: Lemma 1 SPJ (2008)}, we yield
\begin{align}
\left|\sum_{m=1}^{\infty}\frac{1}{m!}F_{\chi}^{\left(m\right)}\left(\mu_{b}z^{2}/\Omega\right)\Xi_{m}z^{2m}\right| & \leq C_{2}\sum_{m=1}^{\infty}\frac{1}{m!}\Xi_{m}\leq C_{2}D\sum_{m=1}^{\infty}\frac{1}{m!}2^{2m}m!C_{\Omega}^{m}\left(\overline{C}_{1}b\right)^{m-1}\label{Eq. (A.94) in SPJ (2008)}\\
 & =C_{2}D\left(\overline{C}_{1}b\right)^{-1}\sum_{m=1}^{\infty}\left(4C_{\Omega}\overline{C}_{1}b\right)^{m},\nonumber 
\end{align}
 where $D_{m}\leq D$ for some $D<\infty$. The right-hand side of
\eqref{Eq. (A.94) in SPJ (2008)} is bounded in view of $b<1/(4C_{\Omega}\overline{C}_{1})$.
This implies  that
\begin{align}
\mathbb{P}\left(\left|\frac{\int_{0}^{1}\Sigma\left(u\right)dW_{1}\left(u\right)}{\sqrt{\mathscr{G}_{b}}}\right|\leq z\right) & =\sum_{m=1}^{\infty}\frac{1}{m!}F_{\chi}^{\left(m\right)}\left(\mu_{b}z^{2}/\Omega\right)\Xi_{m}z^{2m},\label{Eq. (A.95) in SPJ (2008)}
\end{align}
 provided that $b<1/(4C_{\Omega}\overline{C}_{1})$. 

From \eqref{Eq. (46) in SPJ (2008)} we have 
\begin{align}
Z_{T,0}\left(z\right)=\mathbb{P}\left(\left|\frac{\sqrt{T}\left(\widehat{\beta}-\beta_{0}\right)}{\sqrt{\widehat{\Omega}_{b}}}\right|\leq z\right) & =\mathbb{E}\left(F_{\chi}\left(z^{2}\zeta_{b,T}\right)\right)+O\left(T^{-1}\right).\label{Eq. (A.96) in SPJ (2008)}
\end{align}
 Using a similar argument as in \eqref{Eq. (A.93) in SPJ (2008)},
\begin{align}
\mathbb{E}\left(F_{\chi}\left(z^{2}\zeta_{b,T}\right)\right)- & \sum_{m=1}^{\infty}\frac{1}{m!}F_{\chi}^{\left(m\right)}\left(\mu_{b,T}z^{2}\right)\Xi_{m,T}z^{2m}\overset{}{\rightarrow}0,\label{Eq. (A.97) in SPJ (2008)}
\end{align}
 uniformly over $T$ since by Lemma \ref{Lemma 2 in SPJ (2008)}-(iii)
we have 
\begin{align*}
\Xi_{m,T} & =\Xi_{m}+O\left(\frac{2^{2m}m!}{T^{2}}\left(\overline{C}_{1}b\right)^{m-2}\right),
\end{align*}
uniformly in $m$ and $|F_{\chi}^{\left(m\right)}(\mu_{b,T}z^{2})z^{2m}|<C_{2}$
for some constant $C_{2}>0$ for all $m$ so that
\begin{align*}
\left|\sum_{m=1}^{\infty}\frac{1}{m!}F_{\chi}^{\left(m\right)}\left(\mu_{b,T}z^{2}\right)\Xi_{m,T}z^{2m}\right| & \leq C_{2}\sum_{m=1}^{\infty}\frac{1}{m!}\left|\Xi_{m}\right|+\frac{C_{2}}{T^{2}}\sum_{m=1}^{\infty}2^{2m}\left(\overline{C}_{1}b\right)^{m-2}<\infty,
\end{align*}
 provided that $b<1/(4\overline{C}_{1})$. Note that $b<1/(16C_{2,\Omega}\int_{-\infty}^{\infty}|k\left(x\right)|dx)<1/(4\overline{C}_{1})$
by assumption. It follows that
\begin{align}
Z_{T,0}\left(z\right) & =\sum_{m=1}^{\infty}\frac{1}{m!}F_{\chi}^{\left(m\right)}\left(\mu_{b,T}z^{2}\right)\Xi_{m,T}z^{2m}+O\left(T^{-1}\right),\label{Eq. (A.98) in SPJ (2008)}
\end{align}
 uniformly over $z\in\mathbb{R}_{+}.$ 

By Lemma \ref{Lemma 2 in SPJ (2008)}-(i), we have 
\begin{align}
F_{\chi}^{\left(m\right)}\left(\mu_{b,T}z^{2}\right) & =F_{\chi}^{\left(m\right)}\left(\mu_{b}z^{2}/\Omega\right)+O\left(F_{\chi}^{\left(m+1\right)}\left(\mu_{b}z^{2}/\Omega\right)z^{2}T^{-1}\right).\label{Eq. (A.99b)}
\end{align}
 Combining \eqref{Eq. (A.95) in SPJ (2008)} and \eqref{Eq. (A.98) in SPJ (2008)}-\eqref{Eq. (A.99b)}
leads to
\begin{align}
|Z_{T,0}\left(z\right) & -Z_{0}\left(z\right)|\label{Eq. (A.100) in SPJ (2008)}\\
 & =\left|\sum_{m=1}^{\infty}\frac{1}{m!}F_{\chi}^{\left(m\right)}\left(\mu_{b,T}z^{2}\right)\Xi_{m,T}z^{2m}-\sum_{m=1}^{\infty}\frac{1}{m!}F_{\chi}^{\left(m\right)}\left(\mu_{b}z^{2}\right)\Xi_{m}z^{2m}\right|+O\left(\frac{1}{T}\right)\nonumber \\
 & =\left|\sum_{m=1}^{\infty}\frac{1}{m!}F_{\chi}^{\left(m\right)}\left(\mu_{b}z^{2}\right)z^{2m}\left(\Xi_{m,T}-\Xi_{m}\right)\right|+O\left(\frac{1}{T}\right)\nonumber \\
 & =\left|\sum_{m=1}^{\infty}\frac{1}{m!}F_{\chi}^{\left(m\right)}\left(\mu_{b}z^{2}\right)z^{2m}O\left(\frac{m!2^{2m}(\overline{C}_{1}b)^{m-2}}{T^{2}}\right)\right|+O\left(\frac{1}{T}\right)\nonumber \\
 & =O\left(\frac{1}{T^{2}}\sum_{m=1}^{\infty}2^{2m}(\overline{C}_{1}b)^{m-2}\right)+O\left(\frac{1}{T}\right)\nonumber \\
 & =O\left(\frac{1}{T}\right).\nonumber 
\end{align}
uniformly over $z\in\mathbb{R}$ where we have used Lemma \ref{Lemma 2 in SPJ (2008)}-(iii).
Hence, $D_{1}=O\left(T^{-1}\right).$ 

Let $\widehat{\Omega}=\int_{0}^{1}\widehat{\Omega}\left(u\right)du$
where $\widehat{\Omega}\left(u\right)$ was defined in \eqref{Eq. Sigma_hat(u)}.
Note that $\mathscr{\widehat{G}}_{b}=\mathscr{\mathscr{G}}_{b}+O((Th_{1}h_{2})^{-1/2})$
by definition of $\widehat{\Omega}\left(u\right).$ Using this and
proceeding as in \eqref{Eq. (A.93) in SPJ (2008)}, we yield 
\begin{align}
\widehat{Z}_{0}\left(z\right) & =\mathbb{P}\left(\left|\frac{\int_{0}^{1}\widehat{\Sigma}\left(u\right)dW_{p}\left(u\right)}{\sqrt{\mathscr{\widehat{G}}_{b}}}\right|\leq z\right)\label{Eq. (A.93) for Z0_hat}\\
 & =\lim_{C\rightarrow\infty}\mathbb{E}\left(F_{\chi}\left(z^{2}\widehat{\mathscr{G}}_{b}/\widehat{\Omega}\right)\mathbf{1}\left(\left|\mathscr{\widehat{G}}_{b}-\widehat{\mu}_{b}\right|\leq\widehat{\Omega}C\right)\right)\nonumber \\
 & =\lim_{C\rightarrow\infty}\mathbb{E}\sum_{m=1}^{\infty}\frac{1}{m!}F_{\chi}^{\left(m\right)}\left(\widehat{\mu}_{b}z^{2}/\widehat{\Omega}\right)\widehat{\Omega}^{-m}\left(\mathscr{\widehat{G}}_{b}-\widehat{\mu}_{b}\right)^{m}z^{2m}\mathbf{1}\left\{ \left|\mathscr{\widehat{G}}_{b}-\widehat{\mu}_{b}\right|\leq\Omega C\right\} \nonumber \\
 & =\lim_{C\rightarrow\infty}\mathbb{E}\sum_{m=1}^{\infty}\frac{1}{m!}F_{\chi}^{\left(m\right)}\left(\mu_{b}z^{2}/\Omega\right)\Omega^{-m}\left(\mathscr{G}_{b}-\mu_{b}\right)^{m}z^{2m}\mathbf{1}\left\{ \left|\mathscr{G}_{b}-\mu_{b}\right|\leq\Omega C\right\} +O((Th_{1}h_{2})^{-1/2})\nonumber \\
 & =\lim_{C\rightarrow\infty}\sum_{m=1}^{\infty}\frac{1}{m!}F_{\chi}^{\left(m\right)}\left(\mu_{b}z^{2}/\Omega\right)\Xi_{m}z^{2m}\mathbf{1}\left\{ \left|\mathscr{G}_{b}-\mu_{b}\right|\leq\Omega C\right\} +O\left(\left(Th_{1}h_{2}\right){}^{-1/2}\right),\nonumber 
\end{align}
uniformly in $z\in\mathbb{R}_{+}$. This implies $D_{2}=O((Th_{1}h_{2})^{-1/2})$.
$\square$ 

\bibliographystyleReferencesSupp{elsarticle-harv}
\bibliographyReferencesSupp{References_Supp}

\end{singlespace}

\setcounter{page}{14}
\renewcommand{\thepage}{S-\arabic{page}}
\end{document}